\documentclass[lettersize,journal]{IEEEtran}
\usepackage{amsmath,amsfonts}
\usepackage{algorithmic}
\usepackage{algorithm}
\usepackage{array}
\usepackage[caption=false,font=normalsize,labelfont=sf,textfont=sf]{subfig}
\usepackage{textcomp}
\usepackage{stfloats}
\usepackage{url}
\usepackage{verbatim}
\usepackage{graphicx}
\usepackage{cite}
\usepackage{xcolor}
\definecolor{cgcol}{rgb}{0,0,0}

\usepackage{orcidlink}

\DeclareMathOperator*{\argmin}{arg\,min}

\newcommand{\lt}{\left}
\newcommand{\rt}{\right}
\newcommand{\tb}{\textbf}

\begin{document}

\title{\textcolor{cgcol}{Maximum Likelihood based Phase-Retrieval using Fresnel Propagation Forward Models with Optional Constraints}}

\author{K.~Aditya~Mohan\textsuperscript{\orcidlink{0000-0002-0921-6559}},~\IEEEmembership{Senior~Member,~IEEE},~Jean-Baptiste~Forien\textsuperscript{\orcidlink{0000-0002-9229-2455}},~Venkatesh~Sridhar,\textsuperscript{\orcidlink{0000-0002-0394-5830}}~\IEEEmembership{Member,~IEEE},\\ ~Jefferson~Cuadra,~Dilworth~Parkinson\textsuperscript{\orcidlink{0000-0002-1817-0716}},~\IEEEmembership{Member,~IEEE}
\thanks{K. A. Mohan, J. B. Forien, and V. Sridhar are affiliated with 
Lawrence Livermore National Laboratory, Livermore, CA.
D. Parkinson is affiliated with Lawrence Berkeley National 
Laboratory, Berkeley, CA.
J. Cuadra's contributions to this paper was made
when he was affiliated with Lawrence Livermore National Laboratory, Livermore, CA.}}



\maketitle

\begin{abstract}
\textcolor{cgcol}{X-ray phase-contrast tomography (XPCT) is widely used 
for high contrast 3D imaging using either synchrotron 
or laboratory microfocus X-ray sources.}
XPCT enables an order of magnitude improvement
in image contrast of the reconstructed material interfaces
with low X-ray absorption contrast.
The dominant approaches to 3D reconstruction
using XPCT relies on the use of phase-retrieval
algorithms that make one or more limiting approximations 
for the experimental configuration and material properties.
Since many experimental scenarios violate such approximations,
the resulting reconstructions contain blur, artifacts, 
or other quantitative inaccuracies.
Our solution to this problem is to formulate new iterative non-linear
phase-retrieval (NLPR) algorithms that avoid such limiting approximations.
Compared to the widely used state-of-the-art approaches, 
we show that our proposed algorithms result in sharp
and quantitatively accurate reconstruction with reduced artifacts.
Unlike existing NLPR algorithms, 
our approaches avoid the laborious  manual tuning 
of regularization hyper-parameters while still achieving the stated goals.
As an alternative to regularization, we propose explicit constraints
on the material properties to constrain the solution space
and solve the phase-retrieval problem. 
These constraints are easily user-configurable since
they follow directly from the imaged object's dimensions
and material properties.
\end{abstract}

\begin{IEEEkeywords}
Phase-retrieval, phase-contrast, X-ray, CT, reconstruction, tomography, synchrotron.
\end{IEEEkeywords}

\section{Introduction}

\begin{figure}[!thb]
\begin{center}
\includegraphics[width=3.5in]{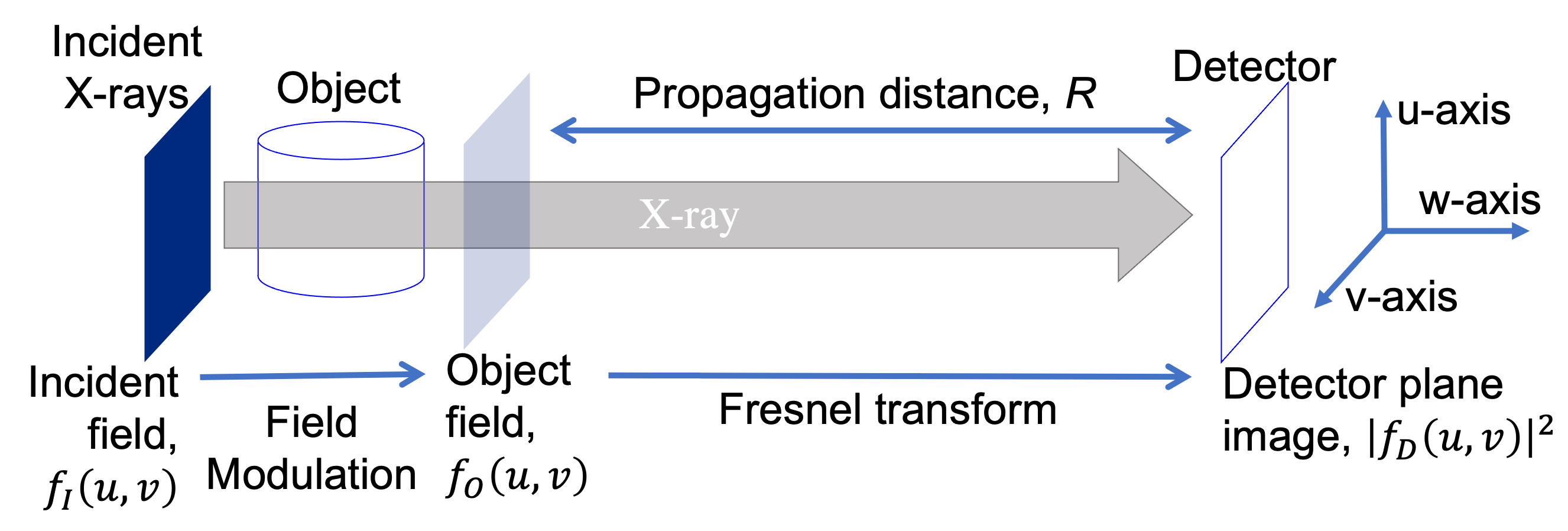}
\vspace{-0.2in}
\end{center}
\caption{\label{fig:sysdiag}
Parallel-beam X-ray phase-contrast tomography (XPCT) experiment 
at a synchrotron user-facility. 
The 3D object is rotated along an axis and 
2D detector measurements are periodically 
acquired at several rotation angles. 
The object-to-detector propagation distance, $R$, 
is \textcolor{cgcol}{adjusted} to produce phase-contrast 
in the measured X-ray images.
The labels for the X-ray fields are defined 
in section \ref{sec:contmod}.
}
\end{figure}

Propagation-based X-ray phase-contrast tomography (XPCT) at synchrotron beamlines
is widely used for 3D imaging
due to the high-intensity, monochromatic, spatially coherent, and parallel-beam properties of synchrotron X-rays.
\textcolor{cgcol}{XPCT is also a popular tool for obtaining higher edge contrast in laboratory micro-focus X-ray CT systems.}
XPCT is used for 3D reconstruction of a wide variety 
of objects in biology \textcolor{cgcol}{\cite{karunakaran_factors_2015, croton_situ_2018, wilkins1996phase}}, 
material science \textcolor{cgcol}{\cite{zielke_degradation_2015, parab_high_2016, sun_morphological_2016,seastwood_three-dimensional_2015, mayo2003x}}, 
medical imaging \cite{pacile_advantages_2018, bravin_x-ray_2012}, and paleontology \cite{fernandez_phase_2012, friis_phase-contrast_2007}. 
In synchrotron XPCT\footnote{Henceforth, XPCT refers to XPCT 
using the monochromatic, parallel, and coherent X-rays at a synchrotron.}, 
the object is exposed to a parallel beam of X-rays
and the X-ray intensity is recorded by a 2D detector 
at several rotation angles of the object (Fig. \ref{fig:sysdiag}).
\textcolor{cgcol}{Phase-contrast is a function of 
the Fresnel number and the phase shift 
induced by the object on the X-ray field.
A detailed discussion of the parameters governing phase-contrast is provided in \cite{gureyev2008some}. 
In our application (Fig. \ref{fig:sysdiag}), 
we adjust the Fresnel number by modifying the
propagation distance to achieve adequate phase-contrast 
in the X-ray images.
XPCT can either be performed at a single object-to-detector distance
or be repeated at several object-to-detector distances.
}

The solution to the inverse problem of object reconstruction 
is a combination of two sequential steps.
First, we perform 2D phase-retrieval at each tomographic view to 
reconstruct the \textcolor{cgcol}{complex-valued images} of the transmission X-ray field, 
$f_O(u,v)$ in Fig. \ref{fig:sysdiag}, 
or its phase shift from the detector images $\lt|f_D(u,v)\rt|^2$
\cite{burvall_phase_2011, langer_quantitative_2008}.
Henceforth, we shall refer to the phase component 
of the recovered field $f_O(u,v)$ as the phase image.
Next, we use a tomographic reconstruction algorithm
to reconstruct the 3D refractive index 
decrement of the object from the phase images \cite{kak_principles_2001}.
\textcolor{cgcol}{In our phase-retrieval algorithms, we reconstruct the 2D phase image at each tomographic view angle from X-ray images at one or more propagation distances.}
Phase-retrieval is a non-linear inverse problem since it relies
on the inversion of a non-linear analytical forward model that relates the phase to the X-ray images.
The tomographic reconstruction step is used to reconstruct
the 3D refractive index decrement from the retrieved phase images 
at all the views. 
Reconstruction of the refractive index decrement 
from the phase images is a linear inverse problem 
that is solved using any analytical or iterative CT 
reconstruction method, including the widely used 
filtered back projection (FBP) algorithm \cite{kak_principles_2001}.

The popular choice for phase-retrieval algorithms in XPCT
relies on the inversion of forward models that express 
the measured X-ray images\footnote{Or, simple functional forms such as affine and/or logarithm transforms of the X-ray images are expressed as a linear transform of the phase.} 
as a linear transformation of the phase image.
For XPCT using X-ray images at a single propagation distance,
phase-retrieval reduces to the application of a digital linear filter on the individual X-ray images \cite{burvall_phase_2011,paganin_simultaneous_2002, beltran_2d_2010, gureyev_optical_2004, bronnikov_reconstruction_1999, wu_x-ray_nodate, Chen2013comparison, chen2011phase, paganin2020boosting}.
These methods also constrain the material composition by 
approximating the absorption index to be zero  
or enforcing a proportionality relation between the 
refractive index decrement and the absorption index.
In the case of multi-distance XPCT, phase-retrieval is performed by the inversion of 
approximate linear relations between the multi-distance 
X-ray images and the phase \cite{ langer_quantitative_2008, yu_evaluation_2018, zabler_optimization_2005, langer_regularization_2010, cloetens_quantitative_2002, guigay_mixed_2007}. 
\textcolor{cgcol}{Some of these methods avoid the approximations of zero-absorption 
and phase-absorption proportionality \cite{cloetens_quantitative_2002, guigay_mixed_2007, langer_quantitative_2008}.}
\textcolor{cgcol}{Gureyev et al. \cite{gureyev2006pcoherence, gureyev2004linear} derives linear analytic formulas for 
phase-retrieval in the Fresnel region for coherent 
and partially-coherent radiation.}
While such linear phase-retrieval (LPR) algorithms are 
computationally fast, the forward modeling approximations 
severely restrict the material composition and experimental design. 
If LPR algorithms are used outside their range of validity, 
they produce reconstructions with quantitative inaccuracies, 
false (unwanted) artifacts, and/or image blur.

\textcolor{cgcol}{To address the limitations of 
linear phase-retrieval (LPR),
several non-linear phase-retrieval (NLPR) algorithms 
have been proposed \cite{mohan_direct_2016, davidoiu_non-linear_2011, davidoiu_non-linear_2012, davidoiu_nonlinear_2012,  davidoiu_nonlinear_2013, Ruhlandt_2014_3DPR, Moosmann_2010_Nonlinear, 
Mom_2022_PrimalDual, Maretzke_2016_RegNewton}.
The algorithms in \cite{davidoiu_non-linear_2011, davidoiu_non-linear_2012, davidoiu_nonlinear_2012,  davidoiu_nonlinear_2013, Maretzke_2016_RegNewton, Mom_2022_PrimalDual, Moosmann_2010_Nonlinear} 
reconstruct the phase images with regularization to
promote sparsity in projection space. 
Alternatively, the algorithms in \cite{mohan_direct_2016, Ruhlandt_2014_3DPR}
directly reconstruct the 3D object 
using tomography consistency conditions to improve phase-retrieval
by leveraging information from across view angles. 
While the latter approach is arguably better to 
constrain the solution space, 
it is also highly compute intensive while also being 
difficult to parallel compute.
Hence, we pursue the former disjoint approach of phase-retrieval 
followed by tomographic reconstruction of the object. }

\textcolor{cgcol}{NLPR algorithms \cite{mohan_direct_2016, davidoiu_non-linear_2011, davidoiu_non-linear_2012, davidoiu_nonlinear_2012,  davidoiu_nonlinear_2013, Ruhlandt_2014_3DPR, Moosmann_2010_Nonlinear, 
Mom_2022_PrimalDual, Maretzke_2016_RegNewton}
are generally more accurate than LPR algorithms since
they avoid the linear approximations made 
in the forward measurement model.
NLPR estimates the phase images by minimizing a 
distance measure between the X-ray images 
and the output of a non-linear forward measurement model. 
They also use certain sparsity constraints
in the form of prior models or regularization functions
 to limit the solution space for the phase-retrieval problem.
The use of such explicit regularization functions  
requires the user to manually
tune one or more parameters to achieve
the claimed performance improvements of artifact reduction, 
improved sharpness, and quantitative accuracy.
In contrast, our approach achieves these benefits while 
avoiding parameter tuning due to the 
absence of regularization functions.
Optionally, our approach also supports implicit constraints 
in our non-linear forward model that 
follow directly from the material properties.}

\textcolor{cgcol}{In recent years, several deep learning (DL) approaches 
\cite{Mom_2023_DeepGaussNewt, Wu_2022_DeepConcatNet, Rucha_2023_RobustDL, Li_22_PhyNNSparse, Wu_2022_PhyNNTIE, Li_2022_U_NetDL, Zhang_2021_PhaseGAN} 
have also been proposed for phase-retrieval in XPCT.}
Using DL, the authors demonstrated improved performance 
without the excessive computational cost of NLPR approaches. 
\textcolor{cgcol}{However, the limitation of DL approaches is the need for representative
training data to train the neural network models.}
In contrast, both LPR and NLPR algorithms do not require
any training data and are broadly applicable to a 
wide range of objects. 

In this paper, we present new non-linear phase-retrieval 
(NLPR) algorithms for both single-distance and multi-distance XPCT. 
A limited version of this manuscript 
with preliminary results was published
in the form of a conference proceedings paper \cite{mohan_constrained_2020}.
Our NLPR algorithms also support constraints 
on the material composition.
In particular, we demonstrate the use of the 
phase-absorption proportionality 
 constraint \textcolor{cgcol}{(mathematically equivalent to 
 the single material constraint)} for single-distance XPCT.
First, we formulate discrete non-linear forward models using 
the Fresnel transform in Fourier frequency space.
This model expresses the measured data as an analytical
non-linear function of the phase images.
Next, we formulate an objective function that is a measure
of the distance between the measured data 
and the forward model output. 
Unlike existing non-linear approaches, 
we do not use prior-models or regularization functions 
in the objective function  
to achieve the desired goals
of artifact-free, sharp, and accurate reconstructions.
Finally, we iteratively solve for the phase images that minimize
the objective function while also satisfying any constraint
on the material composition. 

We initialize our NLPR algorithms with the phase and absorption
images that are estimated using LPR algorithms.
\textcolor{cgcol}{
We perform a comprehensive investigation of initialization
strategies by varying the 
LPR method
used for initialization and its regularization parameter.
The characteristics and contributions of our proposed NLPR algorithms are} -
\begin{itemize}
    \item \textcolor{cgcol}{Maximum likelihood formulation of phase-retrieval by inversion of non-linear forward measurement models.}
    \item \textcolor{cgcol}{Fresnel propagation based non-linear forward models with optional constraints.}
    \item \textcolor{cgcol}{Phase-retrieval by minimization of non-convex objective functions using the LBFGS algorithm.}
    \item Quantitatively accurate and sharp reconstructions with 
    reduced artifacts compared to LPR approaches.
    \item Avoids manual tuning of regularization hyper-parameters
    due to the absence of regularizing prior models.
    \item \textcolor{cgcol}{Our recommended initialization strategy 
    also avoids hyper-parameter tuning.} 
    \item \textcolor{cgcol}{Optional constraints such as single material (homogeneous) object,
    phase-absorption proportionality, or zero-absorption.}
    \item Open-source software with documentation.
\end{itemize}

We implemented our algorithms using the \textit{PyTorch}
framework and python programming language. 
For faster computation, our algorithms can 
also be run on multiple GPUs. 
We have released our phase-retrieval software under an open-source
license along with documentation at the link \url{https://github.com/phasetorch/phasetorch}.
 
\section{X-ray Measurement Physics}
\label{sec:contmod}
In this section, we present a mathematical formulation 
for the physics of measurement in XPCT.

\subsection{Modulation of the X-ray Field}
In XPCT, the measured data is sensitive to the 3D variations
in the absorption index, $\beta(u,v,w)$, and refractive index 
decrement, $\delta(u,v,w)$, of the imaged object.
Here, $(u,v,w)$ represents the 3D Cartesian coordinate 
system such that the X-ray propagation is along the $w-$axis 
and perpendicular to the $u-v$ plane.
The absorption index and refractive index decrement 
are expressed compactly
as the complex refractive index 
$n(u,v,w) = 1-\delta(u,v,w)+i\beta(u,v,w)$,
\textcolor{cgcol}{where $i$ is the imaginary unit.}

As X-rays propagate through an object, 
the X-ray field undergoes a change in both amplitude and phase. 
The reduction in X-ray intensity after propagation through 
an object is a direct measure of the object's absorption 
index, $\beta(u,v,w)$.
The phase shift of the X-ray field 
that is induced by the object is a direct measure 
of its refractive index decrement, $\delta(u,v,w)$.
Ignoring constant phase terms, 
the modulated X-ray field, $f_O(u,v)$, is expressed as,
\begin{equation}
\label{eq:fieldmod}
f_O(u,v) = f_I(u,v) T(u,v),
\end{equation}
\textcolor{cgcol}{where $f_I(u,v)$ is the complex-valued 
incident X-ray field,} 
$T(u,v)$ is the transmission function such that $T(u,v)=\exp\lt\lbrace -A(u,v)-i\phi(u,v) \rt\rbrace$, 
where
\begin{multline}
\label{eq:ampphasemod}
A(u,v) = \frac{2\pi}{\lambda} \int_w \beta\lt(u,v,w\rt) dw 
\text{ and } \\
\phi(u,v) = \frac{2\pi}{\lambda} \int_w \delta\lt(u,v,w\rt) dw.
\end{multline}
Here, $\lambda$ is the wavelength of the monochromatic 
and coherent X-ray field.

\subsection{Measurement of Fresnel Propagated Field}
As the X-ray field propagates from the object to the detector,
the phase shifts induced by the object manifest as Fresnel 
diffraction fringes in the X-ray intensity images that 
are measured at the detector.  
The X-ray field at the detector plane, $f_D(u,v)$, 
is expressed as a 2D convolution of the X-ray field 
at the exit plane, $f_O(u,v)$, downstream of the sample
and the Fresnel impulse response function, \textcolor{cgcol}{$\frac{\exp(ikR)}{i\lambda R}\exp\lt\lbrace \frac{i\pi}{\lambda R} \lt(u^2+v^2\rt)\rt\rbrace$}, 
where $R$ is the object to detector distance
and $k=\frac{2\pi}{\lambda}$ is the wavenumber
\cite{langer_quantitative_2008, mohan_constrained_2020, mohan_direct_2016}.
The X-ray field propagation is 
depicted in Fig. \ref{fig:sysdiag}.

\textcolor{cgcol}{The} Fresnel transform is more efficiently computed in 
Fourier frequency space.
Let $F_D(\mu,\nu)$ and $F_O(\mu,\nu)$ denote the 
2D Fourier transform of the
X-ray fields $f_D(u,v)$ and $f_O(u,v)$ respectively,
where $\lt(\mu,\nu\rt)$ are the 2D Fourier 
frequency coordinates. Then,
\begin{equation}
\label{eq:contfreqfres}
F_D(\mu,\nu) = F_O(\mu,\nu) \exp\lt\lbrace -i\pi\lambda R \lt(\mu^2+\nu^2\rt)\rt\rbrace.
\end{equation}
 
Information on the phase of the X-ray field is lost 
since detector measurements are only sensitive 
to the intensity of the X-ray field.
\textcolor{cgcol}{Thus, the detector measurements are modeled as,
$|f_D(j\Delta,k\Delta)|^2$,
where $\Delta$ is the width of each detector pixel, 
$|\cdot|$ denotes the magnitude of a complex number, and
$(j, k)$ denote the row and column indices of the detector pixel.
The square-root of the normalized detector measurement is then modeled as,
\begin{equation}
    \label{eq:normdetmeas}
    y(j,k) = \left|\frac{f_D(j\Delta,k\Delta)}{f_I(j\Delta,k\Delta)}\right|,
\end{equation}
where $f_I(u,v)$ is the incident X-ray field from equation \eqref{eq:fieldmod}.}

\section{Forward Model}
\label{sec:forwmod}
The forward models formulated in this section 
will be used in the phase-retrieval algorithms 
that estimate the transmission function
from the measurements \textcolor{cgcol}{$y(j,k)$}.
To formulate a forward model, 
we first translate the continuous space
expressions in section \ref{sec:contmod} to discrete space.
Let $x(j,k)$ denote a sampling of the transmission 
function $T(u,v)$ in equation \eqref{eq:fieldmod}. 
Let $g_D(j,k)$, $g_O(j,k)$, and $g_I(j,k)$ denote the 
sampled discrete representation of the continuous 
space X-ray fields $f_D(u,v)$,
$f_O(u,v)$, and $f_I(u,v)$ respectively\footnote{
$g_D(j,k)$, $g_O(j,k)$, and $g_I(j,k)$ are a sampling 
of the X-ray fields $f_D(u,v)$, $f_O(u,v)$, and $f_I(u,v)$  
in discrete space such that $u=j\Delta$ and $v=k\Delta$.}.
Let $G_D(p,q)$, $G_O(p,q)$, and $G_I(p,q)$
represent the discrete Fourier transform (DFT) coefficients
of $g_D(j,k)$, $g_O(j,k)$, and $g_I(j,k)$ respectively.

\subsection{Discretization of Measurement Model}
We derive the discretized relation 
between the incident X-ray field 
$g_I(j,k)$ and the X-ray field $g_O(j,k)$ 
at the exit plane of the object. 
Given that $x(j,k)$ represents the transmission function 
in discrete space, we have,
\begin{equation}
\label{eq:objtran}
g_O(j,k) = g_I(j,k) x(j,k).
\end{equation}

Next, we discretize the relation between the \textcolor{cgcol}{square-root normalized detector measurements} $y(j,k)$ in equation \eqref{eq:normdetmeas} and
the X-ray field $g_O(j,k)$ at the exit plane of the object. 
To sample the Fresnel transform in equation \eqref{eq:contfreqfres},
we substitute $\mu=p\Delta_{\mu}$ and $\nu=q\Delta_{\nu}$, 
where $(p,q)$ are the discrete frequency coordinates,
$\Delta_{\mu}=\frac{1}{N_u \Delta}$ is the sampling width 
along the $\mu$-axis, 
and $\Delta_{\nu}=\frac{1}{N_v \Delta}$ is the sampling 
width along the $\nu$-axis. Note that $N_u$ and $N_v$ 
are the number of discrete coordinates along the 
$u$-axis and $v$-axis respectively.
Thus, a discrete sampling of Fourier space 
Fresnel transform is given by,
\begin{equation}
\label{eq:discfreqfres}
H\lt(p,q\rt) = \exp\lt\lbrace -i\pi\lambda R \lt(p^2\Delta_{\mu}^2+q^2\Delta_{\nu}^2\rt)\rt\rbrace.
\end{equation}
Given equation \eqref{eq:discfreqfres}, we can now express 
the relation between the DFT $G_D(p,q)$ of $g_D(i,j)$ 
and DFT $G_O(p,q)$ of $g_O(i,j)$ as,
\begin{equation}
\label{eq:discfrestran}
G_D(p,q) = H(p,q) G_O(p,q).
\end{equation} 
We use edge padding in the space domain for $g_O(j,k)$ 
to avoid circular convolution artifacts. 
\textcolor{cgcol}{In edge padding, the pixel values along an edge are copied to the neighboring padded regions.}
Since the detector only measures the intensity of the X-ray field, 
the square root of the \textcolor{cgcol}{normalized measurement 
in the absence of noise} is expressed as,
\textcolor{cgcol}{
\begin{equation}
\label{eq:detavgmod}
y(j,k) = \lt|\frac{g_D(j,k)}{g_I(j,k)}\rt|,
\end{equation}}
where $g_D(j,k)$ is the discrete X-ray field in the detector plane
given by the inverse discrete Fourier transform (IDFT) of $G_D(p,q)$
and $|\cdot|$ denotes the magnitude of a complex number.

In practice, the noise in detector measurements is modeled 
using Poisson statistics. 
Due to the variance stabilizing property of the square 
root transformation of Poisson random variables \cite{johnson_univariate_nodate}, 
we can model the noise statistics of the square root 
\textcolor{cgcol}{normalized} measurement $y(j,k)$ (experimentally determined using equation \eqref{eq:sqrtmeas} in the supplementary document)
as a Gaussian distribution with constant variance. 
Thus, the forward model for the square root \textcolor{cgcol}{normalized} 
measurement is given by,
\textcolor{cgcol}{
\begin{equation}
\label{eq:addnoise}
y(j,k) = \lt|\frac{g_D(j,k)}{g_I(j,k)}\rt| + n(j,k),
\end{equation}}
where $n(j,k)$ is additive Gaussian noise with a 
constant variance for all $j$ and $k$.

\subsection{Non-Linear Forward Models for XPCT}
The forward model that relates $y(j,k)$ to $x(j,k)$ is
obtained by combining equations \eqref{eq:objtran}, \eqref{eq:discfrestran}, and \eqref{eq:addnoise}.
For convenience of notation, we will use matrix-vector 
notation for mathematical formulation of the forward model in discrete space.
In section \ref{sssec:unconx}, we present a forward model
where $x(j,k)$ is unconstrained, which does not impose restrictions on the material
composition of the imaged object.
In this case, the $x(j,k)$ is a complex valued number that uniquely 
encodes the X-ray phase shift and total X-ray absorption by the object.
In section \ref{sssec:conx}, we present a forward model 
where $x(j,k)$ is constrained, which restricts the 
material composition of the imaged object.
In this case, the feasible solution space of the complex valued $x(j,k)$ is restricted such that it can only be expressed 
as the function of a real valued $z(j,k)$.
This constraint is useful to restrict the 
feasible solution space for 
the phase shift and X-ray absorption.

\subsubsection{Unconstrained $\tb{x}$}
\label{sssec:unconx}
Let $\tb{y}$, $\tb{x}$, and $\tb{n}$ be column vectors whose elements include a raster ordering of the discrete samples $y(j,k)$, $x(j,k)$, and $n(j,k)$ respectively.
The matrix $\tb{H}$ is an operator that when left-multiplied to 
$\tb{x}$ returns the Fresnel transform of $\tb{x}$. 
The forward model term 
\textcolor{cgcol}{$\lt|\tb{H}\tb{x}\rt|$} computes the magnitude of the IDFT of the Fresnel transform (equation \eqref{eq:discfrestran}) of the DFT of $x(j,k)$\footnote{
\textcolor{cgcol}{Note that the matrix $\tb{H}$ includes the forward and
inverse Fourier transform operations in discrete space
unlike the continuous space transform $H(p,q)$ in equation
\eqref{eq:discfrestran}.}
}. 
The vector form of the unconstrained forward model is,
\begin{equation}
\label{eq:mulforwmod}
\tb{y} = \lt|\tb{H}\tb{x}\rt| + \tb{n}.
\end{equation}
 where $\lt|\cdot\rt|$ denotes element-wise magnitude.
The DFT and IDFT
operations are implemented using fast Fourier transform algorithms.
Equation \eqref{eq:mulforwmod} 
expresses the dependence of the real valued measurement
vector $\tb{y}$ on the complex valued transmission
vector $\tb{x}$.

\subsubsection{Constrained $\tb{x}$}
\label{sssec:conx}
In the absence of data $\tb{y}$ at multiple  propagation distances,
unique reconstruction of $\tb{x}$ is achieved by imposing constraints on the imaged sample.
For single-distance XPCT, 
it is typical to use constraints such as single material \cite{paganin_simultaneous_2002},
phase-absorption proportionality \cite{wu_x-ray_nodate}, 
or zero absorption \cite{bronnikov_reconstruction_1999}.
However, such constraints may introduce inaccuracies and artifacts 
for objects that do not satisfy these constraints.

The single-material and phase-absorption proportionality 
constraints are mathematically expressed as,
\begin{equation}
\label{eq:phaseattendual}
\delta(u,v,w) \propto \beta(u,v,w) \Rightarrow \phi(u,v)  \propto A(u,v). 
\end{equation}
Alternatively, under the zero-absorption constraint, we have,
\begin{equation}
\label{eq:zeroabs}
\beta(u,v,w) = 0 \Rightarrow A(u,v) = 0. 
\end{equation}

The two constraints in equations \eqref{eq:phaseattendual} and \eqref{eq:zeroabs} are equivalent to expressing the complex valued $\tb{x}$
in terms of a real valued $\tb{z}$ such that,
\begin{equation}
\label{eq:cmplxasrealvec}
\tb{x} = \tb{z}^{\alpha+i\gamma},
\end{equation}
where $\alpha$ and $\gamma$ are real valued constants \textcolor{cgcol}{that are dependent on the 
material composition.}
For the phase-absorption proportionality constraint (equation \eqref{eq:phaseattendual}),
we set the ratio of $\gamma$ over $\alpha$
equal to the ratio of the refractive index decrement $\delta$ to the absorption index $\beta$.
For the zero-absorption constraint in equation \eqref{eq:zeroabs}, we set $\alpha=0$.
We discuss the possible choices for $\alpha$ and $\gamma$
and the associated trade-offs in section \ref{ssec:conxnlpr}.

Equation \eqref{eq:cmplxasrealvec} halves the dimensionality 
of the phase-retrieval problem by expressing the complex 
valued vector $\tb{x}$ in equation \eqref{eq:mulforwmod} 
as a function of a real valued vector $\tb{z}$. 
Given the constraint in equation \eqref{eq:cmplxasrealvec},
the forward model that expresses $\tb{y}$ in terms of $\tb{z}$
is given by,
\begin{equation}
\label{eq:sinforwmod}
\tb{y} =  \lt|\tb{H}\tb{z}^{\alpha+i\gamma}\rt| + \tb{n}.
\end{equation}

\section{Non-Linear Phase-Retrieval (NLPR)}
\label{sec:nlprall}
In this section, we will formulate  phase-retrieval algorithms 
using the maximum likelihood (ML) estimation framework. 
The reconstruction $\tb{x}$ is such that it minimizes the 
negative log-likelihood function $l(\tb{y}; \tb{x})$, 
which is a measure of the statistical likelihood 
of the measurement data $\tb{y}$ 
given the transmission function $\tb{x}$. 
Under the ML framework, we perform phase-retrieval by solving the optimization problem of 
$\hat{\tb{x}} = \argmin_{\tb{x}} l(\tb{y}; \tb{x})$.
First, we formulate an approach to estimate the unconstrained transmission function $\tb{x}$. 
Next, we present an approach to estimate $\tb{x}$ 
such that it also satisfies the constraint in 
equation \eqref{eq:cmplxasrealvec}.
Finally, we solve for the X-ray absorption and phase shift at the exit plane of the object from the estimated $\tb{x}$. We do not use a prior likelihood model to enforce sparsity in 
reconstruction of $\tb{x}$.
\textcolor{cgcol}{Instead, we initialize $\tb{x}$ 
using conventional linear phase-retrieval algorithms,}
which mimics the role of regularization 
and leads to improved solutions \cite{Gureyev_2003_composite}.

\subsection{Unconstrained NLPR (U-NLPR)}
\label{ssec:unconxnlpr}
To reconstruct $\tb{x}$ without any constraints, 
we utilize X-ray images at several 
propagation distances. 
Let $\tb{y}_l$ denote the square root \textcolor{cgcol}{normalized} measurements
at a propagation distance of $R_l$.
Then, the negative log-likelihood function is
$l(\tb{y}; \tb{x}) = \frac{1}{\sigma^2}\sum_{l=1}^{L} 
\lt|\lt|\tb{y}_l-\lt|\tb{H}_l\tb{x}\rt|\rt|\rt|_2^2$,
where $L$ is the total number of distances.
Here, the noise in each element of $\tb{y}_l$ 
is approximated to be additive Gaussian with variance $\sigma^2$
and $\lt|\lt|\cdot\rt|\rt|^2_2$ denotes the squared 
$l^2$ norm of a vector.
Thus, the unconstrained $\tb{x}$ is estimated 
by solving the following optimization problem,
\begin{equation}
\label{eq:unconxnlpr}
\textcolor{cgcol}{\hat{\tb{x}}} = \arg\min_{\tb{x}}l(\tb{y}; \tb{x}) = \arg\min_{\tb{x}} \sum_{l=1}^{L} \lt|\lt|\tb{y}_l-\lt|\tb{H}_l\tb{x}\rt|\rt|\rt|_2^2.
\end{equation}
We call this approach as unconstrained NLPR (U-NLPR).

\subsection{Constrained NLPR (C-NLPR)}
\label{ssec:conxnlpr}
\textcolor{cgcol}{
Using measurements at a single propagation distance and in the absence of sparsity constraints, 
it is difficult to independently reconstruct both the
phase and absorption of the X-ray field after propagation 
through the imaged objects.}
Information on the phase and absorption 
are entangled in the X-ray intensity measurements.
Hence, we impose restrictions on the composition
of the object to halve the number
of unknowns during phase-retrieval using equation \eqref{eq:cmplxasrealvec}.

Let $\tb{y}_l$ denote the square root \textcolor{cgcol}{normalized} measurements
at a propagation distance of $R_l$.
Then, the negative log-likelihood function is
$l(\tb{y}; \tb{z}) = \frac{1}{\sigma^2}\sum_{l=1}^{L} \lt|\lt|\tb{y}_l-\lt|\tb{H}_l\tb{z}^{\alpha+i\gamma}\rt|\rt|\rt|_2^2$. We estimate $\tb{z}$ 
by solving the following 
optimization problem,
\begin{equation}
\label{eq:conxnlpr}
\hat{\tb{z}} = \arg\min_{\tb{z}}l(\tb{y}; \tb{z}) = \arg\min_{\tb{z}} \sum_{l=1}^{L}\lt| \lt|\tb{y}_l-\lt|\tb{H}_l\tb{z}^{\alpha+i\gamma}\rt|\rt|\rt|_2^2.
\end{equation}  
The complex valued transmission function is then
estimated as $\hat{\tb{x}} = \hat{\tb{z}}^{\alpha+i\gamma}$.
This method is called constrained NLPR (C-NLPR).

The scalar constraint parameters of 
$\alpha$ and $\gamma$ are dependent on the material composition.
They also influence the speed of convergence 
of C-NLPR.
Hence, it is important to intelligently set
the parameters of $\alpha$ and $\gamma$.
For a pure-phase object with zero absorption,
we can choose $\alpha=0$ and $\gamma=1$.
However, reconstruction of pure-phase objects will not be investigated in this paper.
If a sample is homogeneous, i.e., 
consists of a single material \cite{paganin_simultaneous_2002}
or satisfies the phase-absorption 
proportionality constraint
\cite{wu_x-ray_nodate},
then we have several choices for setting $\alpha$ and $\gamma$.
Let $\delta$ and $\beta$ denote the scalar values
of the refractive index decrement and absorption index
under these constraints, i.e., we assume 
$\delta(u,v,w)/\beta(u,v,w)=\delta/\beta \,\, \forall u,v,w$. 

\subsubsection{$\text{C-NLPR / One-}\alpha$}
\label{sssec:alpone}
For this constraint, we set $\alpha=1$ such that
$\tb{x}=\tb{z}^{1+i\gamma}$ from equation \eqref{eq:cmplxasrealvec}.
From equations \eqref{eq:ampphasemod} and \eqref{eq:cmplxasrealvec},
we see that $\lt|\tb{x}\rt|=\tb{z}$
is a discretization of the absolute value for the transmission function
$\lt|T(u,v)\rt|=\exp\lt\lbrace-A(u,v)\rt\rbrace$. 
The dynamic range of $\tb{z}$ is determined by the corresponding
dynamic ranges for $A(u,v)$ that is in-turn determined by $\beta(u,v,w)$.
For low X-ray absorption materials, 
if $z_n$ denotes the $n^{th}$ element of $\tb{z}$,
then $z_n\approx 1$ since $\beta\approx 0$. 
\textcolor{cgcol}{The precision of 32-bit floating point numbers
becomes progressively inadequate to accurately 
represent the dynamic range of $z_n$ 
as it approaches $1$.}
Hence, $\alpha=1$ may lead to very slow convergence or
numerical instabilities at very low values of $\beta$.
Since $\tb{z}$ is the discretized representation of $\exp\lt\lbrace-A(u,v)\rt\rbrace$, 
$\tb{z}^{\gamma}$ is a discretization of 
$\exp\lt\lbrace-\phi(u,v)\rt\rbrace$ 
only when $\gamma=\delta/\beta$.
Hence, $\alpha=1$ and $\gamma=\delta/\beta$ leads to the 
parameterization of $\tb{x}=\tb{z}^{1+i\delta/\beta}$, 
which only requires knowledge of the ratio of $\delta/\beta$. 
This method will be referred to as $\text{C-NLPR/One-}\alpha$.

\subsubsection{$\text{C-NLPR / One-}\gamma$}
\label{sssec:gamone}
For this constraint, we set $\gamma=1$ 
such that $\tb{x}=\tb{z}^{\alpha+i}$ 
from equation \eqref{eq:cmplxasrealvec}.
From equations \eqref{eq:ampphasemod} and \eqref{eq:cmplxasrealvec},
we see that $\tb{z}^i$ 
is a discretization of the
phase component of the transmission function 
$\exp\lt\lbrace-i\phi(u,v)\rt\rbrace$.
The dynamic range of $\tb{z}$ is determined by the corresponding
dynamic range of $\phi(u,v)$ or $\delta(u,v,w)$.
For objects with a high refractive index decrement, 
if $z_n$ denotes the $n^{th}$ element of $\tb{z}$, 
then $z_n\approx 0$ if $\delta$ is very large. 
Hence, $\gamma=1$ may lead to very slow convergence or
numerical instabilities for highly refractive materials.
Since $\tb{z}$ is a discretization of $\exp\lt\lbrace -\phi(u,v)\rt\rbrace$, 
$\tb{z}^{\alpha}$ will represent $\exp\lt\lbrace-A(u,v)\rt\rbrace$ 
only when $\alpha=\beta/\delta$.
Hence, $\gamma=1$ and $\alpha=\beta/\delta$ leads to the 
parameterization of $\tb{x}=\tb{z}^{\beta/\delta+i}$, 
which only requires knowledge of the ratio of $\delta/\beta$.
This method will be referred to as $\text{C-NLPR/One-}\gamma$.

\subsubsection{$\text{C-NLPR / TrOpt-}\alpha,\gamma$}
\label{sssec:knowndb}
For this constraint, we control the dynamic range of 
$\tb{z}$ depending on the individual known values of $\delta$ and $\beta$.
We set $\alpha$ and $\gamma$ such that
$z_n$ approximately varies between a preset 
low value of $T_l$ and a high value of $1$. 
\textcolor{cgcol}{
For $T_l$, we simply choose a value that is greater than 
$0$ but also an order of magnitude less than $1$.}
In this paper, we choose $T_l=0.01$.
Then, the values for $\alpha$ and $\gamma$ are such that,
\begin{itemize}
    \item $z_n=1$ in the absence of any material 
    along the ray path for the $n^{th}$ pixel.
    \item $z_n=T_l$ when a material with refractive index 
    decrement $\delta$ and absorption index $\beta$ 
    lies along the entirety of the $n^{th}$ ray.
\end{itemize}
Ideally, $z_n$ should only vary between $T_l$ and $1$ in 
the absence of measurement non-idealities and a perfect
choice for $\delta$ and $\beta$.
In practice, $z_n$ varies approximately within 
this chosen transmission range, which reduces the risk of
numerical instabilities.
Thus, our choice for $\alpha$ and $\gamma$ are,
\begin{align}
    \alpha & = -\frac{2\pi}{\lambda\log\lt(T_l\rt)} \beta \Delta \max(N_u, N_v), \label{eq:alphaopt}\\
    \gamma & = -\frac{2\pi}{\lambda\log\lt(T_l\rt)} \delta \Delta \max(N_u, N_v), \label{eq:gammaopt}
\end{align}
where $\Delta$ is the pixel width. The number of pixels 
in the X-ray images along the $u-$axis and $v-$axis are $N_u$
and $N_v$ respectively.
However, this particular choice of $\alpha$ and $\gamma$ 
requires \textcolor{cgcol}{approximate} knowledge of both $\delta$ and $\beta$. 
In contrast, the previous two choices that are described in 
sections \ref{sssec:alpone} and \ref{sssec:gamone} 
only require knowledge of the ratio $\delta/\beta$.
The constraint obtained using equations 
\eqref{eq:alphaopt} and \eqref{eq:gammaopt}
will be referred to as $\text{C-NLPR/TrOpt-}\alpha,\gamma$ 
(TrOpt indicates optimized transmission).

In section \ref{sec:results}, 
we demonstrate that $\alpha=1, \gamma=\frac{\delta}{\beta}$ 
provides the best trade-off
between stable optimization and good performance while 
only requiring knowledge of the ratio $\delta/\beta$.

\subsection{Optimization Algorithm}
We implemented our NLPR algorithms using python programming language 
and \textit{PyTorch} framework \cite{NEURIPS2019_9015}.
\textit{PyTorch} uses algorithmic differentiation 
(or automatic differentiation)
to compute gradients of the objective functions that are used for minimization. 
In equation \eqref{eq:unconxnlpr}, 
algorithmic differentiation is used to compute the gradient of $l(\tb{y};\tb{x})$ with respect to $\tb{x}$. 
Similarly, algorithmic differentiation is used 
to compute the gradient of $l(\tb{y};\tb{z})$ 
with respect to $\tb{z}$ in equation \eqref{eq:conxnlpr}.

We use the LBFGS algorithm \cite{liu_limited_nodate, nocedal_updating_nodate} to solve the
optimization problems in equations \eqref{eq:unconxnlpr} and \eqref{eq:conxnlpr}.
The gradients computed using algorithmic differentiation 
are used by the LBFGS optimization algorithm 
to reconstruct $\tb{x}$ in \eqref{eq:unconxnlpr} and $\tb{z}$ in \eqref{eq:conxnlpr}.
In this paper, we use the LBFGS implementation in \cite{lbfgs_hjmshi}
with a history size of $64$.
The maximum number of iterations is capped at $10^4$
and we use Wolfe line search for optimal selection of step-size.
We use a convergence criteria that automatically stops the
LBFGS iterations based on the convergence of reconstruction
and objective function values.
We stop LBFGS when the following conditions are met for $M$ 
consecutive iterations,
\begin{itemize}
    \item Average of the absolute differences 
    in the reconstruction ($\tb{z}$ in C-NLPR or $\tb{x}$ in U-NLPR),
    expressed as a percentage, is less than $L_r$.
    The difference is computed between the reconstructions 
    at consecutive two iterations.
    \item The absolute difference 
    in the objective function ($l\lt(\tb{y}; \tb{z}\rt)$ 
    in C-NLPR or $l\lt(\tb{y}; \tb{x}\rt)$ in U-NLPR),
    expressed as a percentage, is less than $L_c$.
\end{itemize}
In this paper, we choose $M=5$, $L_c=1\%$, and $L_r=0.5\%$ for 
all the simulated and experimental results in section \ref{sec:results}.
We recognize that this particular setting for the convergence criteria 
may appear overly stringent.
Our particular choice was designed to ensure sufficient
convergence for all data presented in this paper.
However, for any given data set, 
the convergence criteria may be relaxed for reduced run-time.

\subsection{Initialization}
\subsubsection{Multi-distance linear phase-retrieval}
The performance of U-NLPR is dependent on the 
initial estimate for $\tb{x}$ that is used to initialize the optimization
in equation \eqref{eq:unconxnlpr}.
If the phase and absorption in equation \eqref{eq:ampphasemod}
are set to zero, then each element of the vector $\tb{x}$ is $1$.

For multi-distance phase-retrieval using U-NLPR, 
we show in section \ref{sec:results} that zero-initialization for the phase/absorption leads to improved results compared to the state-of-the-art conventional approaches to multi-distance phase-retrieval.
However, we also demonstrate  that
initialization using conventional phase-retrieval methods 
can further improve the performance of U-NLPR.
In this paper, we explore the use of the 
following phase-retrieval (PR) methods for 
initialization of U-NLPR.
\textcolor{cgcol}{
\begin{enumerate}
    \item Contrast Transfer Function (CTF) PR: Derived by Taylor expansion of the X-ray 
    transmission field under the assumptions of weak absorption and slowly varying phase shift \cite{langer_quantitative_2008, zabler_optimization_2005}. 
    \item Transport of Intensity Equation (TIE) PR: 
    Derived by Taylor expansion of the 
    X-ray transmission field under the assumption
    of small propagation distances \cite{langer_quantitative_2008, paganin1998noninterferometric}.
    \item Mixed PR: Extends the validity of the approximations in CTF and TIE by forming
    a new hybrid PR that combines CTF and TIE \cite{langer_quantitative_2008, guigay_mixed_2007}.
\end{enumerate}
}

Unlike U-NLPR, these conventional phase-retrieval methods use a regularization parameter $\alpha^{\prime}$ that must be fine-tuned for  acceptable performance \cite{langer_quantitative_2008}. 
The value of this regularization parameter $\alpha^{\prime}$ 
will impact both the performance of the 
conventional phase-retrieval method used for initialization
of U-NLPR and the U-NLPR algorithm.
To achieve the best performance for U-NLPR without the need for parameter tuning, 
we use the Contrast Transfer Function (CTF) phase-retrieval \cite{langer_quantitative_2008}
with a sufficiently
low value for the regularization given by, 
\begin{equation}
\label{eq:ctfregchoice}
\alpha^{\prime} = 2\nu (BC-A^2),
\end{equation}
where $\nu$ is a very small value and 
the term $(BC-A^2)$ is defined\footnote{\textcolor{cgcol}{The terms $A$, $B$, 
and $C$ are defined in \cite{langer_quantitative_2008} 
and are not related to any terms that are defined in this paper.
The regularization parameter $\alpha^{\prime}$ in our paper
is equivalent to the $\alpha$ in equation (14) of the paper
\cite{langer_quantitative_2008}.}} in the reference \cite{langer_quantitative_2008}. 
In this paper, we set $\nu=10^{-8}$. 

\textcolor{cgcol}{As a pre-processing step before running multi-distance phase-retrieval algorithms including U-NLPR, 
the X-ray images must be registered such that the object
appears at the same location in the X-ray images at all the propagation distances.}

\subsubsection{Single-distance linear phase-retrieval}
\textcolor{cgcol}{For single-distance phase-retrieval using C-NLPR, 
we investigate the use of zero initialization 
and Paganin PR \cite{paganin_simultaneous_2002} for the phase images.
Paganin PR solves the transport of intensity equation 
while assuming a single-material object (phase-absorption 
proportionality).}

\subsection{Estimation of Phase and Absorption}
To perform tomographic reconstruction,
we need to estimate discrete sampled representations of $A(u,v)$ and $\phi(u,v)$ (from equation \eqref{eq:ampphasemod}).
Let the vectors $\tb{A}$ and $\boldsymbol{\phi}$ be the discrete 
representations of $A(u,v)$ and $\phi(u,v)$ in raster order.

\subsubsection{U-NLPR}
The X-ray absorption, $\hat{\tb{A}}$, and phase shift, $\hat{\boldsymbol{\phi}}$, that is induced by the object 
on the incident X-ray field is,
\begin{equation}
\label{eq:uestabsphase}
\hat{\tb{A}} = -\log\lt(\lt|\hat{\tb{x}}\rt|\rt) \text{ and } \hat{\boldsymbol{\phi}} = -\tan^{-1}\lt(\frac{\hat{\tb{x}}^{(I)}}{\hat{\tb{x}}^{(R)}}\rt),
\end{equation}
where $\hat{\tb{x}}^{(I)}$ and $\hat{\tb{x}}^{(R)}$ are the imaginary
and real parts of $\hat{\tb{x}}$ respectively.
Here, $\log\lt(\cdot\rt)$ and $\tan^{-1}(\cdot)$\footnote{We use $numpy.arctan2(\cdot)$ function of numpy \cite{harris2020array}. 
It is the signed angle between the ray
from the origin to $(1,0)$ and 
the ray from the origin to the point-of-interest.} are 
element-wise vector operators.
\textcolor{cgcol}{The phase estimated using U-NLPR is wrapped if
the dynamic range for the phase exceeds $2\pi$.}
Hence, we use phase unwrapping \cite{herraez_fast_2002, scikit-image} 
to unwrap the phase images $\hat{\boldsymbol{\phi}}$
prior to tomographic reconstruction.

\subsubsection{C-NLPR}
The X-ray absorption and phase images for C-NLPR
are computed as,
\begin{equation}
\label{eq:cestabsphase}
\hat{\tb{A}} = -\alpha\log\lt(\hat{\tb{z}}\rt) \text{ and } \hat{\boldsymbol{\phi}} = -\gamma\log\lt(\hat{\tb{z}}\rt).
\end{equation}
\textcolor{cgcol}{Since the phase $\hat{\boldsymbol{\phi}}$ obtained using  
equation \eqref{eq:cestabsphase} is already unwrapped, 
we do not need to apply an explicit phase unwrapping procedure.} 

Note that equations \eqref{eq:uestabsphase} and \eqref{eq:cestabsphase}
are for a single angular view of the CT scan. 
Hence, we need to repeatedly apply
equations \eqref{eq:uestabsphase} and \eqref{eq:cestabsphase}
to compute the absorption 
and phase images for each view independently.

\section{Tomographic Reconstruction}
In XPCT, measurement data is acquired at several 
rotation angles of the object. 
Let $\boldsymbol\phi^{(n)}$ and $\tb{A}^{(n)}$ 
denote the phase and absorption images
at view index $n$.
The $\hat{{\boldsymbol\phi}}^{(n)}$ and $\hat{\tb{A}}^{(n)}$ 
from equations \eqref{eq:uestabsphase}
and \eqref{eq:cestabsphase}
are \textcolor{cgcol}{proportional to the projections} of the refractive index decrement 
and absorption index at view $n$.
From equation \eqref{eq:ampphasemod}, 
we see that the phase shift and absorption 
\textcolor{cgcol}{terms 
divided by the wavenumber} 
are linear projections of the refractive index decrement 
and absorption index respectively.
Filtered back projection (FBP) \cite{kak_principles_2001} 
is a popular algorithm that is widely used 
for reconstruction of X-ray absorption 
index from its linear projections. 
\textcolor{cgcol}{Thus, we use FBP to also reconstruct the refractive index decrement from its linear 
projections, $\boldsymbol\phi^{(n)}$ divided by the wavenumber, at all the views.}
In this paper, we do not investigate 
reconstruction of the absorption index.

\subsection{Low Frequency Information Loss}
\label{ssec:lowfreqloss}
The reconstruction of refractive index decrement produced by 
U-NLPR and FBP may contain low frequency artifacts. 
\textcolor{cgcol}{These artifacts 
have been well-documented in the research literature \cite{langer_quantitative_2008,langer_priors_2014,petruccelli_transport_2013,gureyev2004linear} on
multi-distance phase-retrieval algorithms.}
Low frequency artifacts refer to slowly varying spurious
artifacts in the reconstructions due to the loss
of low frequency information in the measurements.
Multi-distance phase-retrieval does not use material
constraints such as the phase-absorption proportionality.
Instead, they rely on measurements at a wide range of 
propagation distances for inversion.

We will investigate the loss of low frequency information
using the transfer function in equation \eqref{eq:contfreqfres}.
Equation \eqref{eq:contfreqfres} expresses the X-ray field
$F_D(\mu,\nu)$ at the detector as a function of the X-ray field
at the exit-plane of the object $F_O(\mu,\nu)$ in Fourier space.
In the limit as the frequency components $\mu$ and $\nu$
approach zero, we have,
\begin{equation}
    \mu\rightarrow 0 \text{ and } \nu\rightarrow 0
    \implies F_D(\mu, \nu) \rightarrow F_O(\mu, \nu).
\end{equation}
Thus, the value for $F_D(\mu, \nu)$, in the limit of zero
frequencies, does not change 
with the propagation distance $R$ since $R$ is a parameter
of only the Fresnel transfer function in 
equation \eqref{eq:contfreqfres}.
The detector only measures X-ray intensities
(equation \eqref{eq:normdetmeas}).
To compensate for this loss in phase information, 
we acquire measurements at varying propagation 
distances $R$ such that the phase-contrast fringes
vary in magnitude and thickness.
The phase information that is encoded in the phase-contrast
fringes can be retrieved by acquiring data at multiple distances.
However, since $F_D(\mu, \nu)$ does not vary sufficiently 
with distance $R$ at the low frequencies, 
we obtain low frequency artifacts in the reconstruction due to insufficient information.

During phase-retrieval, we cannot reconstruct the 
average value of the phase irrespective of the number
of multi-distance measurements. 
Let us represent the phase $\phi(u,v)$ 
as the sum of a constant $\phi_0$ and a
zero-mean phase term $\tilde{\phi}(u,v)$, i.e.,
\begin{equation}
    \label{eq:phaseavgoff}
    \phi(u,v)=\phi_0+\tilde{\phi}(u,v) \text{ s.t. } \int_{u,v} \tilde{\phi}(u,v) du dv = 0.
\end{equation}
The constant phase term $\phi_0$ factors out as
the scalar multiple $\exp\lt\lbrace-i\phi_0\rt\rbrace$ 
in equations \eqref{eq:fieldmod} and \eqref{eq:contfreqfres}.
\textcolor{cgcol}{Since the square root normalized measurements $y(j,k)$
in equation \eqref{eq:normdetmeas} measure only the 
X-ray intensity}, information on $\exp\lt\lbrace-i\phi_0\rt\rbrace$
is lost since $\exp\lt\lbrace-i\phi_0\rt\rbrace$ is 
a multiplying factor for $f_D(j\Delta,k\Delta)$.

\begin{figure*}[!htb]
\begin{center}
\begin{tabular}{cccc}
\hspace{-0.1in}
\includegraphics[width=1.7in]{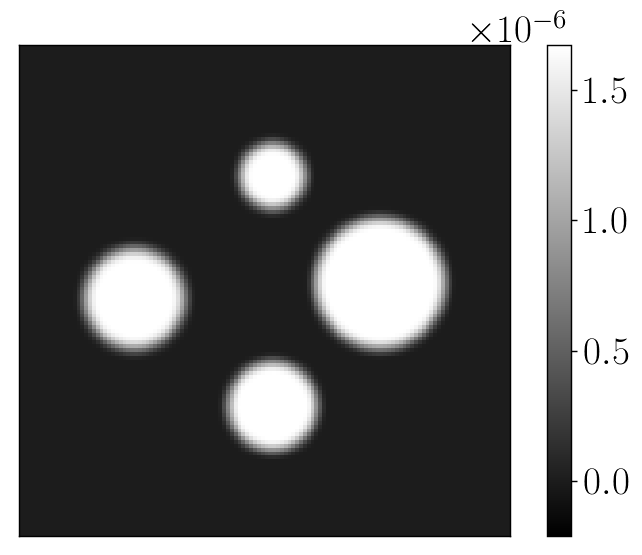} &
\hspace{-0.1in}
\includegraphics[width=1.7in]{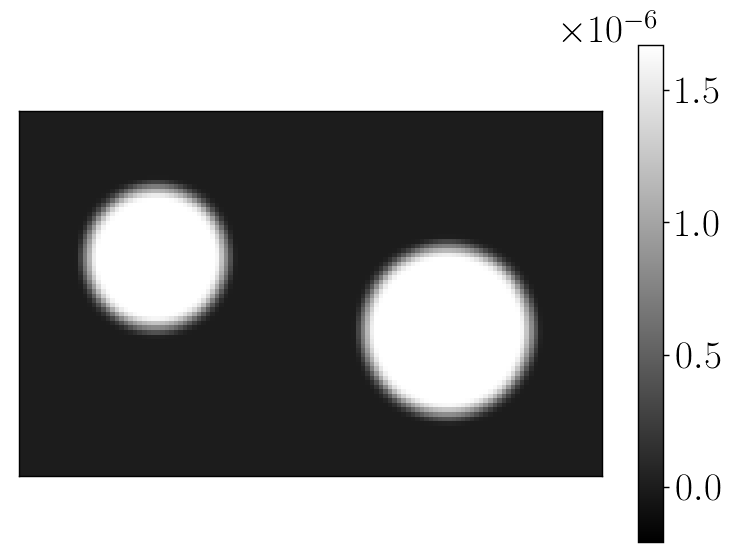} &
\hspace{-0.1in}
\includegraphics[width=1.7in]{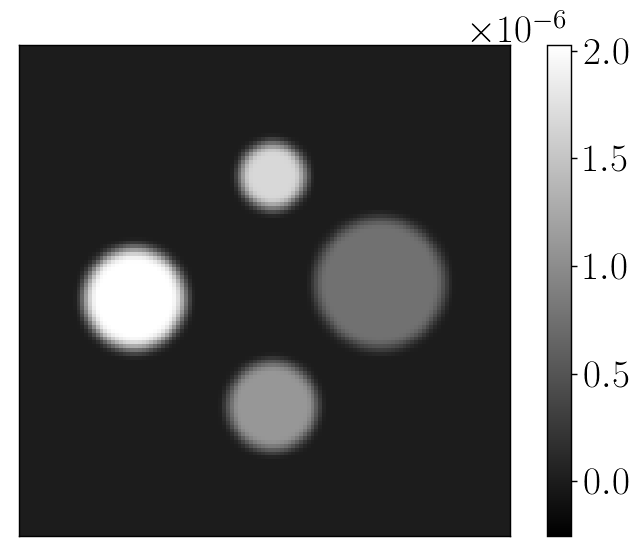} &
\hspace{-0.1in}
\includegraphics[width=1.7in]{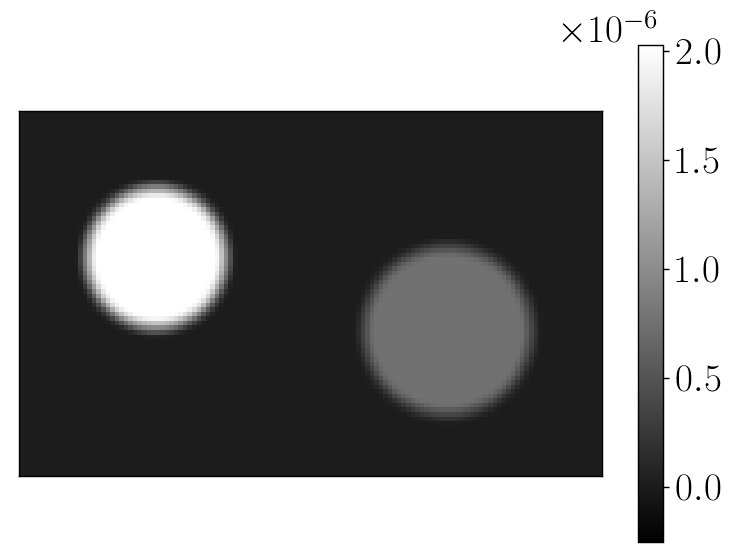} \vspace{-0.03in}\\
\hspace{-0.1in} (a) Single-Material & 
\hspace{-0.1in} (b) Single-Material & 
\hspace{-0.1in} (c) Multi-Material & 
\hspace{-0.1in} (d) Multi-Material \\
 $u-v$ axial slice & $u-w$ axial slice & $u-v$ axial slice & $u-w$ axial slice\\
\hspace{-0.1in}
\includegraphics[width=1.7in]{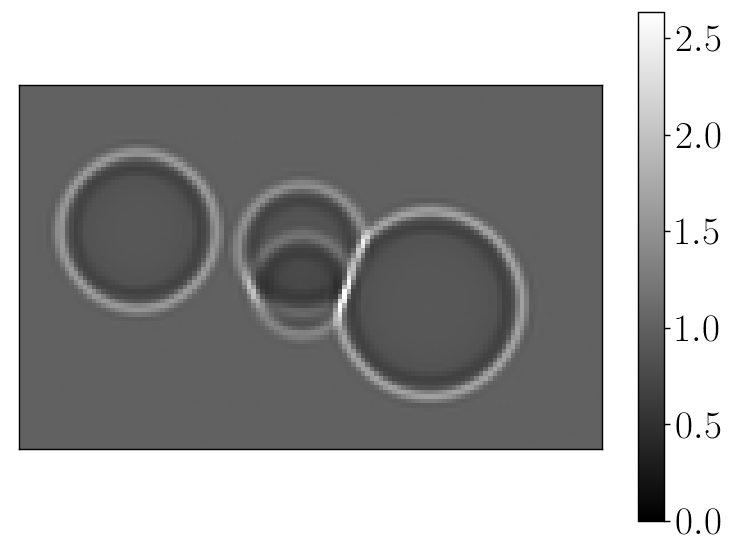} &
\hspace{-0.1in}
\includegraphics[width=1.7in]{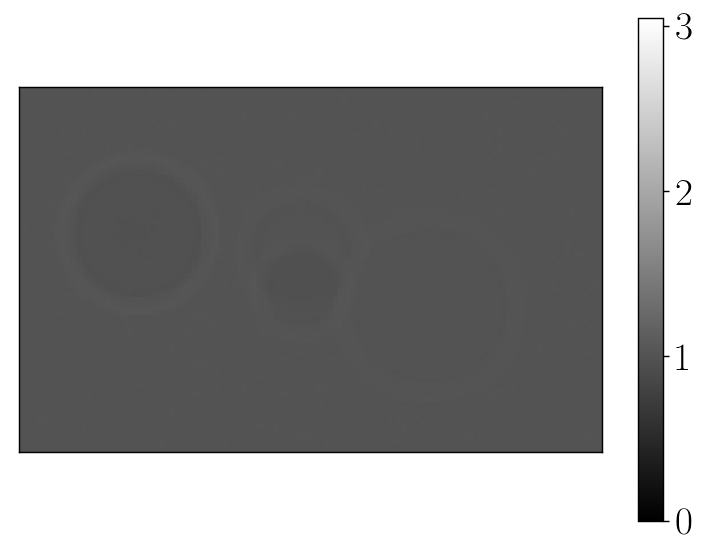} &
\hspace{-0.1in}
\includegraphics[width=1.7in]{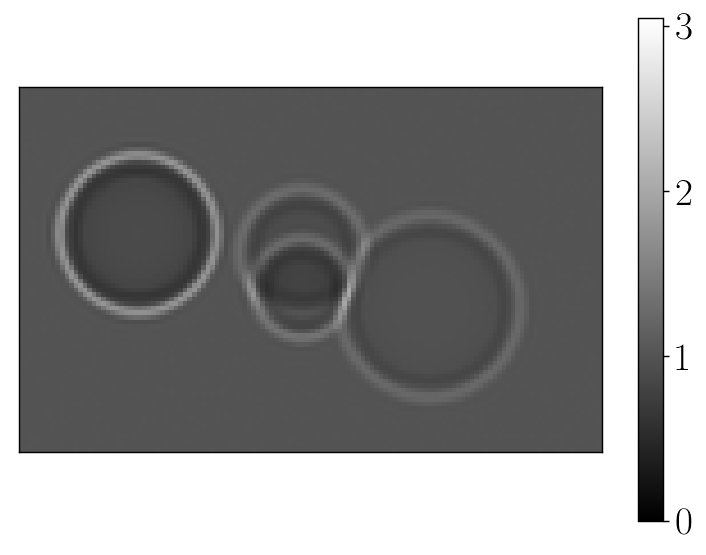} &
\hspace{-0.1in}
\includegraphics[width=1.7in]{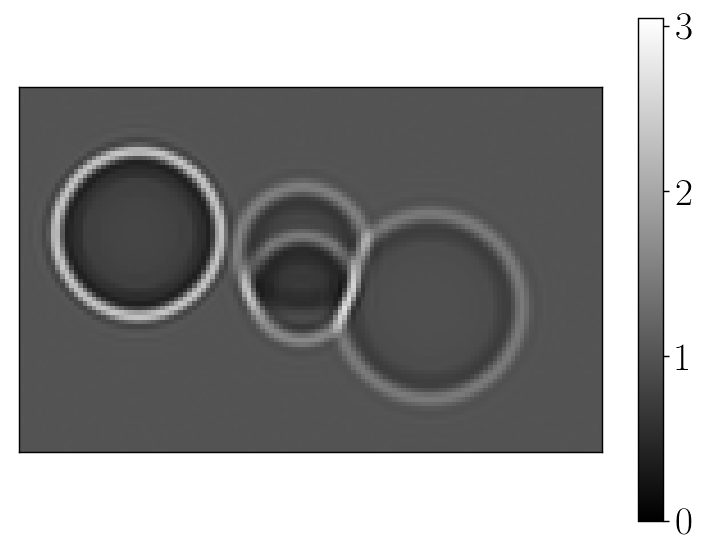} \vspace{-0.1in}\\
\hspace{-0.1in} (e) Single-Material & 
\hspace{-0.1in} (f) Multi-Material & 
\hspace{-0.1in} (g) Multi-Material & 
\hspace{-0.1in} (h) Multi-Material \\
X-ray image $\vert$ $R=200\,mm$ & X-ray image $\vert$ $R=10\,mm$ & X-ray image $\vert$ $R=200\,mm$ & X-ray image $\vert$ $R=400\,mm$ \vspace{-0.03in}\\
\end{tabular}
\end{center}
\caption{\label{fig:simgtrad} Simulation of phase-contrast CT data.
(a, b) and (c, d) show the simulated single-material homogeneous 
object and the multi-material heterogeneous object respectively.
(a, c) show a planar slice ($u-v$ axes) perpendicular to the 
rotation axis and (b, d) show a planar slice ($u-w$ axes) 
parallel to the rotation axis.  
The slices in (a-d) pass through the center of the object.
For the single-material object, the normalized X-ray image
at a propagation distance, $R$, of $200\,mm$ is shown in (e). 
For the multi-material object, the normalized X-ray images 
at propagation distances of $10\,mm$ \textcolor{cgcol}{($FN=2.68$)}, $200\,mm$ \textcolor{cgcol}{($FN=0.13$)}, and $400\,mm$ \textcolor{cgcol}{($FN=0.07$)}
are shown in (f), (g), and (h) 
respectively. \textcolor{cgcol}{$FN$ is the Fresnel number that is
defined as $FN=\frac{\Delta^2}{\lambda R}$.}
All X-ray images in (e-h) are at the first tomographic view.
Phase-contrast fringes are visible in (e), (g), and (h).
From (f-h), we observe that increasing $R$ also increases 
the strength of phase-contrast.}
\end{figure*}

\begin{figure*}[bht!]
\begin{center}
\begin{tabular}{cc}
\includegraphics[height=1.9in]{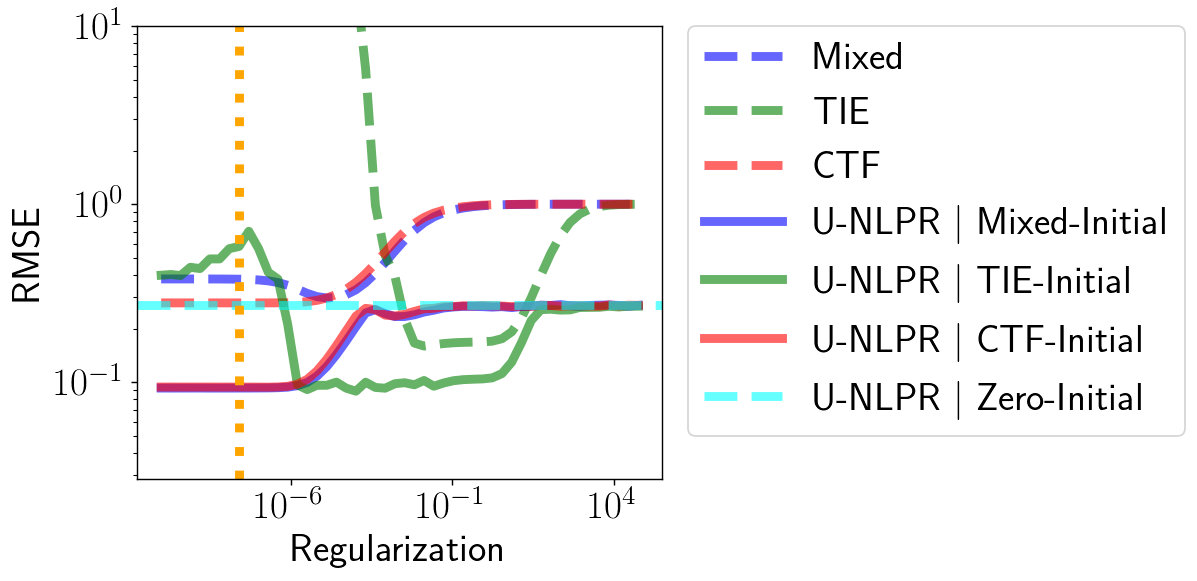} &
\includegraphics[height=1.9in]{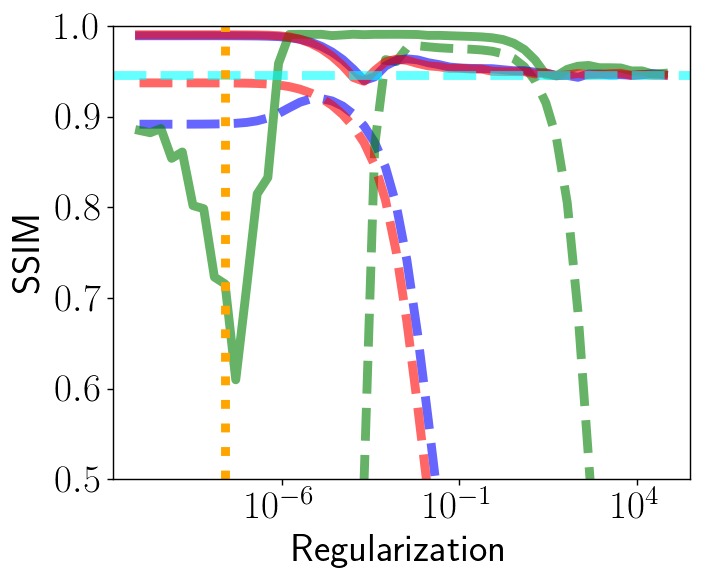} \\
(a) NRMSE vs. Regularization of Conventional PR  &
(b) SSIM vs. Regularization of Conventional PR \\
\end{tabular}
\end{center}
\caption{\label{fig:simswpmulti} 
Quantitative comparison of the refractive index decrement
reconstructions using various phase-retrieval (PR) algorithms. 
(a) and (b) shows the normalized root mean squared error (NRMSE) 
and structural similarity index measure (SSIM) as a 
function of the regularization parameter, $\alpha^{\prime}$, 
for the various PR methods.
Lower is better for NRMSE and higher is better for SSIM.
Both NRMSE and SSIM are computed for 
the entire volume of the foreground spheres 
after background subtraction (section \ref{ssec:quanteval}). 
The dashed lines show the NRMSE and SSIM 
for the conventional methods \cite{langer_quantitative_2008}
of Transport of Intensity Equation (TIE), Contrast Transfer Function (CTF), and Mixed phase-retrieval. 
The solid lines show the NRMSE and SSIM 
for U-NLPR that is initialized using
either TIE, CTF, Mixed, or zero phase/absorption values
(indicated in the legend of (a)). 
Since U-NLPR does not use any regularization, 
the regularization parameter 
along the horizontal axis of (a, b) belongs to the
conventional PR methods used for initialization.
The vertical dotted orange line indicates 
the fixed choice for the regularization $\alpha^{\prime}$
as defined in equation \eqref{eq:ctfregchoice}. 
Irrespective of the regularization,
U-NLPR produces lower NRMSE and higher SSIM when compared
to the phase-retrieval method used for initialization. 
\textcolor{cgcol}{The point of intersection of the vertical orange dotted line with the solid red line indicates the performance for U-NLPR with CTF initialization using the regularization from equation (19), which achieves the best performance while avoiding manual tuning of the regularization hyper-parameter.}
}
\end{figure*}

\makeatletter
\define@key{Gin}{simsz}[true]{%
    \edef\@tempa{{Gin}{width=1.4in,keepaspectratio=true}}%
    \expandafter\setkeys\@tempa
}
\makeatother

\newcommand{\simprhspo}{\hspace{-0.17in}}
\newcommand{\simprhspt}{\hspace{-0.15in}}

\begin{figure*}[htb!]
\begin{center}
\begin{tabular}{ccccc}
\multicolumn{5}{c}{Multi-distance Phase-Retrieval (PR) followed by FBP Reconstruction of the Multi-Material Object} \\
\simprhspo
\includegraphics[simsz]{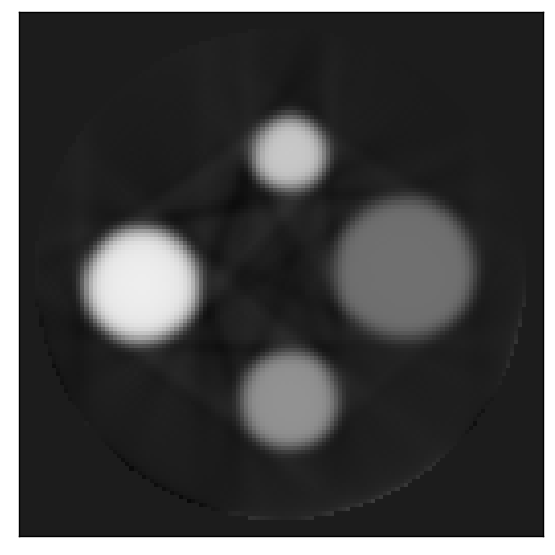} &
\simprhspt
\includegraphics[simsz]{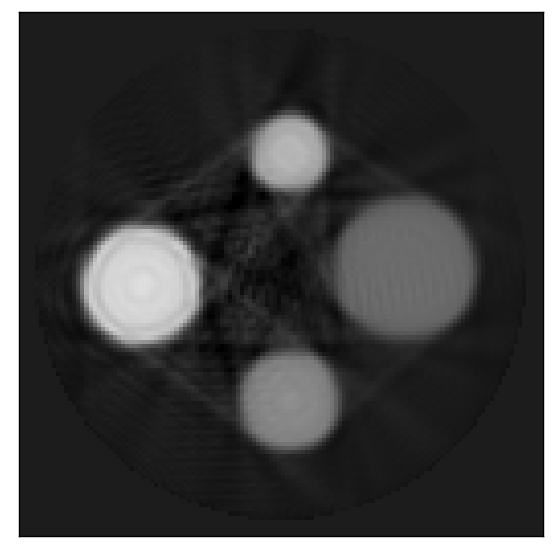} & 
\simprhspt
\includegraphics[simsz]{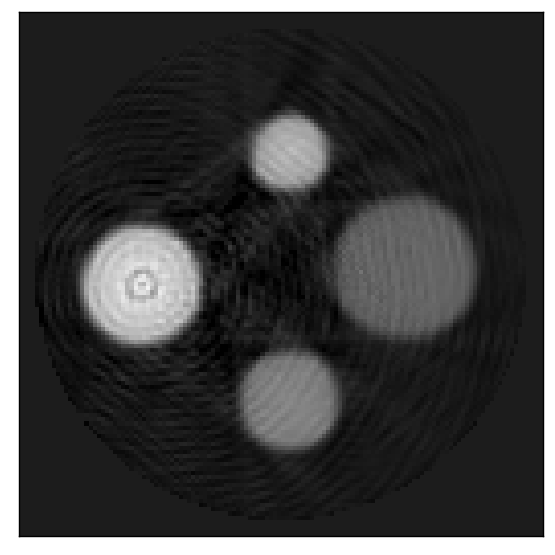} & 
\simprhspt
\includegraphics[simsz]{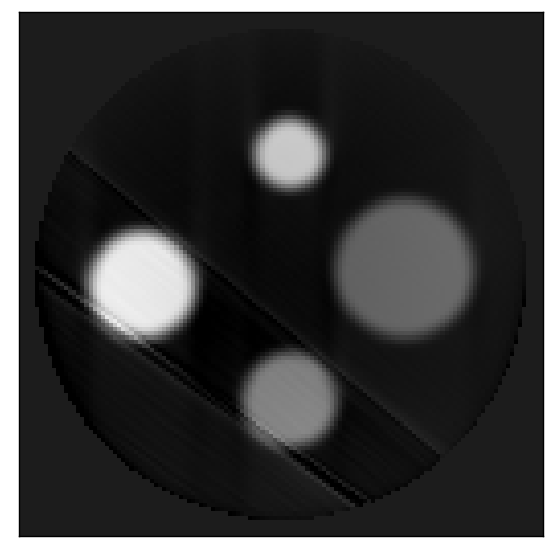} &
\simprhspt
\includegraphics[simsz]{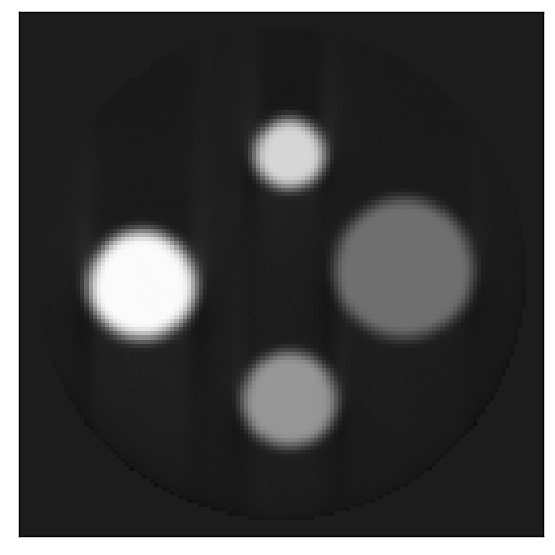} \\ 
\simprhspo (a) TIE & 
\simprhspt (b) CTF & 
\simprhspt (c) Mixed & 
\simprhspt (d) U-NLPR/Zero-Initial & 
\simprhspt (e) U-NLPR/CTF-Initial\\
\end{tabular}
\end{center}
\caption{\label{fig:simmultipr} 
Tomographic reconstructions of the refractive index decrement
from the phase images produced by the CTF, TIE, Mixed, and 
U-NLPR (proposed) phase-retrieval (PR) algorithms. 
(a-e) show planar slices along the $u-v$ axes
that pass through the center 
of the reconstruction volume.
(a), (b), and (c) are using the
TIE, CTF, and Mixed PR algorithms respectively.
For the conventional PR methods of TIE, CTF, and Mixed,
we present the best performing reconstruction at the regularization
with the highest SSIM (from Fig. \ref{fig:simswpmulti}).
(d) shows the reconstruction using U-NLPR that
is initialized with zeros for the phase and absorption.
(e) shows the reconstruction using U-NLPR that is
initialized with CTF at the 
predetermined fixed regularization in equation \eqref{eq:ctfregchoice}.
The gray values in (a-e) are scaled between 
$-2.53\times 10^{-7}$ and $2.03\times 10^{-6}$.
Compared to the conventional PR reconstructions in (a-c), 
U-NLPR reduces streak artifacts and noise as shown in (d, e). Reconstruction slices along the $u-w$ axes
are shown in Fig \ref{fig:supsimmultipr} of the 
supplementary document. 
}
\end{figure*}

\begin{figure}[htb!]
\begin{center}
\begin{tabular}{cc}
\hspace{-0.15in}
\includegraphics[width=1.7in]{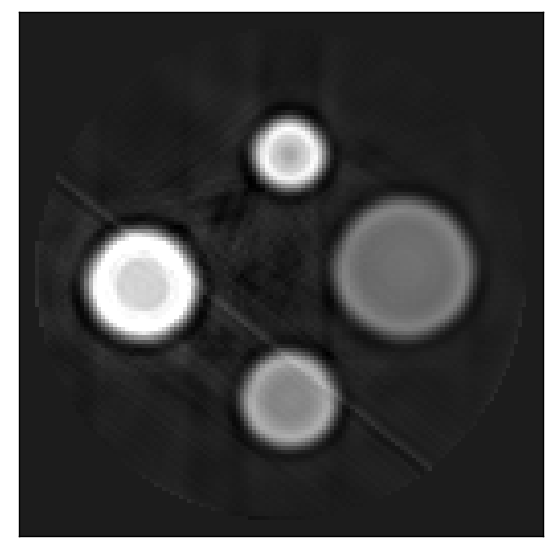} &
\hspace{-0.2in}
\includegraphics[width=1.9in]{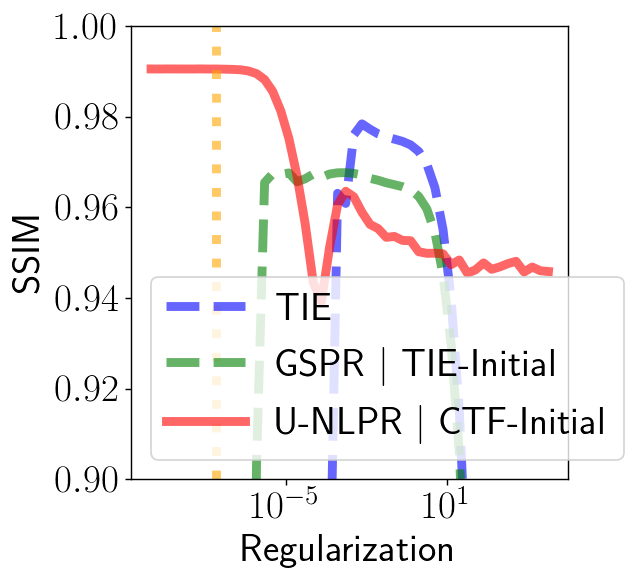} 
\\
(a) GSPR $\vert$ TIE-Initial & 
(b) SSIM vs. Regularization
\end{tabular}
\end{center}
\caption{\label{fig:gsprrec}
\textcolor{cgcol}{(a) is the tomographic reconstruction ($u-v$ axes) 
using Gerchberg-Saxton PR (GSPR) that is initialized 
with TIE PR at the best regularization parameter.
(b) is the SSIM as a function of the regularization parameter
for the PR that is used as initialization.
GSPR produces artifacts that resemble 
 Fresnel diffraction fringes in (a). 
U-NLPR with CTF initialization at the fixed regularization of
equation \eqref{eq:ctfregchoice} (intersection of the
orange dotted line and the red solid line)
has higher SSIM than GSPR at any regularization value.
}}
\end{figure}

\begin{figure*}[htb!]
\begin{center}
\begin{tabular}{ccccc}
\multicolumn{3}{c}{Single Distance PR + FBP of Single-Material Object} & \multicolumn{2}{c}{Single Distance PR + FBP of Multi-Materials} \\
\simprhspo \includegraphics[simsz]{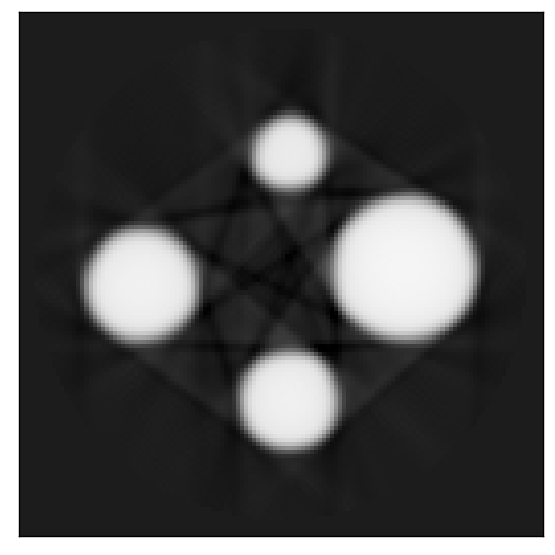} & 
\simprhspt \includegraphics[simsz]{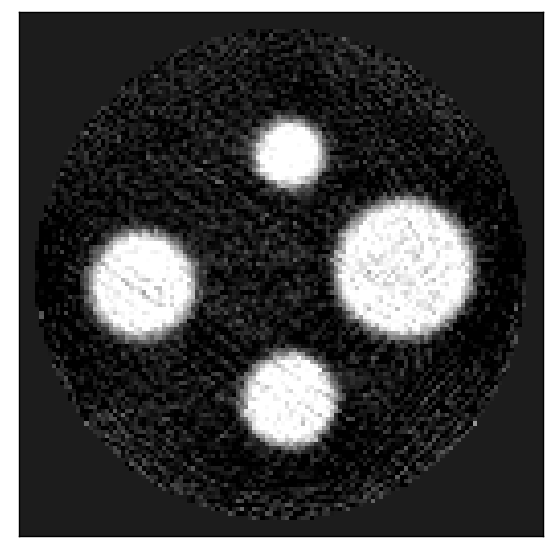} &
\simprhspt \includegraphics[simsz]{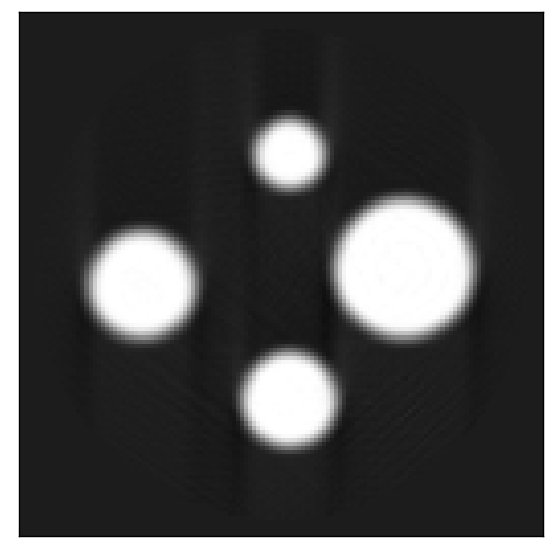} &
\simprhspt \includegraphics[simsz]{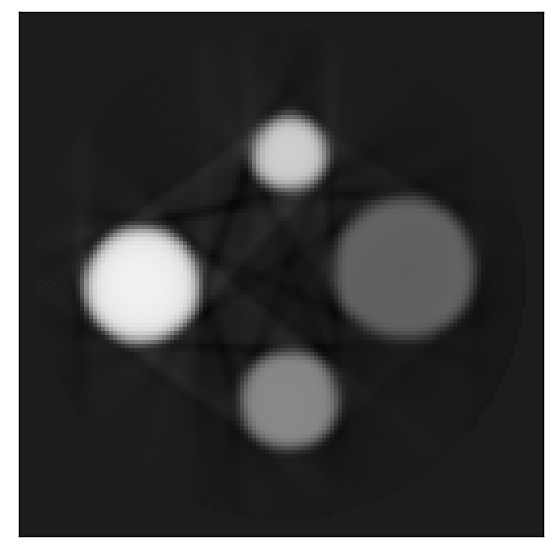} & 
\simprhspt \includegraphics[simsz]{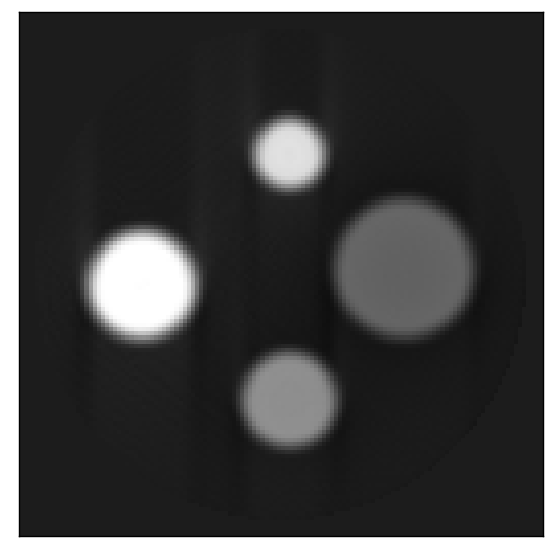} \\
\simprhspo (a) Paganin & 
\simprhspt (b) C-NLPR/0-Initial & 
\simprhspt (c) C-NLPR/Pag-Initial & 
\simprhspt (d) Paganin & 
\simprhspt (e) C-NLPR/Pag-Initial \\
\end{tabular}
\end{center}
\caption{\label{fig:simsinglepr} 
Tomographic reconstructions of the refractive index decrement from 
the phase images produced by the single-distance 
phase-retrieval (PR) algorithms of Paganin and C-NLPR/One-$\alpha$ (proposed).
(a-c) show reconstructions of the single material object.
(d, e) show reconstructions of the multi-material object.
(a-e) show planar slices along the 
$u-v$ axes that pass through 
the center of the reconstruction volume.
(a) and (d) show reconstructions using Paganin PR.
(b) shows the reconstruction using C-NLPR that is 
initialized with zeros for the phase image 
(label C-NLPR/0-Initial). 
(c) and (e) show reconstructions using C-NLPR 
that is initialized using Paganin PR (label C-NLPR/Pag-Initial). 
The gray values in (a-e) are scaled between
$-2.09\times 10^{-7}$ and $1.67\times 10^{-6}$.
C-NLPR with Paganin initialization produces the 
best reconstruction that minimize noise and artifacts.
}
\end{figure*}

\begin{figure}[htb!]
\begin{center}
\begin{tabular}{cc}
\hspace{-0.2in}
\includegraphics[width=1.6in]{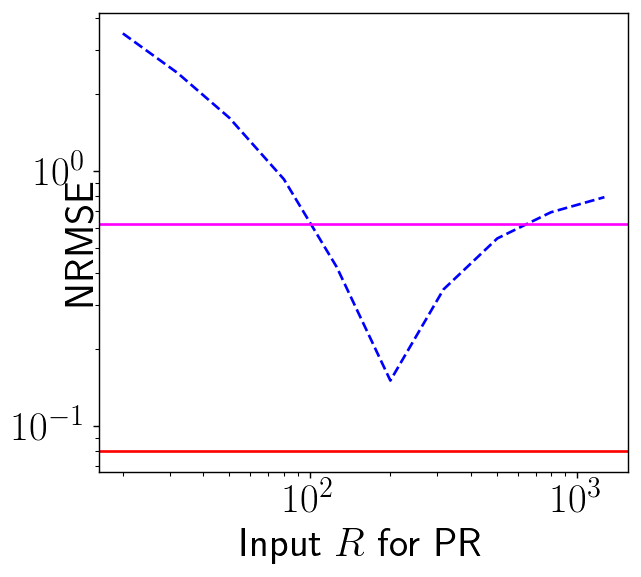} & 
\hspace{-0.2in}
\includegraphics[width=1.6in]{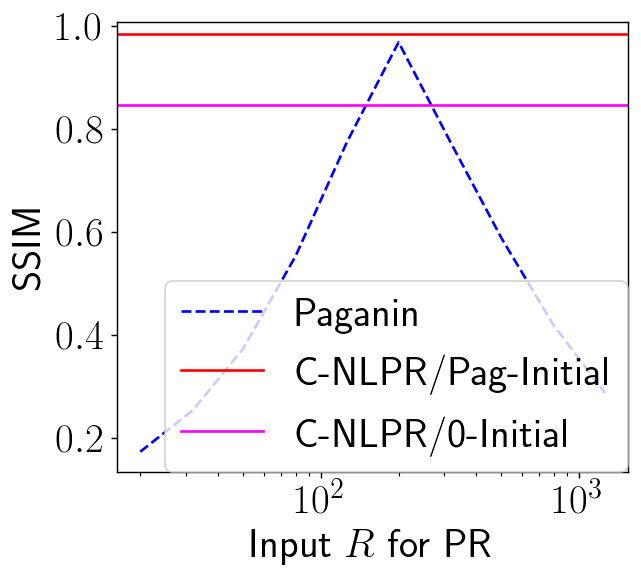} \\
(a) NRMSE & (b) SSIM \\
\end{tabular}\\
\end{center}
\caption{\label{fig:sweepsingledist} 
NRMSE and SSIM between reconstruction and ground-truth for 
the refractive index decrement as a function of 
the (inaccurate) propagation distance $R$ that is input 
to Paganin phase-retrieval (PR).
The true propagation distance $R$ used for simulation
is $200\,mm$.
The NRMSE is lowest and SSIM is highest when $R$ 
 is set equal to its true value of $200\,mm$.
Initializing C-NLPR with zeros for the phase results
in sub-optimal performance that is worse than Paganin PR.
C-NLPR initialized with Paganin PR (at the correct $R$) has the best performance.
Here, C-NLPR refers to C-NLPR/One-$\alpha$.
}
\end{figure}


\newcommand{\consinhsp}{\hspace{-0.1in}}
\begin{figure*}[htb!]
\begin{center}
\begin{tabular}{cccc}
\consinhsp
\includegraphics[width=1.7in]{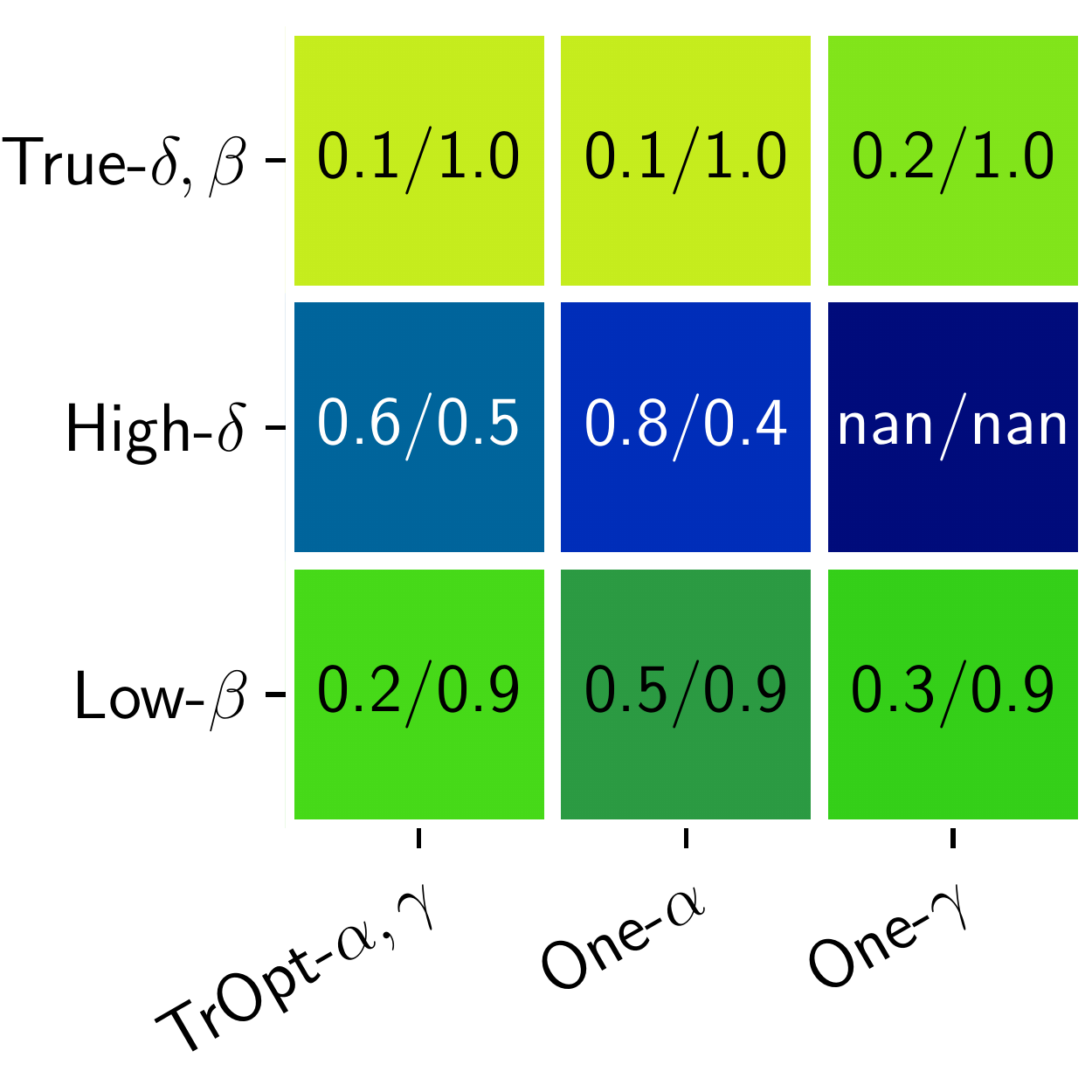} & 
\consinhsp
\includegraphics[width=1.7in]{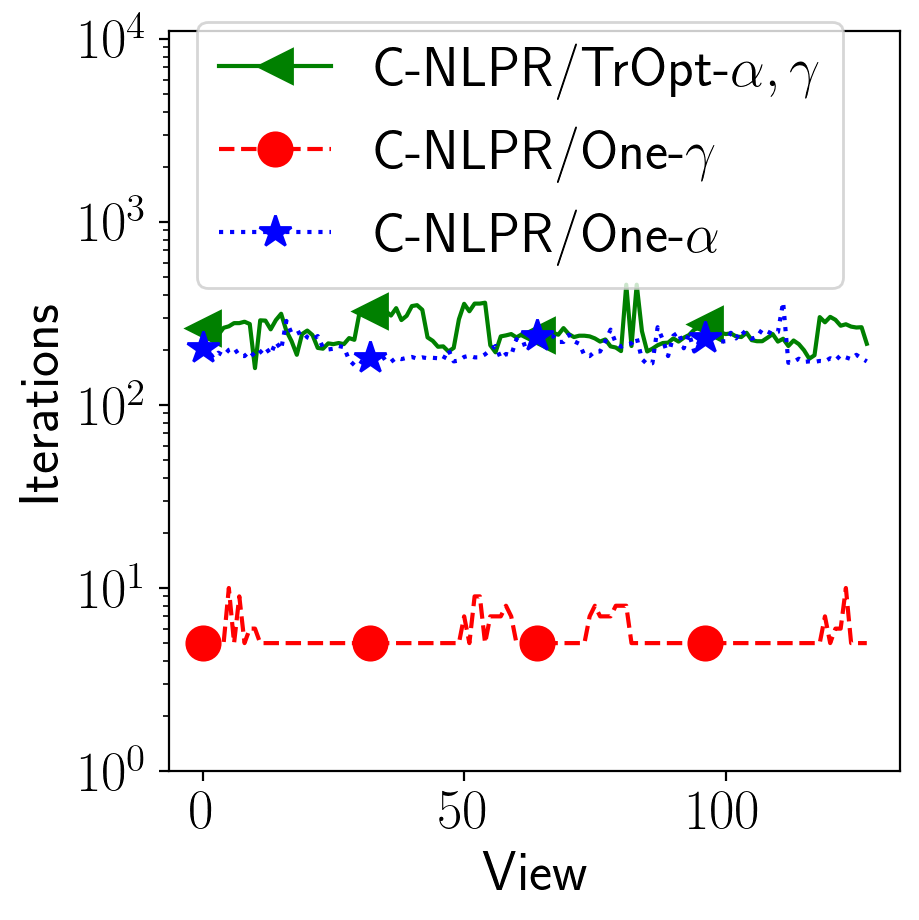} &
\consinhsp
\includegraphics[width=1.7in]{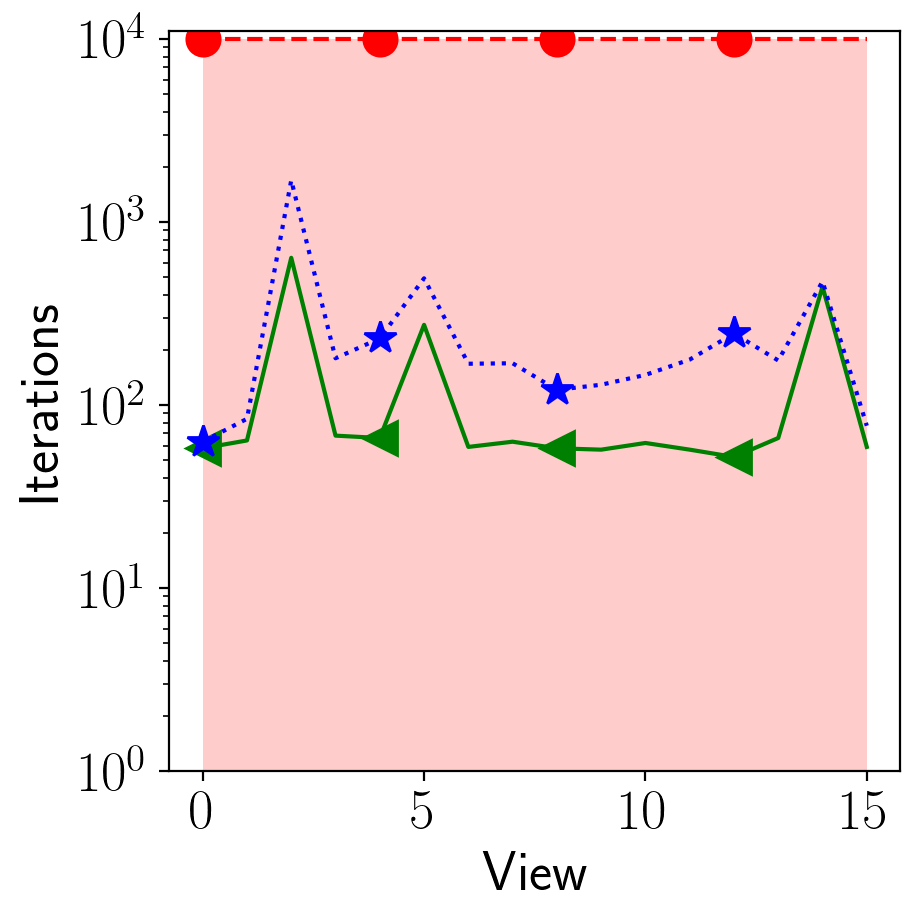} & 
\consinhsp
\includegraphics[width=1.7in]{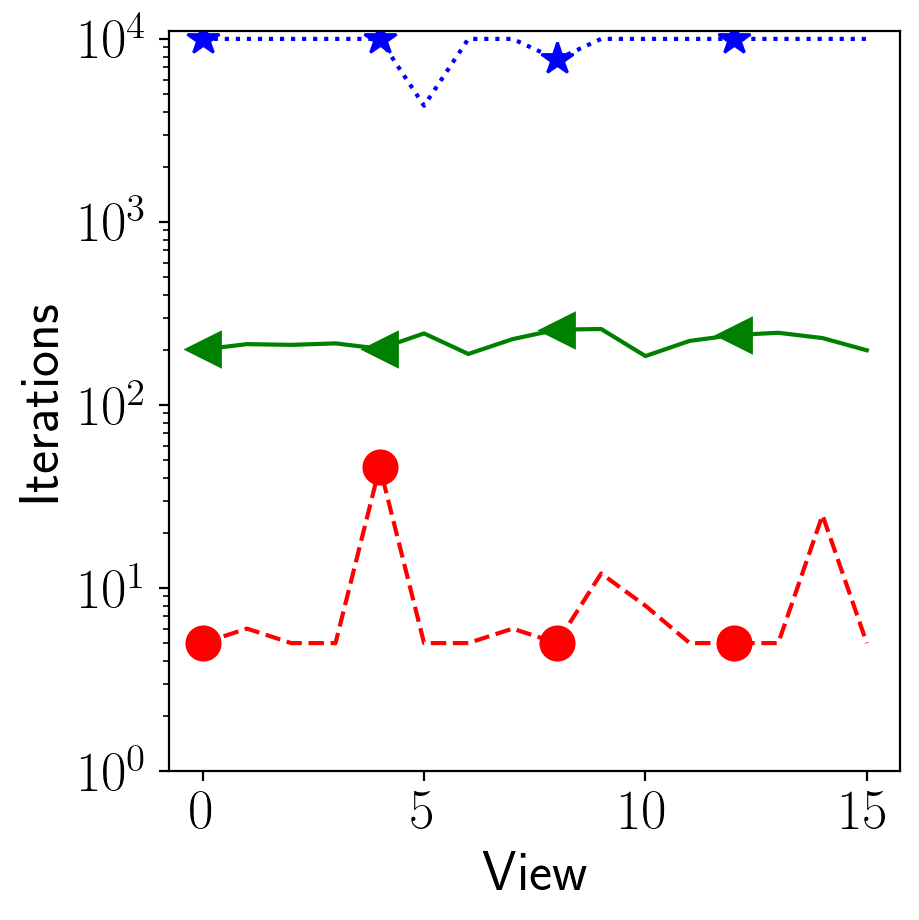} \\
(a) RMSE/SSIM &
(b) True-$\delta,\beta$ &
(c) High-$\delta$ & 
(d) Low-$\beta$
\end{tabular}\\
\end{center}
\caption{\label{fig:simconvgsingle}
Quantification of performance and speed 
of C-NLPR for various choices of $\alpha$ and $\gamma$ as 
described in sections \ref{sssec:alpone}, 
\ref{sssec:gamone}, and \ref{sssec:knowndb}.
Our analysis is repeated for various $\delta, \beta$ values.
True-$\delta,\beta$ uses the ground-truth 
$\delta,\beta$ of SiC.
High-$\delta$ refers to a $\delta$ that is $10\times$ 
the $\delta$ of SiC.
Low-$\beta$ refers to a $\beta$ that is $0.02\times$ 
the $\beta$ of SiC.
\textcolor{cgcol}{(a) shows the performance metrics of RMSE/SSIM within each square block.
(a) uses a linear colormap where yellow indicates the best, green is better, blue is bad, and dark blue is the worst performance.
(b, c, d) show the number of LBFGS iterations 
as a function of the view index.
Corrupted reconstructions caused by numerical instabilities 
is indicated by ``nan'' (not-a-number) in (a) 
and shaded with a translucent color in (c). }
From (a, c), we see that C-NLPR/One-$\gamma$ 
results in ``nan'' due to numerical instabilities 
for High-$\delta$. \textcolor{cgcol}{
Hence, we recommend avoiding C-NLPR/One-$\gamma$
for large ratios of $\delta/\beta$.}
With C-NLPR/One-$\alpha$, 
while we achieve good overall performance in (a), 
the number of iterations may reach large numbers
for Low-$\beta$ as shown in (d).
\textcolor{cgcol}{With C-NLPR/TrOpt-$\alpha,\gamma$, 
we achieve good overall performance (see (a)) 
without the need for a large number of iterations 
(see (b, c, d)).}
Legend for (c, d) is same as in (b).
}
\end{figure*}


\makeatletter
\define@key{Gin}{expsz}[true]{%
    \edef\@tempa{{Gin}{width=1.4in, keepaspectratio=true}}%
    \expandafter\setkeys\@tempa
}
\makeatother

\newcommand{\expprhspo}{\hspace{-0.1in}}
\newcommand{\expprhspt}{\hspace{-0.15in}}

\begin{figure*}[htb!]
\begin{center}
\begin{tabular}{ccccc}
\expprhspo
\includegraphics[expsz]{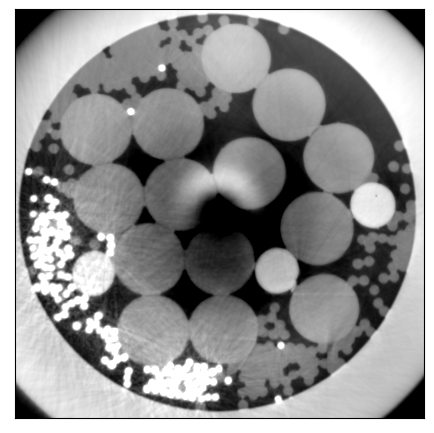} & 
\expprhspt
\includegraphics[expsz]{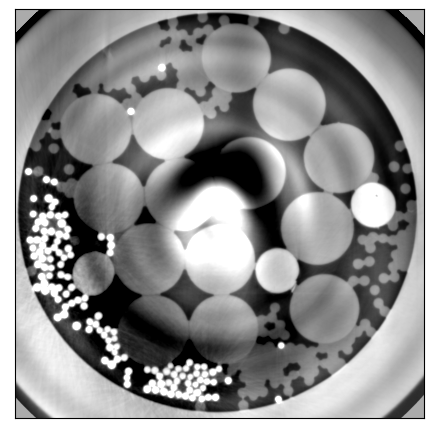} &
\expprhspt
\includegraphics[expsz]{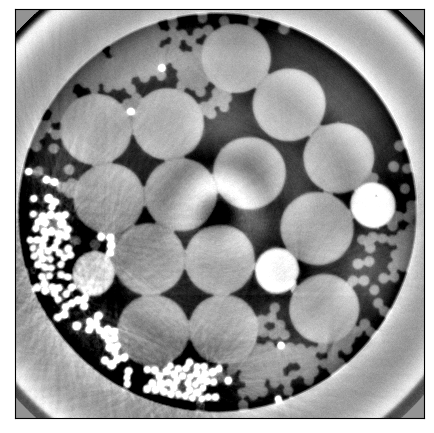} & 
\expprhspt
\includegraphics[expsz]{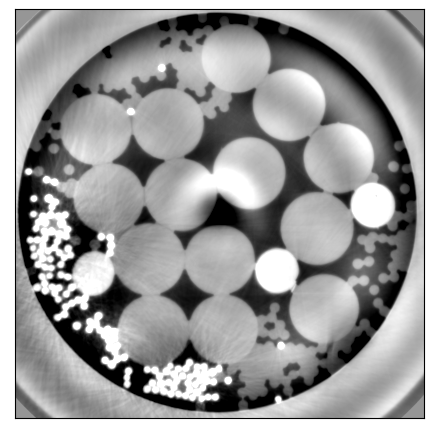} &
\expprhspt
\includegraphics[expsz]{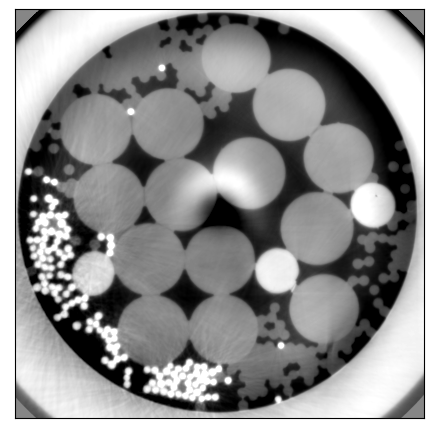} \vspace{-0.07in}\\
\expprhspo (a) CTF & 
\expprhspt (b) TIE & 
\expprhspt (c) Mixed & 
\expprhspt (d) U-NLPR/0-Initial & 
\expprhspt (e) U-NLPR/CTF-Initial \\
\expprhspo
\includegraphics[expsz]{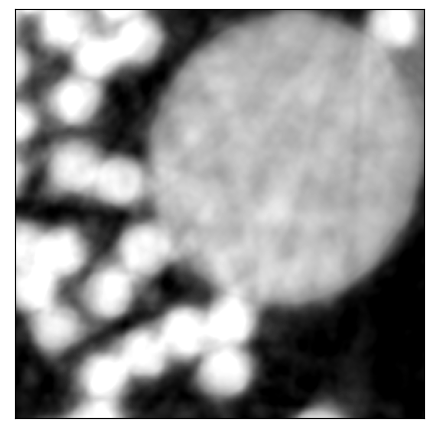} & 
\expprhspt
\includegraphics[expsz]{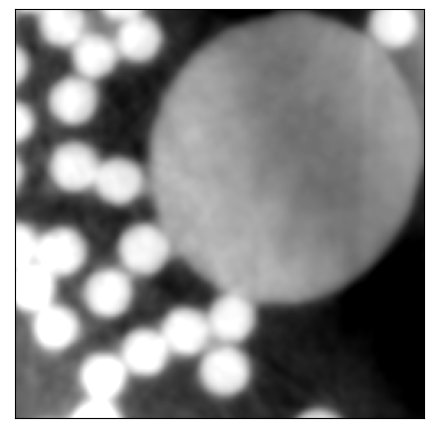} &
\expprhspt
\includegraphics[expsz]{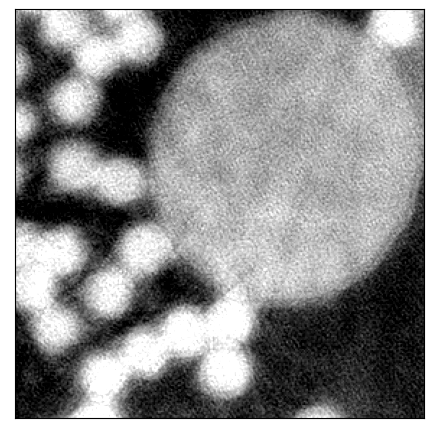} &
\expprhspt
\includegraphics[expsz]{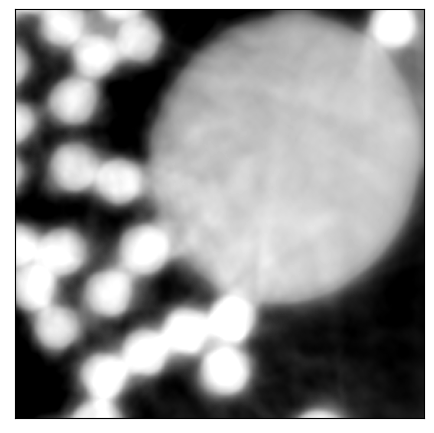} &
\expprhspt
\includegraphics[expsz]{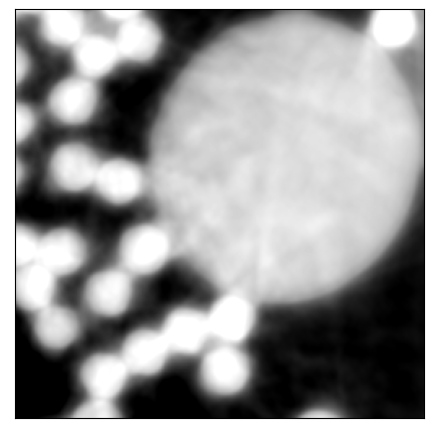} \vspace{-0.07in}\\
\expprhspo (f) CTF & 
\expprhspt (g) TIE & 
\expprhspt (h) Mixed & 
\expprhspt (i) U-NLPR/0-Initial & 
\expprhspt (j) U-NLPR/CTF-Initial \\
\expprhspo 
\includegraphics[expsz]{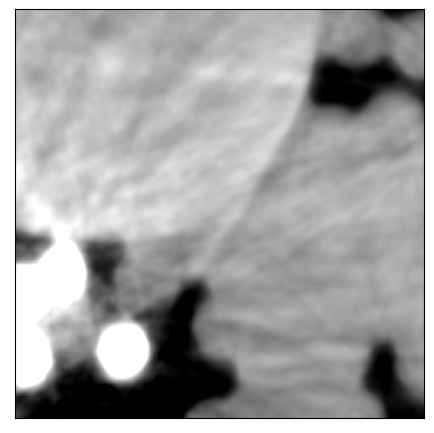} & 
\expprhspt
\includegraphics[expsz]{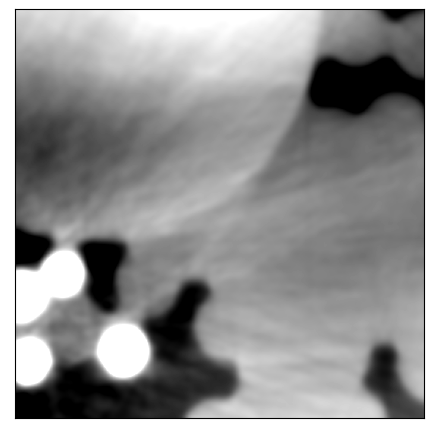} &
\expprhspt
\includegraphics[expsz]{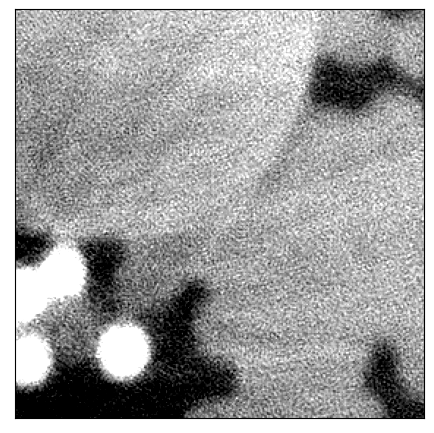} & 
\expprhspt
\includegraphics[expsz]{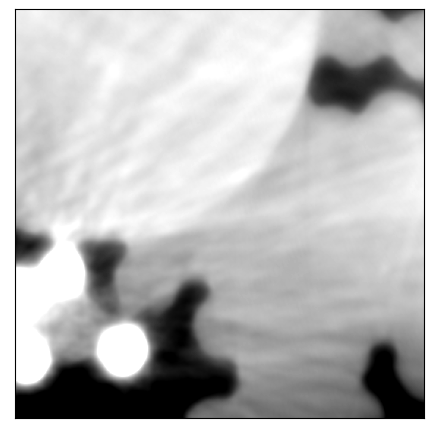} &
\expprhspt
\includegraphics[expsz]{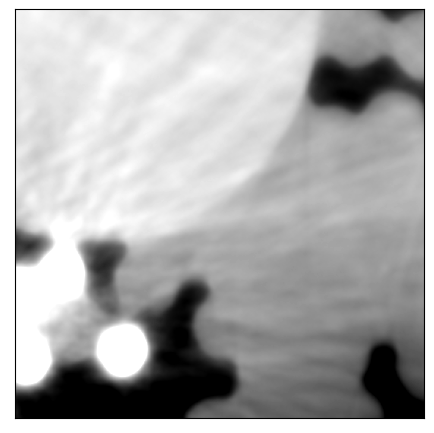} \vspace{-0.07in}\\
\expprhspo (k) CTF & 
\expprhspt (l) TIE & 
\expprhspt (m) Mixed & 
\expprhspt (n) U-NLPR/0-Initial & 
\expprhspt (o) U-NLPR/CTF-Initial 
\end{tabular}
\end{center}
\caption{\label{fig:expmultipr} 
Experimental data reconstruction comparison 
of the refractive index decrement for various
multi-distance phase-retrieval (PR) algorithms. 
(a-e) show planar reconstruction slices along the $u-v$ axes 
passing through the center of the volume. 
The reconstruction in (a-e) is cropped to show the region  
within the interior of the sample holder.
(f-j) and (k-o) zooms into two different regions 
of the reconstructions in (a-e).
Since the quantitative values vary substantially between 
PR methods (see section \ref{ssec:lowfreqloss}),
we scale the gray values of each image individually 
between the value percentiles of $5\%$ and $95\%$.
Both CTF and TIE produce substantial 
low frequency artifacts in (a), (b), (g), and (l).
Mixed reduces the artifacts but has increased noise as shown in (h, m).
U-NLPR, irrespective of the initialization, produces the 
best reconstructions that minimize noise and artifacts. 
In particular, U-NLPR with CTF initialization 
(using the fixed regularization in equation \eqref{eq:ctfregchoice})
produce less intense artifacts at the center of the images 
in (d, e) when compared to zero-initialization.
\textcolor{cgcol}{Unlike U-NLPR, 
the CTF, TIE, and Mixed PR results 
are at optimal regularization values
that were manually chosen to achieve the best visual quality of reconstructions.}
}
\end{figure*}


\makeatletter
\define@key{Gin}{expsinsz}[true]{%
    \edef\@tempa{{Gin}{width=1.5in, keepaspectratio=true}}%
    \expandafter\setkeys\@tempa
}
\makeatother

\begin{figure*}[htb!]
\begin{center}
\begin{tabular}{cccc}
\expprhspo
\includegraphics[expsinsz]{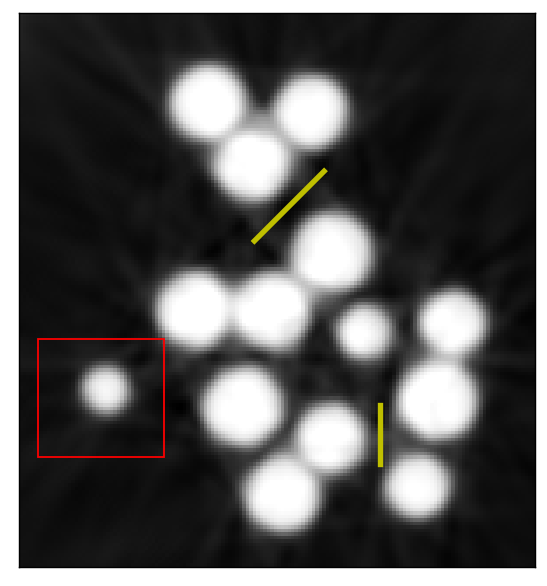} &
\expprhspt
\includegraphics[expsinsz]{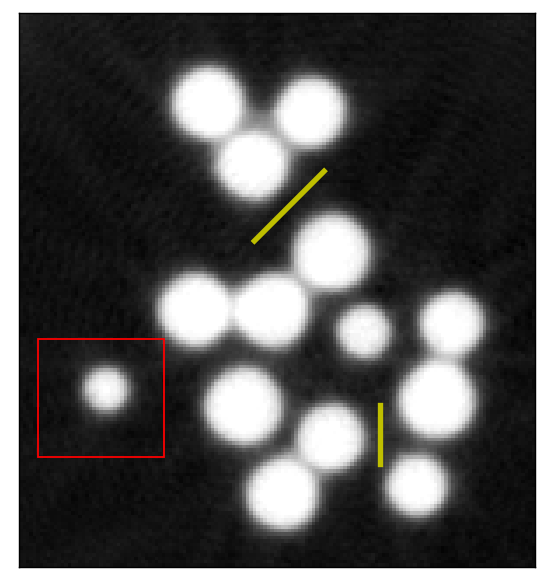} & 
\expprhspt
\includegraphics[width=1.6in]{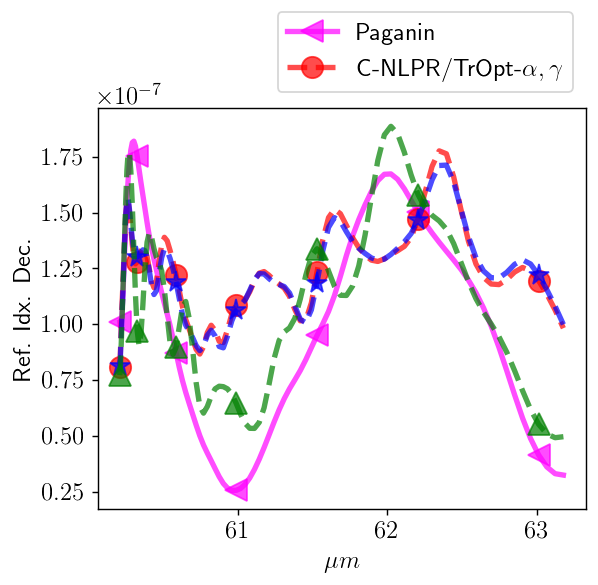} & 
\expprhspt
\includegraphics[width=1.6in]{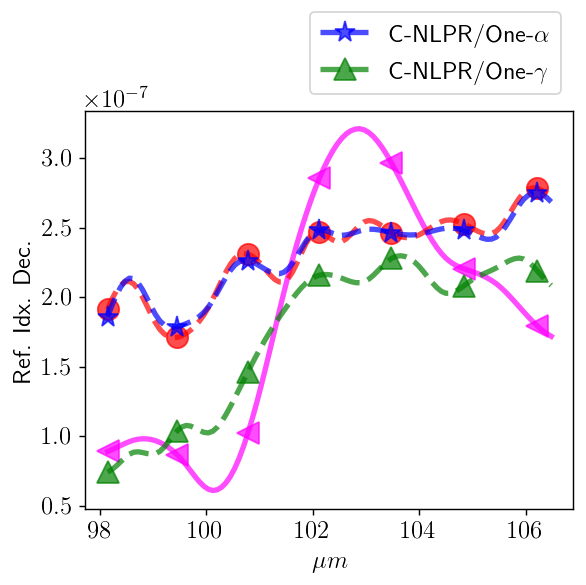} \vspace{-0.07in}\\
\expprhspo (a) Paganin PR & 
\expprhspt (b) C-NLPR/One-$\alpha$& 
\expprhspt (c) Top Line Profile & 
\expprhspt (d) Bottom Line Profile\\
\expprhspt
\includegraphics[expsinsz]{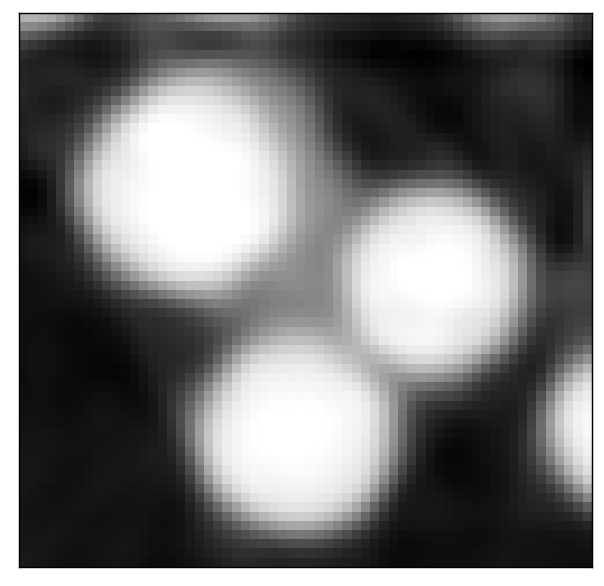} &
\expprhspt
\includegraphics[expsinsz]{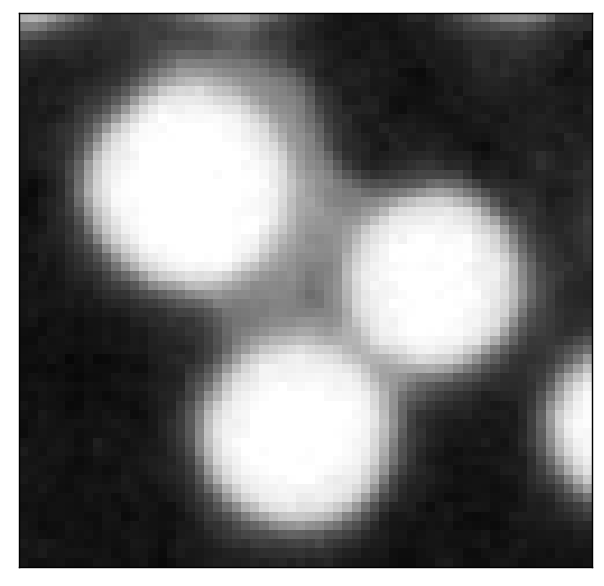} \vspace{-0.07in} &
\multicolumn{2}{c}{\includegraphics[width=3.2in]{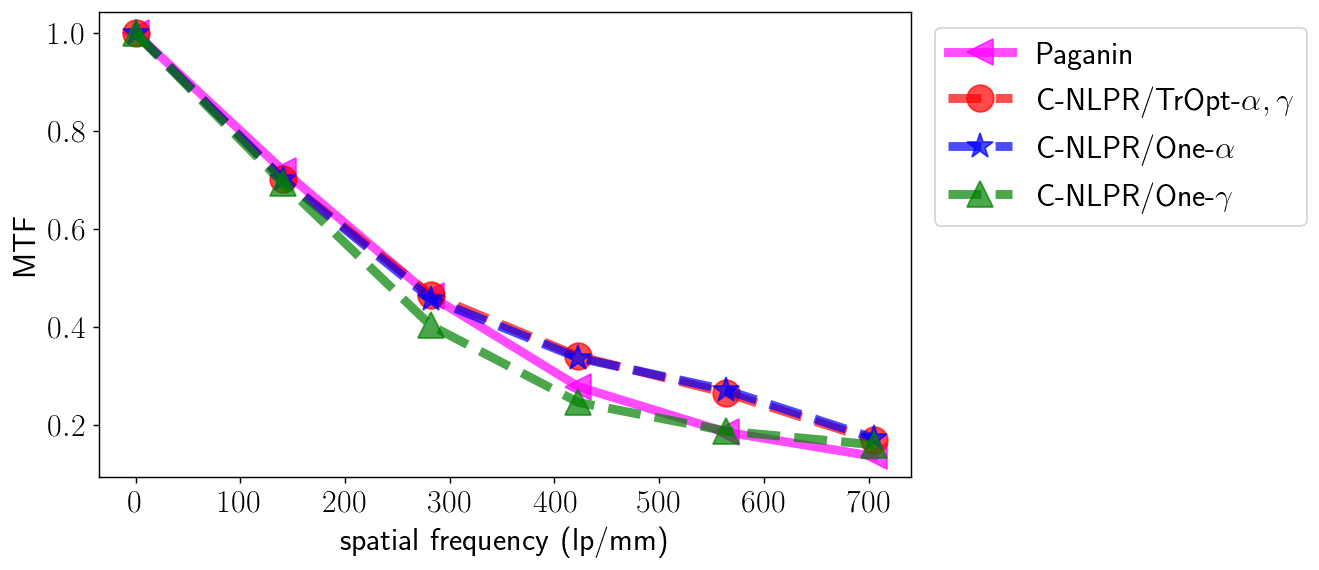}}\\
\expprhspo (e) Paganin PR Zoomed & 
\expprhspt (f) C-NLPR/One-$\alpha$ Zoomed & 
 \multicolumn{2}{c}{(g) 
Modulation Transfer Function (MTF)} \\
\end{tabular}
\end{center}
\caption{\label{fig:expsinrec}
Tomographic reconstruction of the refractive index decrement
from phase images retrieved using 
single-distance phase-retrieval (PR) methods.
(a) and (b) show the center slice ($u-v$ axes) of the reconstruction 
using Paganin PR and C-NLPR/One-$\alpha$ respectively.
(c) and (d) show line profile comparisons between 
Paganin and C-NLPR along the yellow marked lines in (a) and (b).
In (c, d), the large dynamic range for Paganin and 
C-NLPR/One-$\gamma$ indicate the presence of streak artifacts
since these line-profiles are in the background region of (a, b).
Both C-NLPR/TrOpt-$\alpha,\gamma$ and C-NLPR/One-$\alpha$ 
reduce streaking artifacts as illustrated in (c, d). 
(e, f) zooms into a region along a different slice
in the presence of a second material between the SiC fibers,
which demonstrates the sharper reconstruction using C-NLPR/One-$\alpha$
when compared to Paganin PR.
Images in (a, b, e, f) are scaled between $-1.79\times 10^{-7}$
and $1.43\times 10^{-6}$.
(c, d) use a common legend that is indicated at the top of the plots.
\textcolor{cgcol}{(g) is the modulation transfer function (MTF) 
for the sharpness of the disc inside the red square in (a, b).
(g) indicates sharper reconstructions 
using C-NLPR/One-$\alpha$ and C-NLPR/TrOpt-$\alpha,\gamma$
since the corresponding curves are above the curve for Paganin PR at the higher frequencies.
Our recommendation is to use C-NLPR/One-$\alpha$
that only requires knowledge of $\delta/\beta$.}
}
\end{figure*}

\begin{figure}[htb!]
\begin{center}
\begin{tabular}{c}
\hspace{-0.2in}
\includegraphics[width=3.5in]{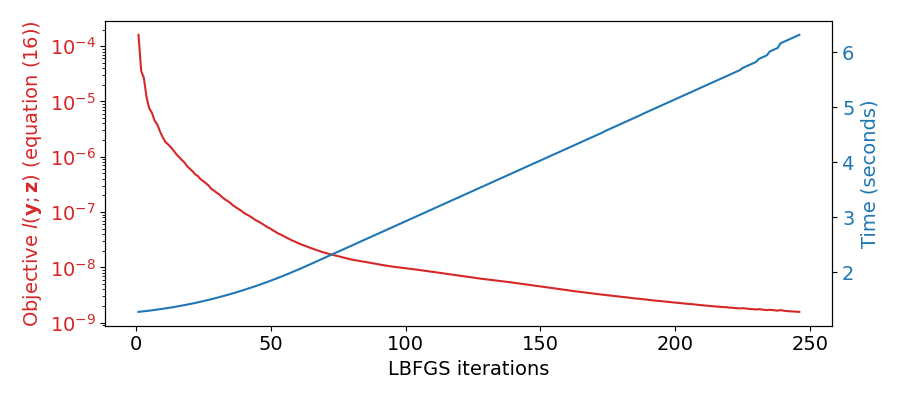} 
\vspace{-0.1in} \\
(a) Objective $l(\tb{y}; \tb{z})$ and run time \\
\hspace{-0.2in}
\includegraphics[width=3.5in]{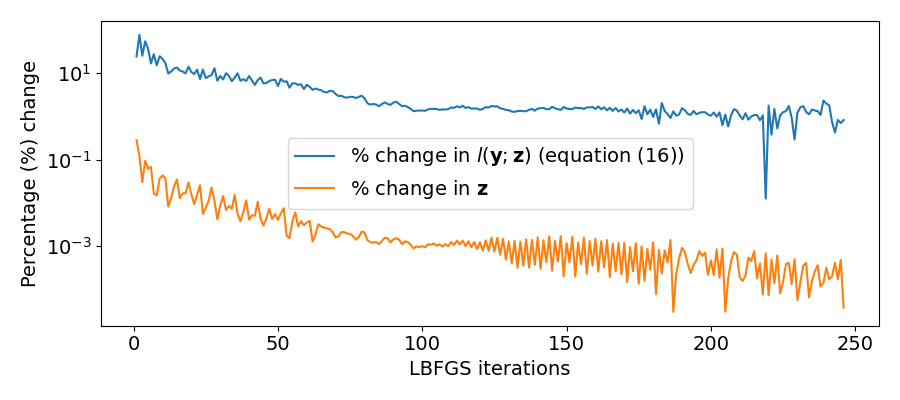} 
\vspace{-0.1in} \\
(b) \% change in $l(\tb{y}; \tb{z})$ and $\tb{z}$\\
\end{tabular}\\
\end{center}
\caption{\label{fig:expsinconvplot} 
\textcolor{cgcol}{
(a) is a plot of the objective function, $l(\tb{y}; \tb{z})$, 
and the total time elapsed for C-NLPR/One-$\alpha$
as a function of the iteration number of LBFGS. 
(b) is a plot of the percentage (\%) change in 
the objective function, $l(\tb{y}; \tb{z})$, and the 
percentage change in the estimated reconstruction, $\tb{z}$, 
as a function of the iteration number of LBFGS.
The total run time for C-NLPR/One-$\alpha$ was $6.3$ seconds.
This analysis is for C-NLPR/One-$\alpha$ at the first view angle for the experimental data results in 
Fig. \ref{fig:expsinrec}.}
}
\end{figure}


\subsection{Quantitative Evaluation}
\label{ssec:quanteval}
Our approach to extracting quantitative information from the
refractive index decrement reconstructions 
is to compare the reconstructed values
for the material-of-interest with the reconstructed values
of the background air. 
We assume that the theoretical refractive index decrement of the background air is $0$.
Unfortunately, this process of
quantitative evaluation cannot be automated 
since determination of the background air region is non-trivial
and application dependent. 
First, we compute the average value of the 
reconstruction in the background, denoted as $\delta^{(bg)}$.
Next, we compute the average value of the reconstruction
inside the material-of-interest, denoted as $\delta^{(m)}$.
Finally, we compute the background subtracted refractive index
decrement of the material-of-interest as,
\begin{equation}
    \label{eq:diffdelta}
    \delta^{(m:dif)} = \delta^{(m)}-\delta^{(bg)}.
\end{equation}
Background subtraction is necessary due to the loss of 
the average value for the refractive index in the reconstruction. 
After background subtraction, $\delta^{(m:dif)}$
is the quantitatively accurate refractive index decrement
that is comparable to the theoretical values.

\section{Results}
\label{sec:results}
\subsection{Simulated Data}
\label{ssec:simresults}
In this section, we compare the performance of our 
new NLPR algorithms against existing approaches using
simulated phase-contrast CT data of a 3D object 
at a monochromatic X-ray energy of $20\,keV$.
To avoid using the same forward model for both 
simulation and inversion,
we simulate X-ray images from
finely sampled objects
at very high resolutions
using equation \eqref{eq:detavgmod}.
The simulated phase-contrast CT data, 
the stack of X-ray images over all views, 
is then sub-sampled to a size of $128\times 80\times 128$ 
by block-averaging the simulated images over 
non-overlapping square-shaped windows of pixels.
Finally, we simulate Poisson-like noise in the measurements 
by adding zero-mean Gaussian noise with a standard deviation 
that is $0.1\%$ of the simulated data (equation \eqref{eq:addnoise}).

Cross-sectional slices through the simulated
object along with the corresponding simulated X-ray images 
with phase-contrast are shown in Fig. \ref{fig:simgtrad}.
First, we simulate a single-material object 
(Fig. \ref{fig:simgtrad} (a, b)) that consists of several spheres 
with differing diameters but made using the same material. 
The material used for the spheres is SiC 
with a refractive index decrement ($\delta$) 
of $1.67\times 10^{-6}$ and absorption index 
($\beta$) of $4.77\times 10^{-9}$ at an X-ray energy of $20\,keV$.
 For this object, the normalized X-ray image with 
 phase-contrast at a propagation distance of $R=200\,mm$ 
 is shown in Fig. \ref{fig:simgtrad} (e).
 Next, we simulate a multi-material object 
 (Fig. \ref{fig:simgtrad} (c, d))
 that consists of four spheres with varying diameters 
 and differing atomic composition. 
 For the spheres, we chose SiC ($\delta=1.67\times 10^{-6},
 \beta=4.77\times 10^{-9}$), Teflon ($\delta=1.1\times 10^{-6},
 \beta=9.09\times 10^{-10}$), Alumina ($\delta=2.03\times 10^{-6}, 
 \beta=3.97\times 10^{-9}$), and Polyimide ($\delta=7.61\times 
 10^{-7}, \beta=3.21\times 10^{-10}$) as the materials.
 The X-ray images with phase-contrast for this heterogeneous object 
 are shown in Fig. \ref{fig:simgtrad} (f-h) 
 at propagation distances, $R$, of $10\,mm$, $200\,mm$, and $400\,mm$.
 For both the single and multi-material objects, 
 the radii of the spheres were chosen to be
$12\,\mu m$, $16\,\mu m$, $20\,\mu m$, and $24\,\mu m$ respectively.
We simulate X-ray images of size $80 \times 128$ with 
pixel width of $1.29\,\mu m$ at $128$ tomographic views 
equally spaced over an angular range of $180^0$.
For phase-retrieval (PR), 
we use the square root of the normalized X-ray images 
(equation \eqref{eq:sqrtmeas}
in supplementary document).

A quantitative comparison of the refractive index decrement 
reconstructions using various multi-distance phase-retrieval 
algorithms is presented in Fig. \ref{fig:simswpmulti}.
At each tomographic view, 
we use phase-retrieval to estimate the phase images 
that are a measure of the phase shift induced
by the object on the X-ray field. 
Then, we use filtered back projection (FBP) to perform 
a tomographic reconstruction of the refractive index 
decrement from the phase images.
Fig. \ref{fig:simswpmulti} (a) is the normalized root mean 
squared error (NRMSE)\footnote{NRMSE is normalized using 
the averaged $l^2$ norm of the ground-truth.} between the 
reconstructions and the ground-truth.
Fig. \ref{fig:simswpmulti} (b) is the structural similarity 
index measure (SSIM)\footnote{For SSIM, the inputs are
linearly scaled such that the minimum and maximum values 
of ground-truth are mapped to $-1$ and $1$ respectively.
We use Gaussian weights with standard deviation of $8$ pixels.}.
We use the NRMSE and SSIM implementations from 
\textit{scikit-image} \cite{scikit-image}. 
We compute NRMSE and SSIM for the 
entire volume of the spheres after background
subtraction as described in the section
\ref{ssec:quanteval}.
For the conventional phase-retrieval 
methods \textcolor{cgcol}{\cite{langer_quantitative_2008, cloetens_quantitative_2002, guigay_mixed_2007}} 
of CTF, TIE, and Mixed,
we plot (dashed lines) the NRMSE and SSIM measures 
as a function of its regularization parameter.
To adequately reduce NRMSE and increase SSIM 
using TIE phase-retrieval, 
we must carefully tune its regularization hyper-parameter 
to lie within a narrow range of parameter values.
An arbitrary choice of either a very low or high parameter 
value for regularization will lead to sub-optimal 
results with TIE.
For CTF and Mixed phase-retrieval, 
it is possible to achieve low NRMSE and high SSIM by 
choosing a sufficiently small value for the regularization. 
As the regularization parameter is reduced, 
the performance of CTF does not degrade 
while Mixed leads to a marginal reduction in performance.
For CTF, we can minimize the NRMSE and maximize SSIM 
by choosing a sufficiently small fixed value for the 
regularization parameter.
In particular, we fix the regularization parameter value 
using \textcolor{cgcol}{equation} \eqref{eq:ctfregchoice} to avoid the need for parameter tuning.

We use U-NLPR from section \ref{ssec:unconxnlpr} for 
multi-distance phase-retrieval without any material constraints.
The performance of U-NLPR is influenced by the estimates of the phase and absorption that is used for initialization
of the optimization in equation \eqref{eq:unconxnlpr}. 
In Fig. \ref{fig:simswpmulti}, 
we investigate the use of phase images from CTF, TIE, or Mixed 
phase-retrieval as initial estimates for initialization of U-NLPR.
The performance of these conventional methods is dependent
on the chosen value of the regularization parameter.
In Fig. \ref{fig:simswpmulti}, 
we plot (solid lines) the NRMSE and SSIM
for U-NLPR as a function of the regularization parameter 
of the conventional method (i.e., CTF, TIE, and Mixed) 
used for initialization. 
We observe that U-NLPR always produces lower NRMSE 
and higher SSIM when compared to the estimates 
from the method used for initialization.
Importantly, we observe that the best performance of U-NLPR 
is achieved by initialization using the CTF algorithm 
with a sufficiently low value for the regularization as
specified in equation \eqref{eq:ctfregchoice}.
While the Mixed approach may also be suitable for 
initialization, we prefer the CTF approach due to the ease
of choosing the regularization using equation \eqref{eq:ctfregchoice}.
Note that the SSIM with Mixed PR decreases slightly at
very low regularization values. 

The reconstructions at the \textcolor{cgcol}{optimal regularization} with the highest SSIM using the 
conventional phase-retrieval algorithms
of TIE, CTF, and Mixed approaches are shown in 
Fig. \ref{fig:simmultipr} (a-c). 
\textcolor{cgcol}{While such an optimal selection is perhaps an unfair advantage to the conventional algorithms,
we did this to compare against the
best possible reconstructions.}
In Fig. \ref{fig:simmultipr} (d), we show the reconstruction
using U-NLPR that is initialized with zero values 
for the phase and absorption. 
In Fig. \ref{fig:simmultipr} (e), we show 
the U-NLPR reconstruction
that is initialized using CTF phase-retrieval that used the 
regularization of equation \eqref{eq:ctfregchoice}. 
Since U-NLPR does not use any regularization, 
we avoid the need for the tedious manual tuning of  
regularization hyper-parameters.
From Fig. \ref{fig:simmultipr} (a-c) 
and Fig. \ref{fig:simgtrad} (c), 
we observe that conventional phase-retrieval methods 
produce spurious streak artifacts. 
While the artifacts are faint in the case of TIE, 
both CTF and Mixed produce strong streak artifacts. 
When initialized with zeros for the phase and absorption, 
U-NLPR in Fig. \ref{fig:simmultipr} (d)
produce a better reconstruction with significantly 
fewer streak artifacts than the conventional methods.
When initialized using CTF with the fixed regularization
of equation \eqref{eq:ctfregchoice}, 
U-NLPR produces the best reconstruction with reduced artifacts
as shown in Fig. \ref{fig:simmultipr} (e). 

\textcolor{cgcol}{
We also compared our U-NLPR algorithm with a composite approach that uses TIE phase-retrieval followed by a Gerchberg-Saxton phase-retrieval (GSPR) algorithm in 
Fig. \ref{fig:gsprrec}.
While GSPR \cite{Gureyev_2003_composite, fienup1982phase} is primarily used for far-field diffraction 
imaging, it has also been successfully applied 
for imaging in the Fresnel region.
Fig. \ref{fig:gsprrec} (a) is the refractive index 
reconstruction using GSPR and TIE initialization. 
We observe artifacts that are similar to Fresnel diffraction
fringes in Fig. \ref{fig:gsprrec} (a).
U-NLPR with CTF initialization at 
the regularization from equation \eqref{eq:ctfregchoice} 
has higher SSIM than GSPR at any regularization.}

We visually compare the tomographic reconstruction performance
of single-distance phase-retrieval algorithms in Fig. \ref{fig:simsinglepr}.
Qualitative comparisons of phase-retrieval methods 
for single-material object and 
multi-material objects are shown in 
Fig. \ref{fig:simsinglepr} (a-c) 
and Fig. \ref{fig:simsinglepr} (d, e) respectively.
Paganin phase-retrieval \textcolor{cgcol}{\cite{paganin_simultaneous_2002}} produces streak artifacts 
as shown in Fig. \ref{fig:simsinglepr} (a, d).
We investigate the performance of C-NLPR from 
section \ref{ssec:conxnlpr} for non-linear phase-retrieval 
using the single-material constraint.
If C-NLPR is initialized with zeros for the phase images, 
C-NLPR reduces artifacts but significantly enhances 
the noise in Fig. \ref{fig:simsinglepr} (b).
C-NLPR using Paganin phase-retrieval for initialization 
provides the best quality reconstruction as shown in 
Fig. \ref{fig:simsinglepr} (c, e).
We use the C-NLPR/One-$\alpha$ method from section
\ref{sssec:alpone} in Fig. \ref{fig:simsinglepr}.
Surprisingly, we also observe that the qualitative performance 
does not degrade for \textcolor{cgcol}{the} multi-material object. 

Fig. \ref{fig:sweepsingledist}
shows a quantitative analysis of the reconstruction performance
for various single-distance phase-retrieval algorithms.
With Paganin phase-retrieval, it is common to achieve sharper 
reconstructions by artificially lowering the
propagation distance, $R$, that is input to the algorithm. 
However, from Fig. \ref{fig:sweepsingledist}, we see that an inaccurate setting for $R$ compared to its true value 
leads to sub-optimal NRMSE and SSIM values. 
The best performance for Paganin is achieved when $R$ 
is set equal to the true propagation distance of $200\,mm$.
Initializing C-NLPR with Paganin phase-retrieved images results 
in the lowest NRMSE and highest SSIM among all approaches. 
However, initializing C-NLPR with zero phase images produces 
sub-optimal reconstructions with large amounts of noise
as shown in Fig. \ref{fig:simsinglepr} (b).
In Fig. \ref{fig:simsinglepr} and Fig. \ref{fig:sweepsingledist}, we used 
the C-NLPR/One-$\alpha$ method 
described in section \ref{sssec:alpone}.

The convergence speed and reconstruction quality 
of C-NLPR are strongly influenced
by the choice of $\alpha$ and $\gamma$ 
as evidenced in Fig. \ref{fig:simconvgsingle}.
Fig. \ref{fig:simconvgsingle} (a) compares the 
NRMSE and SSIM for different constraint choices of 
C-NLPR/TrOpt-$\alpha,\gamma$ (section \ref{sssec:knowndb}), 
C-NLPR/One-$\alpha$ (section \ref{sssec:alpone}),
and C-NLPR/One-$\gamma$ (section \ref{sssec:gamone}).
The rows of Fig. \ref{fig:simconvgsingle} (a) correspond to 
different choices for the material properties of $\delta,\beta$ 
while the columns cycle through the various $\alpha, \gamma$ settings
used for imposing the single-material constraint.
The row label ``True-$\delta,\beta$'' indicates that the 
simulated object used the true $\delta,\beta$ values 
for SiC at $20\,keV$.
The label ``High-$\delta$'' indicates a very large value 
for $\delta$ that is $10\times$ higher than the $\delta$ of SiC.
The label ``Low-$\beta$'' indicates a very low value for $\beta$ 
that is $1/50\times$ the $\beta$ of SiC.
For the various materials characterized by its 
$\delta,\beta$ values, we see that C-NLPR/TrOpt-$\alpha,\gamma$ 
consistently produces the lowest NRMSE.
Using the SSIM measure, the results are a
tie between C-NLPR/TrOpt-$\alpha,\gamma$ 
and C-NLPR/One-$\alpha$.
However, the latter only uses knowledge of the ratio $\delta/\beta$.

In Fig. \ref{fig:simconvgsingle} (b-d), 
we plot the number of iterations for 
the LBFGS optimization algorithm as a 
function of the tomographic view index.
Fig. \ref{fig:simconvgsingle} (b), Fig. \ref{fig:simconvgsingle} (c),
and Fig. \ref{fig:simconvgsingle} (d) are 
plots of the number of iterations (to meet our convergence criteria)
for ``True-$\delta,\beta$'', ``High-$\delta$'', and
``Low-$\beta$'' respectively.
For the true values of $\delta,\beta$ of SiC, 
C-NLPR/One-$\gamma$ has the fastest convergence 
in Fig. \ref{fig:simconvgsingle} (b) 
but increased NRMSE in the first row of 
Fig. \ref{fig:simconvgsingle} (a).
In contrast, both C-NLPR/TrOpt-$\alpha,\gamma$
and C-NLPR/One-$\alpha$ converge slower but achieve lower NRMSE.
From Fig. \ref{fig:simconvgsingle} (c) and second row of 
Fig. \ref{fig:simconvgsingle} (a), 
we see that numerical instabilities associated 
with C-NLPR/One-$\gamma$ result in slow convergence and 
corrupted reconstructions due to the high $\delta$ value 
(see section \ref{sssec:gamone}).
From Fig. \ref{fig:simconvgsingle} (c),
C-NLPR/One-$\alpha$ is slow to converge and has a high NRMSE as indicated in the 
third row of Fig. \ref{fig:simconvgsingle} (a)
(see section \ref{sssec:alpone}).
The maximum number of iterations was fixed at $10^4$ 
for the LBFGS optimization. 
While C-NLPR/TrOpt-$\alpha,\gamma$ has the best convergence 
for all cases, it requires knowledge of both $\delta$ and $\beta$.
Alternatively, C-NLPR/One-$\alpha$ only uses knowledge of 
the ratio $\delta/\beta$ while also matching
C-NLPR/TrOpt-$\alpha,\gamma$ in convergence speed for 
``True-$\delta,\beta$'' and ``High-$\delta$''. 
Our recommendation is to use C-NLPR/One-$\alpha$
since it achieves the best overall performance
while only requiring knowledge of the ratio $\delta/\beta$.
\textcolor{cgcol}{Here, our objective is primarily to select 
feasible $\alpha$ and $\gamma$ values 
that impact the convergence speed of the algorithm.
Thus, small errors in the ratio of $\delta/\beta$ 
or the individual values of $\delta,\beta$ 
are inconsequential.
Only an order-of-magnitude change in 
$\alpha,\gamma$ will have a noticeable
impact on convergence.}

\subsection{Experimental Data}
\label{sec:expdatares}
For comparison of phase-retrieval methods using experimental data, 
a bundle of fibers comprising PolyEthylene Terephthalate (PET), 
PolyPropylene (PP), Aluminum ($Al$), and Aluminum Oxide ($Al_2O_3$)
with respective diameters of $200\,\mu m$, $28\,\mu m$, $125\,\mu m$, and $20\,\mu m$ 
were enclosed in a borosilicate capillary glass and scanned for 
tomography at the X-ray imaging beamline 8.3.2 of the Advanced Light
Source, Berkeley, California. Monochromator was set to deliver X-ray
beam with energy of $22\,keV$. A PCO edge camera combined with $10\times$
lens was used to collect X-ray images,
resulting in an effective pixel
size of approximately $0.65\,\mu m$. 
A total number of $1312$ X-ray images were recorded over $180^{\circ}$ 
range and using a $500\,ms$ exposure time. 
\textcolor{cgcol}{The object was placed at 
approximately $21\,m$ from the X-ray source 
and the detector was placed at
object-to-detector propagation distances of $30\,mm$, $100\,mm$, and $250\,mm$ 
for each scan.} 
The size of each X-ray image was $1686\times 2532$.
The normalized X-ray images 
are shown in Fig. \ref{fig:supexpmuldata}
of the supplementary document.

Tomographic reconstructions of the refractive index 
decrement from phase images reconstructed using
various multi-distance phase-retrieval algorithms 
are shown in Fig. \ref{fig:expmultipr}.
The regularization parameters for the conventional phase-retrieval
methods of CTF, TIE, and Mixed are manually tuned to obtain the best
visual quality of reconstructions. 
Substantial low frequency reconstruction artifacts are
observed in Fig. \ref{fig:expmultipr} (a, b, g, l) for 
CTF and TIE phase-retrieval.
Mixed phase-retrieval produces large amounts of reconstruction noise 
as shown in the zoomed images of Fig. \ref{fig:expmultipr} (h, m).
We note that the noise with Mixed can be reduced by adjusting 
the regularization but it also severely reduces 
the quality of reconstruction. 
U-NLPR produces the best reconstructions in 
Fig. \ref{fig:expmultipr} (d, e, i, j, n, o) 
that substantially reduces both low frequency artifacts and noise
when compared to the conventional methods.
Surprisingly, U-NLPR with zero-initialization results in a 
similar level of reconstruction quality as U-NLPR that is initialized
using CTF with regularization from equation \eqref{eq:ctfregchoice}.
Note that some of the artifacts in the centers of
Fig. \ref{fig:expmultipr} (a-e) are ring artifacts \cite{mohan_timbir_2015}
and we do not explore the correction of ring artifacts in this paper.
The complicated interaction between ring-artifact removal 
and phase-retrieval algorithms necessitate further investigation.
Fig. \ref{fig:supexpregsweep} in the 
supplementary document presents a 
quantitative comparison of the phase-retrieval
performance. 

For comparison of single-distance phase-retrieval algorithms, we obtained experimental X-ray CT data with phase-contrast
from the Advanced Light Source (ALS) Beamline 8.3.2.
CT data of SiC fibers was acquired at an 
object-to-detector distance of $R=98\,mm$
and X-ray energy of $20\,keV$.
The size of each X-ray image was $324\times 320$ and 
the pixel width was $0.645\,\mu m$.
X-ray images were acquired at $256$ different
views equally spaced over an angular range of $180$ degrees.
A normalized X-ray image
is shown in Fig. \ref{fig:supexpsindata}
of the supplementary document.

Tomographic reconstructions of the refractive index decrement
from phase images reconstructed using various single-distance
phase-retrieval algorithms are shown in Fig. \ref{fig:expsinrec}.
Fig. \ref{fig:expsinrec} (a, e) and (b, f) show the reconstructions
using Paganin and C-NLPR/One-$\alpha$ respectively.
The line profiles in Fig. \ref{fig:expsinrec} (c, d)
are along the yellow lines in Fig. \ref{fig:expsinrec} (a, b).
They highlight the reduction in streak artifacts 
using C-NLPR/One-$\alpha$ when compared to Paganin phase-retrieval.
The zoomed images in Fig. \textcolor{cgcol}{\ref{fig:expsinrec} (e, f)}
also demonstrate the significant enhancement in sharpness
using C-NLPR/One-$\alpha$ even in the presence of a second
material that is not modeled by the constraints of C-NLPR. 
\textcolor{cgcol}{Fig. \ref{fig:expsinrec} (g) is a plot
of the modulation transfer function (MTF) \cite{mtf_stan}
that indicates sharper reconstructions
from C-NLPR/One-$\alpha$ and
C-NLPR/TrOpt-$\alpha,\gamma$.}
\textcolor{cgcol}{A convergence analysis for the reconstructions in 
Fig. \ref{fig:expsinrec} is shown in Fig. \ref{fig:expsinconvplot}.
Fig. \ref{fig:expsinconvplot} (a) is a plot of the 
time elapsed and the objective function as a function 
of the LBFGS iterations.
Fig. \ref{fig:expsinconvplot} (b) is a plot
of the percentage change in the objective function
and the estimated values as a function 
of the iteration number.
The total time for phase-retrieval 
at one view was $6.3$ seconds
on a NVIDIA Tesla V100 GPU.}

\section{Conclusion}
For propagation-based X-ray phase-contrast tomography (XPCT),
we presented new non-linear phase-retrieval algorithms (NLPR)
to reconstruct the phase shift induced by the object
on the X-ray field. 
Then, we demonstrated reconstruction of the refractive index
decrement in 3D from the phase shift images
using tomographic reconstruction algorithms.
Our approaches do not require any manual tuning of image quality related hyper-parameters 
such as regularization. 
Our NLPR algorithms are suitable for both single-distance
and multi-distance XPCT while also supporting 
constraints on the material composition.
For single-distance XPCT, 
we demonstrated phase-retrieval (PR) under the constraint
of single-material or phase-absorption proportionality.
Our NLPR algorithms produced the best reconstructions
based on both quantitative metrics of accuracy
as well as qualitative evaluation of artifact and noise reduction. 
The superior performance of NLPR is a result of employing non-linear measurement models that are more
accurate than the linear approximate models used by 
existing linear PR approaches.
For multi-distance XPCT, we show that zero-initialization
of the phase for NLPR may be sufficient, but the 
best performance is achieved by initializing with 
the Contrast Transfer Function (CTF) PR.
For single-distance XPCT, we show that NLPR produces the
best reconstruction when initialized with Paganin PR.

\section*{Acknowledgments}
LLNL-JRNL-847272. This work was performed under the auspices of the U.S. Department of Energy by Lawrence Livermore National Laboratory under Contract DE-AC52-07NA27344.
LDRD 22-ERD-011 was used to fund
the research in this paper.
This research used resources of the Advanced Light Source, which is a DOE Office of Science User Facility under contract no. DE-AC02-05CH11231.


\bibliographystyle{IEEEtran}
\bibliography{IEEEabrv,paper}

\begin{thebibliography}{10}
\providecommand{\url}[1]{#1}
\csname url@samestyle\endcsname
\providecommand{\newblock}{\relax}
\providecommand{\bibinfo}[2]{#2}
\providecommand{\BIBentrySTDinterwordspacing}{\spaceskip=0pt\relax}
\providecommand{\BIBentryALTinterwordstretchfactor}{4}
\providecommand{\BIBentryALTinterwordspacing}{\spaceskip=\fontdimen2\font plus
\BIBentryALTinterwordstretchfactor\fontdimen3\font minus
  \fontdimen4\font\relax}
\providecommand{\BIBforeignlanguage}[2]{{%
\expandafter\ifx\csname l@#1\endcsname\relax
\typeout{** WARNING: IEEEtran.bst: No hyphenation pattern has been}%
\typeout{** loaded for the language `#1'. Using the pattern for}%
\typeout{** the default language instead.}%
\else
\language=\csname l@#1\endcsname
\fi
#2}}
\providecommand{\BIBdecl}{\relax}
\BIBdecl

\bibitem{karunakaran_factors_2015}
C.~Karunakaran, R.~Lahlali, N.~Zhu, A.~M. Webb, M.~Schmidt, K.~Fransishyn,
  G.~Belev, T.~Wysokinski, J.~Olson, D.~M.~L. Cooper, and E.~Hallin,
  ``\BIBforeignlanguage{en}{Factors influencing real time internal structural
  visualization and dynamic process monitoring in plants using
  synchrotron-based phase contrast {X}-ray imaging},''
  \emph{\BIBforeignlanguage{en}{Scientific Reports}}, vol.~5, no.~1, Jul. 2015.

\bibitem{croton_situ_2018}
L.~C.~P. Croton, K.~S. Morgan, D.~M. Paganin, L.~T. Kerr, M.~J. Wallace, K.~J.
  Crossley, S.~L. Miller, N.~Yagi, K.~Uesugi, S.~B. Hooper, and M.~J. Kitchen,
  ``\BIBforeignlanguage{en}{In situ phase contrast {X}-ray brain {CT}},''
  \emph{\BIBforeignlanguage{en}{Scientific Reports}}, vol.~8, no.~1, Jul. 2018.

\bibitem{wilkins1996phase}
S.~Wilkins, T.~E. Gureyev, D.~Gao, A.~Pogany, and A.~Stevenson,
  ``Phase-contrast imaging using polychromatic hard x-rays,'' \emph{Nature},
  vol. 384, no. 6607, pp. 335--338, 1996.

\bibitem{zielke_degradation_2015}
L.~Zielke, C.~Barchasz, S.~Waluś, F.~Alloin, J.-C. Leprêtre, A.~Spettl,
  V.~Schmidt, A.~Hilger, I.~Manke, J.~Banhart, R.~Zengerle, and S.~Thiele,
  ``\BIBforeignlanguage{en}{Degradation of {Li}/{S} {Battery} {Electrodes} {On}
  {3D} {Current} {Collectors} {Studied} {Using} {X}-ray {Phase} {Contrast}
  {Tomography}},'' \emph{\BIBforeignlanguage{en}{Scientific Reports}}, vol.~5,
  no.~1, p. 10921, Jun. 2015.

\bibitem{parab_high_2016}
N.~D. Parab, Z.~A. Roberts, M.~H. Harr, J.~O. Mares, A.~D. Casey, I.~E. Gunduz,
  M.~Hudspeth, B.~Claus, T.~Sun, K.~Fezzaa, S.~F. Son, and W.~W. Chen, ``High
  speed {X}-ray phase contrast imaging of energetic composites under dynamic
  compression,'' \emph{Applied Physics Letters}, vol. 109, no.~13, p. 131903,
  Sep. 2016.

\bibitem{sun_morphological_2016}
F.~Sun, L.~Zielke, H.~Markötter, A.~Hilger, D.~Zhou, R.~Moroni, R.~Zengerle,
  S.~Thiele, J.~Banhart, and I.~Manke, ``Morphological {Evolution} of
  {Electrochemically} {Plated}/{Stripped} {Lithium} {Microstructures}
  {Investigated} by {Synchrotron} {X}-ray {Phase} {Contrast} {Tomography},''
  \emph{ACS Nano}, vol.~10, no.~8, pp. 7990--7997, Aug. 2016.

\bibitem{seastwood_three-dimensional_2015}
D.~S. Eastwood, P.~M. Bayley, H.~Jung Chang, O.~O. Taiwo, J.~Vila-Comamala,
  D.~J. L. Brett, C.~Rau, P.~J. Withers, P.~R. Shearing, C.~P. Grey, and
  P.~D. Lee, ``\BIBforeignlanguage{en}{Three-dimensional characterization of
  electrodeposited lithium microstructures using synchrotron {X}-ray phase
  contrast imaging},'' \emph{\BIBforeignlanguage{en}{Chemical Communications}},
  vol.~51, no.~2, pp. 266--268, 2015.

\bibitem{mayo2003x}
S.~C. Mayo, T.~J. Davis, T.~E. Gureyev, P.~R. Miller, D.~Paganin, A.~Pogany,
  A.~W. Stevenson, and S.~Wilkins, ``X-ray phase-contrast microscopy and
  microtomography,'' \emph{Optics express}, vol.~11, no.~19, pp. 2289--2302,
  2003.

\bibitem{pacile_advantages_2018}
S.~Pacilè, P.~Baran, C.~Dullin, M.~Dimmock, D.~Lockie, J.~Missbach-Guntner,
  H.~Quiney, M.~McCormack, S.~Mayo, D.~Thompson, Y.~Nesterets, C.~Hall,
  K.~Pavlov, Z.~Prodanovic, M.~Tonutti, A.~Accardo, J.~Fox, S.~Tavakoli~Taba,
  S.~Lewis, P.~Brennan, D.~Hausermann, G.~Tromba, and T.~Gureyev,
  ``\BIBforeignlanguage{en}{Advantages of breast cancer visualization and
  characterization using synchrotron radiation phase-contrast tomography},''
  \emph{\BIBforeignlanguage{en}{Journal of Synchrotron Radiation}}, vol.~25,
  no.~5, pp. 1460--1466, Sep. 2018.

\bibitem{bravin_x-ray_2012}
A.~Bravin, P.~Coan, and P.~Suortti, ``\BIBforeignlanguage{en}{X-ray
  phase-contrast imaging: from pre-clinical applications towards clinics},''
  \emph{\BIBforeignlanguage{en}{Physics in Medicine and Biology}}, vol.~58,
  no.~1, pp. R1--R35, Dec. 2012.

\bibitem{fernandez_phase_2012}
V.~Fernandez, E.~Buffetaut, E.~Maire, J.~Adrien, V.~Suteethorn, and
  P.~Tafforeau, ``\BIBforeignlanguage{en}{Phase {Contrast} {Synchrotron}
  {Microtomography}: {Improving} {Noninvasive} {Investigations} of {Fossil}
  {Embryos} {In} {Ovo}},'' \emph{\BIBforeignlanguage{en}{Microscopy and
  Microanalysis}}, vol.~18, no.~1, pp. 179--185, Feb. 2012.

\bibitem{friis_phase-contrast_2007}
E.~M. Friis, P.~R. Crane, K.~R. Pedersen, S.~Bengtson, P.~C.~J. Donoghue, G.~W.
  Grimm, and M.~Stampanoni, ``\BIBforeignlanguage{en}{Phase-contrast {X}-ray
  microtomography links {Cretaceous} seeds with {Gnetales} and
  {Bennettitales}},'' \emph{\BIBforeignlanguage{en}{Nature}}, vol. 450, no.
  7169, Nov. 2007.

\bibitem{gureyev2008some}
T.~E. Gureyev, Y.~I. Nesterets, A.~W. Stevenson, P.~R. Miller, A.~Pogany, and
  S.~W. Wilkins, ``Some simple rules for contrast, signal-to-noise and
  resolution in in-line x-ray phase-contrast imaging,'' \emph{Optics express},
  vol.~16, no.~5, pp. 3223--3241, 2008.

\bibitem{burvall_phase_2011}
A.~Burvall, U.~Lundström, P.~A.~C. Takman, D.~H. Larsson, and H.~M. Hertz,
  ``\BIBforeignlanguage{EN}{Phase retrieval in {X}-ray phase-contrast imaging
  suitable for tomography},'' \emph{\BIBforeignlanguage{EN}{Optics Express}},
  vol.~19, no.~11, pp. 10\,359--10\,376, May 2011.

\bibitem{langer_quantitative_2008}
M.~Langer, P.~Cloetens, J.-P. Guigay, and F.~Peyrin,
  ``\BIBforeignlanguage{en}{Quantitative comparison of direct phase retrieval
  algorithms in in-line phase tomography},''
  \emph{\BIBforeignlanguage{en}{Medical Physics}}, vol.~35, no.~10, pp.
  4556--4566, 2008.

\bibitem{kak_principles_2001}
A.~C. Kak and M.~Slaney, \emph{Principles of {Computerized} {Tomographic}
  {Imaging}}.\hskip 1em plus 0.5em minus 0.4em\relax Society of Industrial and
  Applied Mathematics, 2001.

\bibitem{paganin_simultaneous_2002}
D.~Paganin, S.~C. Mayo, T.~E. Gureyev, P.~R. Miller, and S.~W. Wilkins,
  ``\BIBforeignlanguage{en}{Simultaneous phase and amplitude extraction from a
  single defocused image of a homogeneous object},''
  \emph{\BIBforeignlanguage{en}{Journal of Microscopy}}, vol. 206, no.~1, 2002.

\bibitem{beltran_2d_2010}
M.~A. Beltran, D.~M. Paganin, K.~Uesugi, and M.~J. Kitchen,
  ``\BIBforeignlanguage{EN}{{2D} and {3D} {X}-ray phase retrieval of
  multi-material objects using a single defocus distance},''
  \emph{\BIBforeignlanguage{EN}{Optics Express}}, vol.~18, no.~7, pp.
  6423--6436, Mar. 2010.

\bibitem{gureyev_optical_2004}
T.~E. Gureyev, T.~J. Davis, A.~Pogany, S.~C. Mayo, and S.~W. Wilkins,
  ``\BIBforeignlanguage{en}{Optical phase retrieval by use of first {Born}- and
  {Rytov}-type approximations},'' \emph{\BIBforeignlanguage{en}{Applied
  Optics}}, vol.~43, no.~12, p. 2418, Apr. 2004.

\bibitem{bronnikov_reconstruction_1999}
A.~V. Bronnikov, ``\BIBforeignlanguage{en}{Reconstruction formulas in
  phase-contrast tomography},'' \emph{\BIBforeignlanguage{en}{Optics
  Communications}}, vol. 171, no.~4, pp. 239--244, Dec. 1999.

\bibitem{wu_x-ray_nodate}
X.~Wu, H.~Liu, and A.~Yan, ``X-ray phase-attenuation duality and phase
  retrieval,'' \emph{Opt. Lett.}, vol.~30, no.~4, pp. 379--381, Feb 2005.

\bibitem{Chen2013comparison}
\BIBentryALTinterwordspacing
R.~C. Chen, L.~Rigon, and R.~Longo, ``Comparison of single distance phase
  retrieval algorithms by considering different object composition and the
  effect of statistical and structural noise,'' \emph{Opt. Express}, vol.~21,
  no.~6, pp. 7384--7399, Mar 2013. [Online]. Available:
  \url{https://opg.optica.org/oe/abstract.cfm?URI=oe-21-6-7384}
\BIBentrySTDinterwordspacing

\bibitem{chen2011phase}
R.~Chen, H.~Xie, L.~Rigon, R.~Longo, E.~Castelli, and T.~Xiao, ``Phase
  retrieval in quantitative x-ray microtomography with a single
  sample-to-detector distance,'' \emph{Optics Letters}, vol.~36, no.~9, pp.
  1719--1721, 2011.

\bibitem{paganin2020boosting}
D.~M. Paganin, V.~Favre-Nicolin, A.~Mirone, A.~Rack, J.~Villanova, M.~P.
  Olbinado, V.~Fernandez, J.~C. da~Silva, and D.~Pelliccia, ``Boosting spatial
  resolution by incorporating periodic boundary conditions into single-distance
  hard-x-ray phase retrieval,'' \emph{Journal of Optics}, vol.~22, no.~11, p.
  115607, 2020.

\bibitem{yu_evaluation_2018}
B.~Yu, L.~Weber, A.~Pacureanu, M.~Langer, C.~Olivier, P.~Cloetens, and
  F.~Peyrin, ``\BIBforeignlanguage{EN}{Evaluation of phase retrieval approaches
  in magnified {X}-ray phase nano computerized tomography applied to bone
  tissue},'' \emph{\BIBforeignlanguage{EN}{Optics Express}}, vol.~26, no.~9,
  pp. 11\,110--11\,124, Apr. 2018.

\bibitem{zabler_optimization_2005}
S.~Zabler, P.~Cloetens, J.-P. Guigay, J.~Baruchel, and M.~Schlenker,
  ``Optimization of phase contrast imaging using hard x rays,'' \emph{Review of
  Scientific Instruments}, vol.~76, no.~7, p. 073705, Jul. 2005.

\bibitem{langer_regularization_2010}
M.~Langer, P.~Cloetens, and F.~Peyrin, ``Regularization of {Phase} {Retrieval}
  {With} {Phase}-{Attenuation} {Duality} {Prior} for 3-{D} {Holotomography},''
  \emph{IEEE Transactions on Image Processing}, vol.~19, no.~9, pp. 2428--2436,
  Sep. 2010.

\bibitem{cloetens_quantitative_2002}
P.~Cloetens, W.~Ludwig, E.~Boller, L.~Helfen, L.~Salvo, R.~Mache, and
  M.~Schlenker, ``Quantitative phase contrast tomography using coherent
  synchrotron radiation,'' in \emph{Developments in {X}-{Ray} {Tomography}
  {III}}, vol. 4503.\hskip 1em plus 0.5em minus 0.4em\relax SPIE, Jan. 2002,
  pp. 82--91.

\bibitem{guigay_mixed_2007}
J.~P. Guigay, M.~Langer, R.~Boistel, and P.~Cloetens,
  ``\BIBforeignlanguage{en}{Mixed transfer function and transport of intensity
  approach for phase retrieval in the {Fresnel} region},''
  \emph{\BIBforeignlanguage{en}{Optics Letters}}, vol.~32, no.~12, p. 1617,
  Jun. 2007.

\bibitem{gureyev2006pcoherence}
\BIBentryALTinterwordspacing
T.~Gureyev, Y.~Nesterets, D.~Paganin, A.~Pogany, and S.~Wilkins, ``Linear
  algorithms for phase retrieval in the fresnel region. 2. partially coherent
  illumination,'' \emph{Optics Communications}, vol. 259, no.~2, pp. 569--580,
  2006. [Online]. Available:
  \url{https://www.sciencedirect.com/science/article/pii/S0030401805010357}
\BIBentrySTDinterwordspacing

\bibitem{gureyev2004linear}
\BIBentryALTinterwordspacing
T.~Gureyev, A.~Pogany, D.~Paganin, and S.~Wilkins, ``Linear algorithms for
  phase retrieval in the fresnel region,'' \emph{Optics Communications}, vol.
  231, no.~1, pp. 53--70, 2004. [Online]. Available:
  \url{https://www.sciencedirect.com/science/article/pii/S0030401803023320}
\BIBentrySTDinterwordspacing

\bibitem{mohan_direct_2016}
K.~A. Mohan, X.~Xiao, and C.~A. Bouman, ``Direct model-based tomographic
  reconstruction of the complex refractive index,'' in \emph{2016 {IEEE}
  {International} {Conference} on {Image} {Processing} ({ICIP})}, Sep. 2016,
  pp. 1754--1758.

\bibitem{davidoiu_non-linear_2011}
V.~Davidoiu, B.~Sixou, M.~Langer, and F.~Peyrin,
  ``\BIBforeignlanguage{en}{Non-linear iterative phase retrieval based on
  {Frechet} derivative},'' \emph{\BIBforeignlanguage{en}{Optics Express}},
  vol.~19, no.~23, p. 22809, Nov. 2011.

\bibitem{davidoiu_non-linear_2012}
------, ``Non-linear iterative phase retrieval based on {Frechet} derivative
  and projection operators,'' in \emph{2012 9th {IEEE} {International}
  {Symposium} on {Biomedical} {Imaging} ({ISBI})}, May 2012, pp. 106--109.

\bibitem{davidoiu_nonlinear_2012}
------, ``Nonlinear {Phase} {Retrieval} {Using} {Projection} {Operator} and
  {Iterative} {Wavelet} {Thresholding},'' \emph{IEEE Signal Processing
  Letters}, vol.~19, no.~9, pp. 579--582, Sep. 2012.

\bibitem{davidoiu_nonlinear_2013}
------, ``\BIBforeignlanguage{en}{Nonlinear approaches for the single-distance
  phase retrieval problem involving regularizations with sparsity
  constraints},'' \emph{\BIBforeignlanguage{en}{Applied Optics}}, vol.~52,
  no.~17, p. 3977, Jun. 2013.

\bibitem{Ruhlandt_2014_3DPR}
\BIBentryALTinterwordspacing
A.~Ruhlandt, M.~Krenkel, M.~Bartels, and T.~Salditt, ``Three-dimensional phase
  retrieval in propagation-based phase-contrast imaging,'' \emph{Phys. Rev. A},
  vol.~89, p. 033847, Mar 2014. [Online]. Available:
  \url{https://link.aps.org/doi/10.1103/PhysRevA.89.033847}
\BIBentrySTDinterwordspacing

\bibitem{Moosmann_2010_Nonlinear}
\BIBentryALTinterwordspacing
J.~Moosmann, R.~Hofmann, A.~V. Bronnikov, and T.~Baumbach, ``Nonlinear phase
  retrieval from single-distance radiograph,'' \emph{Opt. Express}, vol.~18,
  no.~25, pp. 25\,771--25\,785, Dec 2010. [Online]. Available:
  \url{https://opg.optica.org/oe/abstract.cfm?URI=oe-18-25-25771}
\BIBentrySTDinterwordspacing

\bibitem{Mom_2022_PrimalDual}
\BIBentryALTinterwordspacing
K.~Mom, M.~Langer, and B.~Sixou, ``Nonlinear primal--dual algorithm for the
  phase and absorption retrieval from a single phase contrast image,''
  \emph{Opt. Lett.}, vol.~47, no.~20, pp. 5389--5392, Oct 2022. [Online].
  Available: \url{https://opg.optica.org/ol/abstract.cfm?URI=ol-47-20-5389}
\BIBentrySTDinterwordspacing

\bibitem{Maretzke_2016_RegNewton}
\BIBentryALTinterwordspacing
S.~Maretzke, M.~Bartels, M.~Krenkel, T.~Salditt, and T.~Hohage, ``Regularized
  newton methods for x-ray phase contrast and general imaging problems,''
  \emph{Opt. Express}, vol.~24, no.~6, pp. 6490--6506, Mar 2016. [Online].
  Available: \url{https://opg.optica.org/oe/abstract.cfm?URI=oe-24-6-6490}
\BIBentrySTDinterwordspacing

\bibitem{Mom_2023_DeepGaussNewt}
K.~Mom, M.~Langer, and B.~Sixou, ``Deep gauss-newton for phase retrieval,''
  \emph{Opt. Lett.}, vol.~48, no.~5, pp. 1136--1139, Mar 2023.

\bibitem{Wu_2022_DeepConcatNet}
Y.~Wu, L.~Zhang, S.~Guo, L.~Zhang, F.~Gao, M.~Jia, and Z.~Zhou, ``Enhanced
  phase retrieval via deep concatenation networks for in-line x-ray phase
  contrast imaging,'' \emph{Physica Medica}, vol.~95, pp. 41--49, 2022.

\bibitem{Rucha_2023_RobustDL}
R.~Deshpande, A.~Avachat, F.~J. Brooks, and M.~A. Anastasio, ``Investigating
  the robustness of a deep learning-based method for quantitative phase
  retrieval from propagation-based x-ray phase contrast measurements under
  laboratory conditions,'' \emph{Physics in Medicine \& Biology}, 2023.

\bibitem{Li_22_PhyNNSparse}
F.~Li, Y.~Zhao, S.~Han, D.~Ji, Y.~Li, M.~Zheng, W.~Lv, J.~Jian, X.~Zhao, and
  C.~Hu, ``Physics-informed deep neural network reconstruction framework for
  propagation-based x ray phase-contrast computed tomography with sparse-view
  projections,'' \emph{Opt. Lett.}, vol.~47, no.~16, pp. 4259--4262, Aug 2022.

\bibitem{Wu_2022_PhyNNTIE}
X.~Wu, Z.~Wu, S.~C. Shanmugavel, H.~Z. Yu, and Y.~Zhu, ``Physics-informed
  neural network for phase imaging based on transport of intensity equation,''
  \emph{Opt. Express}, vol.~30, no.~24, pp. 43\,398--43\,416, Nov 2022.

\bibitem{Li_2022_U_NetDL}
S.~Z. Li, M.~G. French, K.~M. Pavlov, and H.~T. Li, ``{Shallow U-Net deep
  learning approach for phase retrieval in propagation-based phase-contrast
  Imaging},'' in \emph{Developments in X-Ray Tomography XIV}, B.~M{\"u}ller and
  G.~Wang, Eds., vol. 12242, International Society for Optics and
  Photonics.\hskip 1em plus 0.5em minus 0.4em\relax SPIE, 2022, p. 122421Q.

\bibitem{Zhang_2021_PhaseGAN}
\BIBentryALTinterwordspacing
Y.~Zhang, M.~A. Noack, P.~Vagovic, K.~Fezzaa, F.~Garcia-Moreno, T.~Ritschel,
  and P.~Villanueva-Perez, ``Phasegan: a deep-learning phase-retrieval approach
  for unpaired datasets,'' \emph{Opt. Express}, vol.~29, no.~13, pp.
  19\,593--19\,604, Jun 2021. [Online]. Available:
  \url{https://opg.optica.org/oe/abstract.cfm?URI=oe-29-13-19593}
\BIBentrySTDinterwordspacing

\bibitem{mohan_constrained_2020}
K.~A. Mohan, D.~Y. Parkinson, and J.~A. Cuadra, ``Constrained {Non}-{Linear}
  {Phase} {Retrieval} for {Single} {Distance} {Xray} {Phase} {Contrast}
  {Tomography},'' \emph{Electronic Imaging}, vol. 2020, no.~14, pp.
  146--1--146--8, Jan. 2020.

\bibitem{johnson_univariate_nodate}
N.~L. Johnson, A.~W. Kemp, and S.~Kotz, \emph{Univariate discrete
  distributions}.\hskip 1em plus 0.5em minus 0.4em\relax John Wiley \& Sons,
  2005, vol. 444.

\bibitem{Gureyev_2003_composite}
\BIBentryALTinterwordspacing
T.~Gureyev, ``Composite techniques for phase retrieval in the fresnel region,''
  \emph{Optics Communications}, vol. 220, no.~1, pp. 49--58, 2003. [Online].
  Available:
  \url{https://www.sciencedirect.com/science/article/pii/S0030401803013531}
\BIBentrySTDinterwordspacing

\bibitem{NEURIPS2019_9015}
A.~Paszke, S.~Gross, F.~Massa, A.~Lerer, J.~Bradbury, G.~Chanan, T.~Killeen,
  Z.~Lin, N.~Gimelshein, L.~Antiga, A.~Desmaison, A.~Kopf, E.~Yang, Z.~DeVito,
  M.~Raison, A.~Tejani, S.~Chilamkurthy, B.~Steiner, L.~Fang, J.~Bai, and
  S.~Chintala, ``Pytorch: An imperative style, high-performance deep learning
  library,'' in \emph{Advances in Neural Information Processing Systems
  32}.\hskip 1em plus 0.5em minus 0.4em\relax Curran Associates, Inc., 2019,
  pp. 8024--8035.

\bibitem{liu_limited_nodate}
D.~C. Liu and J.~Nocedal, ``On the limited memory bfgs method for large scale
  optimization,'' \emph{Mathematical programming}, vol.~45, no. 1-3, pp.
  503--528, 1989.

\bibitem{nocedal_updating_nodate}
J.~Nocedal, ``Updating quasi-newton matrices with limited storage,''
  \emph{Mathematics of computation}, vol.~35, no. 151, pp. 773--782, 1980.

\bibitem{lbfgs_hjmshi}
H.-J.~M. Shi and D.~Mudigere, ``{P}y{T}orch-{LBFGS}: A {P}y{T}orch
  implementation of {L-BFGS},'' \emph{https://github.com/hjmshi/PyTorch-LBFGS},
  2018.

\bibitem{paganin1998noninterferometric}
D.~Paganin and K.~A. Nugent, ``Noninterferometric phase imaging with partially
  coherent light,'' \emph{Physical review letters}, vol.~80, no.~12, p. 2586,
  1998.

\bibitem{harris2020array}
C.~R. Harris, K.~J. Millman, S.~J. van~der Walt, R.~Gommers, P.~Virtanen,
  D.~Cournapeau, E.~Wieser, J.~Taylor, S.~Berg, N.~J. Smith, R.~Kern, M.~Picus,
  S.~Hoyer, M.~H. van Kerkwijk, M.~Brett, A.~Haldane, J.~F. del R{\'{i}}o,
  M.~Wiebe, P.~Peterson, P.~G{\'{e}}rard-Marchant, K.~Sheppard, T.~Reddy,
  W.~Weckesser, H.~Abbasi, C.~Gohlke, and T.~E. Oliphant, ``Array programming
  with {NumPy},'' \emph{Nature}, vol. 585, no. 7825, pp. 357--362, Sep. 2020.

\bibitem{herraez_fast_2002}
M.~A. Herráez, D.~R. Burton, M.~J. Lalor, and M.~A. Gdeisat,
  ``\BIBforeignlanguage{en}{Fast two-dimensional phase-unwrapping algorithm
  based on sorting by reliability following a noncontinuous path},''
  \emph{\BIBforeignlanguage{en}{Applied Optics}}, vol.~41, no.~35, p. 7437,
  Dec. 2002.

\bibitem{scikit-image}
S.~van~der Walt, J.~L. {S}ch\"onberger, J.~{Nunez-Iglesias}, F.~{B}oulogne,
  J.~D. {W}arner, N.~{Y}ager, E.~{G}ouillart, T.~{Y}u, and the scikit-image
  contributors, ``scikit-image: image processing in {P}ython,'' \emph{PeerJ},
  vol.~2, p. e453, 6 2014.

\bibitem{langer_priors_2014}
\BIBentryALTinterwordspacing
M.~Langer, P.~Cloetens, B.~Hesse, H.~Suhonen, A.~Pacureanu, K.~Raum, and
  F.~Peyrin, ``Priors for {X}-ray in-line phase tomography of heterogeneous
  objects,'' \emph{Philosophical Transactions of the Royal Society A:
  Mathematical, Physical and Engineering Sciences}, vol. 372, no. 2010, p.
  20130129, Mar. 2014. [Online]. Available:
  \url{https://royalsocietypublishing.org/doi/10.1098/rsta.2013.0129}
\BIBentrySTDinterwordspacing

\bibitem{petruccelli_transport_2013}
\BIBentryALTinterwordspacing
J.~C. Petruccelli, L.~Tian, and G.~Barbastathis, ``\BIBforeignlanguage{EN}{The
  transport of intensity equation for optical path length recovery using
  partially coherent illumination},'' \emph{\BIBforeignlanguage{EN}{Optics
  Express}}, vol.~21, no.~12, pp. 14\,430--14\,441, Jun. 2013. [Online].
  Available:
  \url{https://www.osapublishing.org/oe/abstract.cfm?uri=oe-21-12-14430}
\BIBentrySTDinterwordspacing

\bibitem{fienup1982phase}
J.~R. Fienup, ``Phase retrieval algorithms: a comparison,'' \emph{Applied
  optics}, vol.~21, no.~15, pp. 2758--2769, 1982.

\bibitem{mohan_timbir_2015}
K.~A. Mohan, S.~V. Venkatakrishnan, J.~W. Gibbs, E.~B. Gulsoy, X.~Xiao,
  M.~De~Graef, P.~W. Voorhees, and C.~A. Bouman, ``{TIMBIR}: {A} {Method} for
  {Time}-{Space} {Reconstruction} {From} {Interlaced} {Views},'' \emph{IEEE
  Transactions on Computational Imaging}, vol.~1, no.~2, pp. 96--111, Jun.
  2015.

\bibitem{mtf_stan}
E.~Committee, \emph{ASTM E1695-95 Standard Test Method for Measurement of
  Computed Tomography (CT) System Performance}.\hskip 1em plus 0.5em minus
  0.4em\relax ASTM International, 1995.

\end{thebibliography}

\clearpage

\setcounter{figure}{0}
\renewcommand{\thefigure}{S\arabic{figure}}

\section{Supplementary Material}
\subsection{Data Normalization}
\label{sec:supdatanorm}
Bright-field (a.k.a. flat-field) and dark-field measurements
are acquired to appropriately normalize the CT scans. 
Bright-field refers to measurements made with the X-ray beam
but without the object and dark-field refers to measurements made in the absence of the X-ray beam.
\textcolor{cgcol}{Typically, bright-field
measurements are made at the same propagation distances as used for the phase-contrast CT scans}.

For normalization, we make certain simplifying approximations on $f_I(u,v)$ that also serve to simplify the subsequent problem of phase-retrieval.
 First, we approximate the incident field $f_I(u,v)$ as a plane 
 wave with constant phase. Without loss of generality, we assume that this constant phase is zero since any information on constant phase terms is lost (equation \eqref{eq:normdetmeas}). 
Next, we normalize the detector measurements using the bright- and dark-fields. 
Lastly, we compute the square root of the normalized detector image before running the algorithms presented in section \ref{sec:nlprall}.
Let $b(j,k)$ and $d(j,k)$ denote the bright-field and dark-field
measurements respectively. 
\textcolor{cgcol}{If $\tilde{y}(j,k)$ denotes the detector image with the object, 
the square root of the normalized detector image is given by,
\begin{equation}
\label{eq:sqrtmeas}
y(j,k) = \sqrt{\frac{\tilde{y}(j,k)-d(j,k)}{b(j,k)-d(j,k)}}.
\end{equation}
Here, 
$(\tilde{y}(j,k)-d(j,k))$ is a measure of
$\lt|f_D(j\Delta,k\Delta)\rt|^2$
and $(b(j,k)-d(j,k))$ is a measure
of $\lt|f_I(j\Delta,k\Delta)\rt|^2$.
}

\subsection{Simulated Data}
\subsubsection{Multi-distance phase-retrieval}
\begin{figure*}[htb!]
\begin{center}
\begin{tabular}{ccccc}
\multicolumn{5}{c}{Multi-distance Phase-Retrieval (PR) followed by FBP Reconstruction (continued from Fig. \ref{fig:simmultipr})} \\
\simprhspo
\includegraphics[simsz]{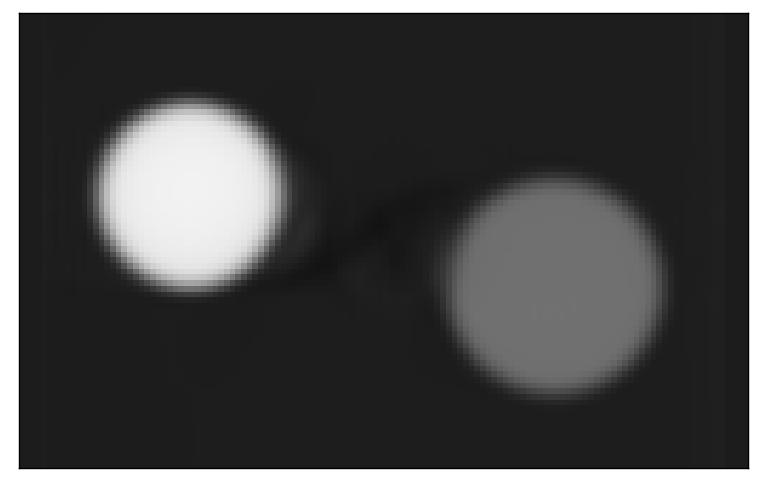} &
\simprhspt
\includegraphics[simsz]{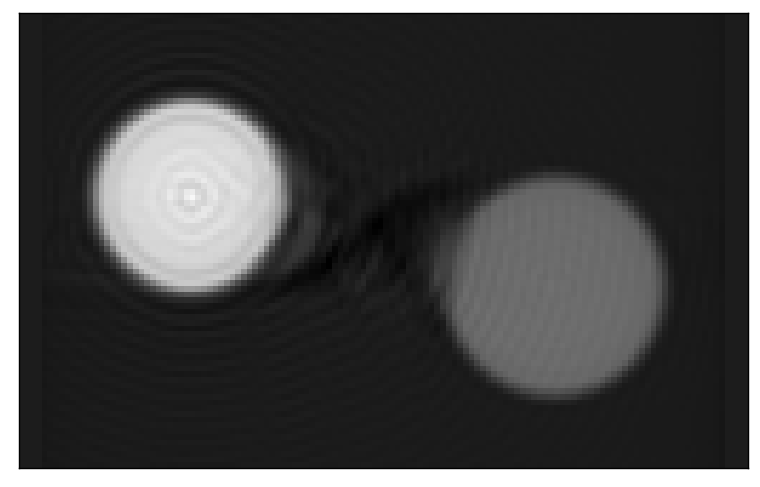} & 
\simprhspt
\includegraphics[simsz]{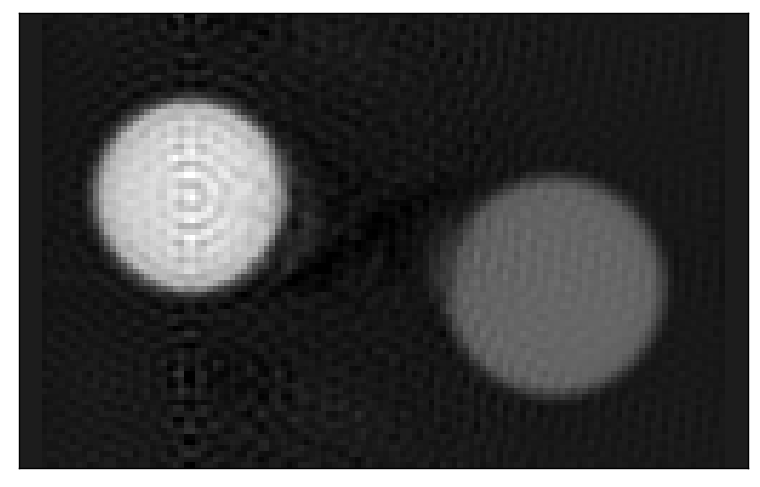} & 
\simprhspt
\includegraphics[simsz]{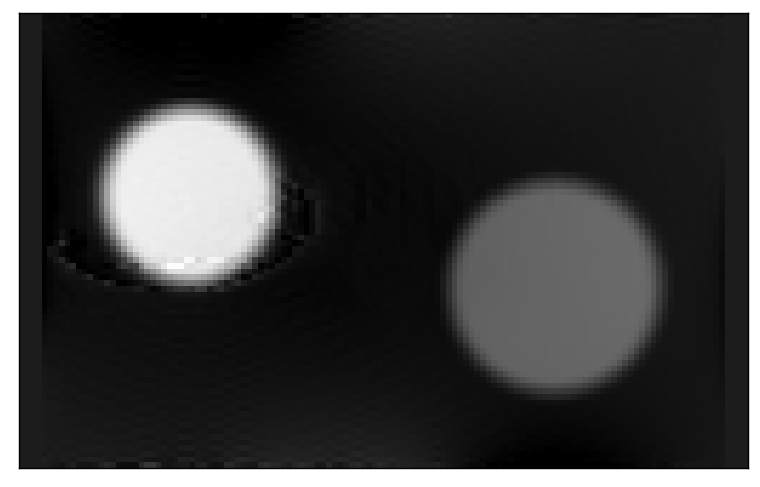} &
\hspace{-0.2in}
\includegraphics[simsz]{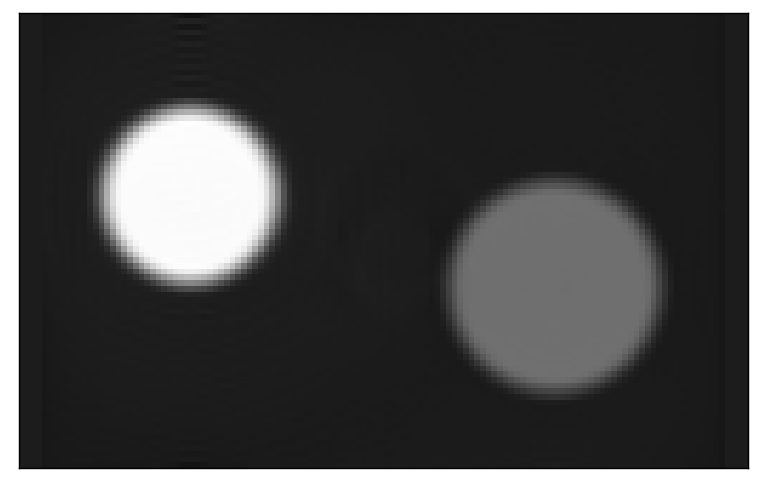} \\ 
\simprhspo (a) TIE & 
\simprhspt (b) CTF & 
\simprhspt (c) Mixed & 
\simprhspt (d) U-NLPR/Zero-Initial & 
\simprhspt (e) U-NLPR/CTF-Initial \\
\end{tabular}
\end{center}
\caption{\label{fig:supsimmultipr} 
Tomographic reconstructions of the refractive index decrement
from the phase images produced by the CTF, TIE, Mixed, and 
U-NLPR (proposed) phase-retrieval (PR) algorithms. 
(a-e) show planar slices along the $u-w$ axes
that pass through the center 
of the reconstruction volume.
(a), (b), and (c) are using the
TIE, CTF, and Mixed PR algorithms respectively.
\textcolor{cgcol}{For the conventional PR methods of TIE, CTF, and Mixed,
we present the best performing reconstruction at the optimal
regularization parameter with the 
highest SSIM (from Fig. \ref{fig:simswpmulti}).}
(d) shows the reconstruction using U-NLPR that
is initialized with zeros for the phase and absorption.
(e) shows the reconstruction using U-NLPR that is
initialized with CTF at the \textcolor{cgcol}{pre-determined}
fixed regularization in equation \eqref{eq:ctfregchoice}.
The gray values in (a-e) are scaled between 
$-2.53\times 10^{-7}$ and $2.03\times 10^{-6}$.
\textcolor{cgcol}{Compared to the conventional PR reconstructions in (a-c), 
U-NLPR reduces artifacts and noise as shown in (d, e) 
without the need for parameter tuning.}
}
\end{figure*}

\textcolor{cgcol}{Fig. \ref{fig:supsimmultipr} is a continuation of the results
from Fig. \ref{fig:simmultipr}.
The simulation scenario for Fig. \ref{fig:supsimmultipr}
and Fig. \ref{fig:simmultipr} are described in
section \ref{ssec:simresults}.
It demonstrates the superior reconstruction performance 
of U-NLPR compared to the conventional PR approaches 
along the $u-w$ slice 
of the refractive index decrement reconstructions.}

\begin{figure*}[!htb]
\begin{center}
\begin{tabular}{ccc}
\hspace{-0.1in}
\includegraphics[width=1.7in]{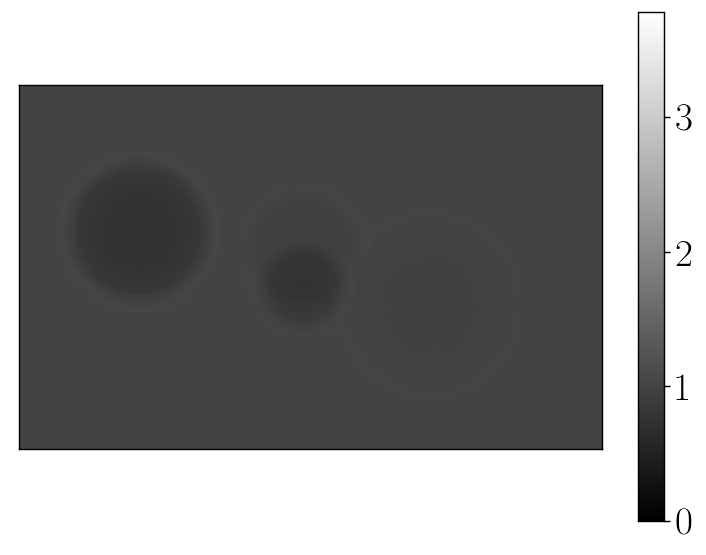} &
\hspace{-0.1in}
\includegraphics[width=1.7in]{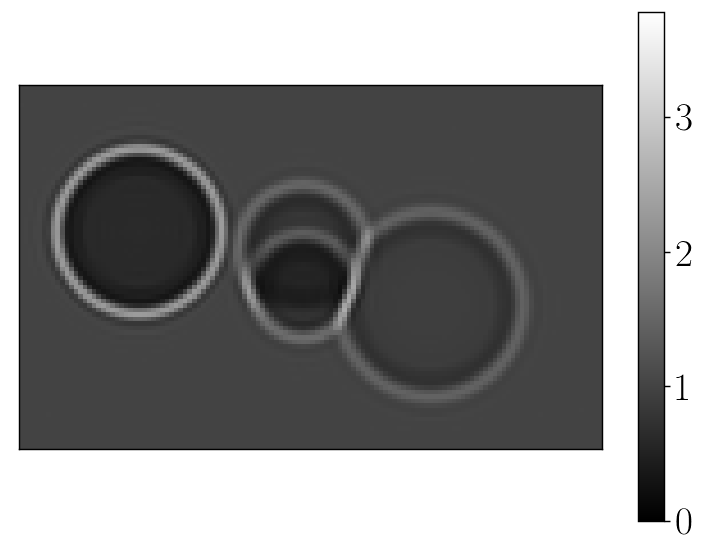} &
\hspace{-0.1in}
\includegraphics[width=1.7in]{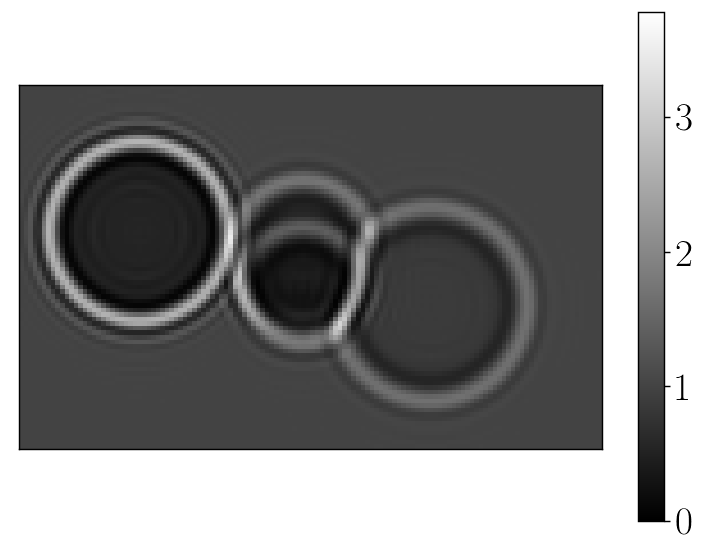} \\
\hspace{-0.1in} (a) $R=10\,mm$ & 
\hspace{-0.1in} (b) $R=400\,mm$& 
\hspace{-0.1in} (c) $R=800\,mm$\\
\end{tabular}
\end{center}
\caption{\label{fig:supsimgtrad-nl}
\textcolor{cgcol}{
Simulated phase-contrast X-ray images
at propagation distances of 
$10\,mm$ ($FN=2.68$), $400\,mm$ ($FN=0.07$), 
and $800\,mm$ ($FN=0.03$)
for a second multi-material object.
This object is multi-material with the same refractive
index decrements as SiC, Teflon, Alumina, and Polyimide 
(section \ref{ssec:simresults}).
Thus, the ground-truth refractive index images
are same as in Fig. \ref{fig:simgtrad}.
The absorption indices were $10\times$ larger 
than those for SiC, Teflon, Alumina, and Polyimide.}
}
\end{figure*}

\begin{figure*}[htb!]
\begin{center}
\begin{tabular}{ccccc}
\multicolumn{5}{c}{$u-v$ axial slice
of the second multi-material object using multi-distance phase-retrieval} \\
\simprhspo
\includegraphics[simsz]{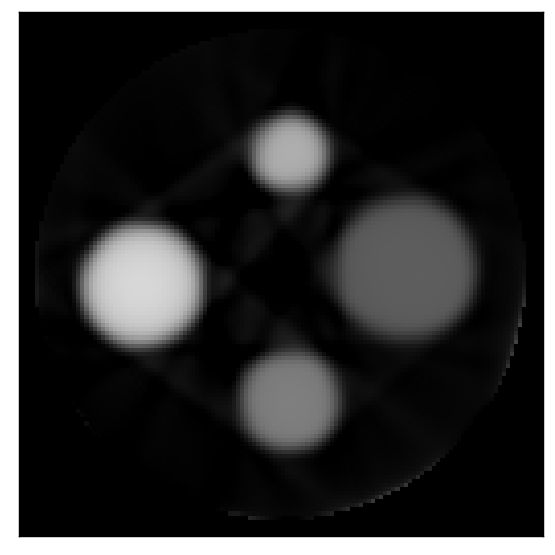} &
\simprhspt
\includegraphics[simsz]{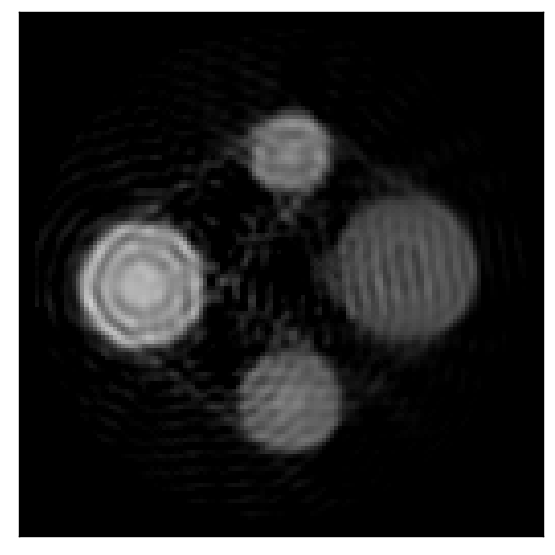} & 
\simprhspt
\includegraphics[simsz]{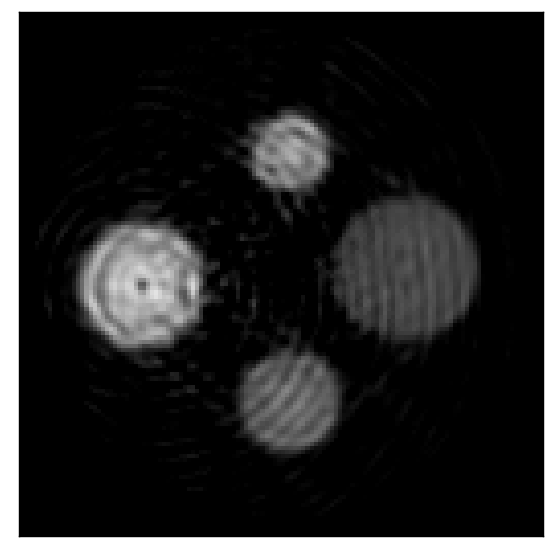} & 
\simprhspt
\includegraphics[simsz]{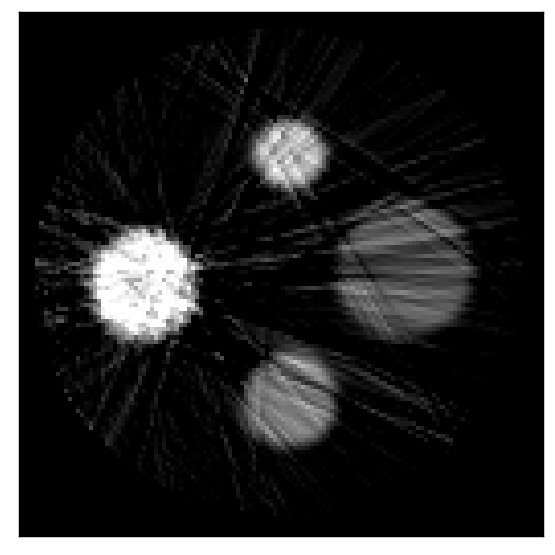} &
\hspace{-0.2in}
\includegraphics[simsz]{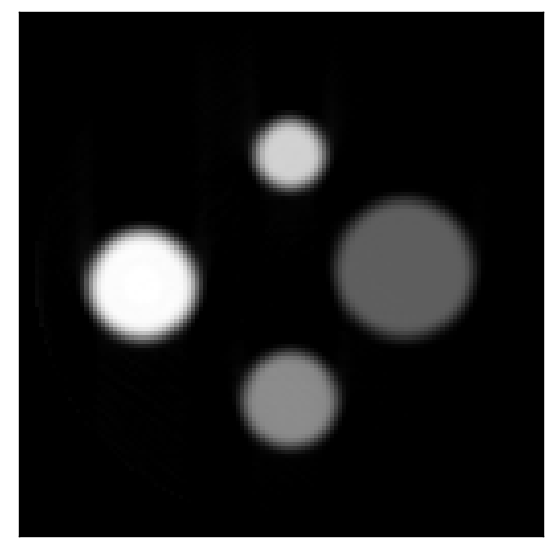} \\ 
\simprhspo (a) TIE & 
\simprhspt (b) CTF & 
\simprhspt (c) Mixed & 
\simprhspt (d) U-NLPR/Zero-Initial & 
\simprhspt (e) U-NLPR/CTF-Initial 
\vspace{0.1in} \\
\multicolumn{5}{c}{$u-w$ axial slice
of the second multi-material object using multi-distance phase-retrieval} \\
\simprhspo
\includegraphics[simsz]{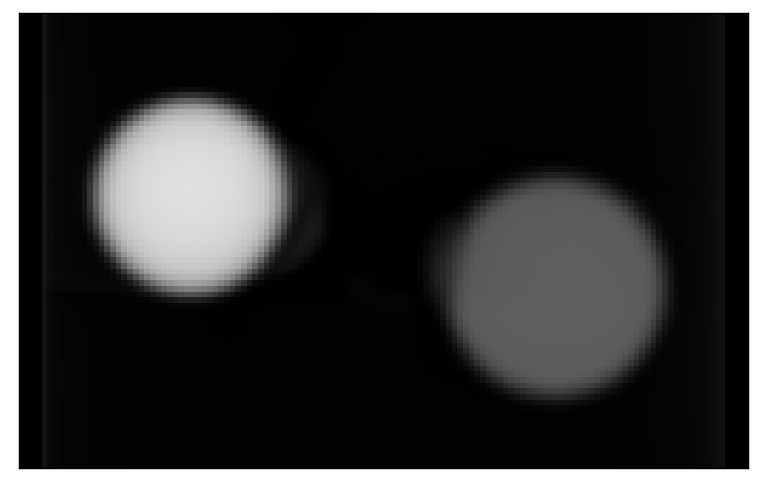} &
\simprhspt
\includegraphics[simsz]{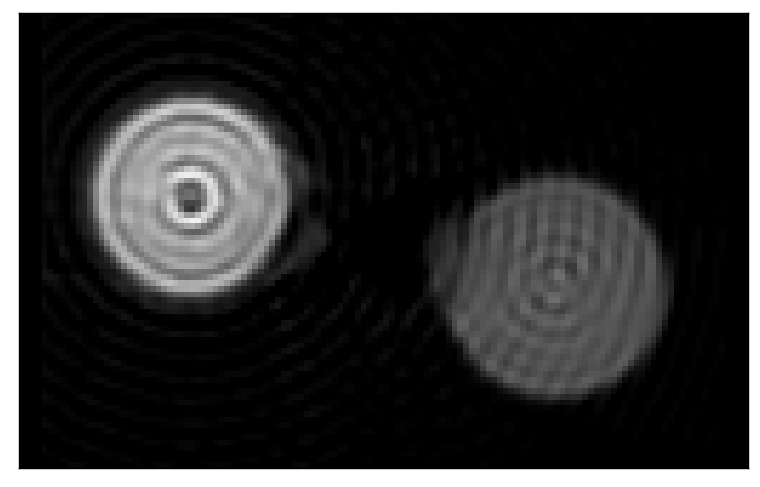} & 
\simprhspt
\includegraphics[simsz]{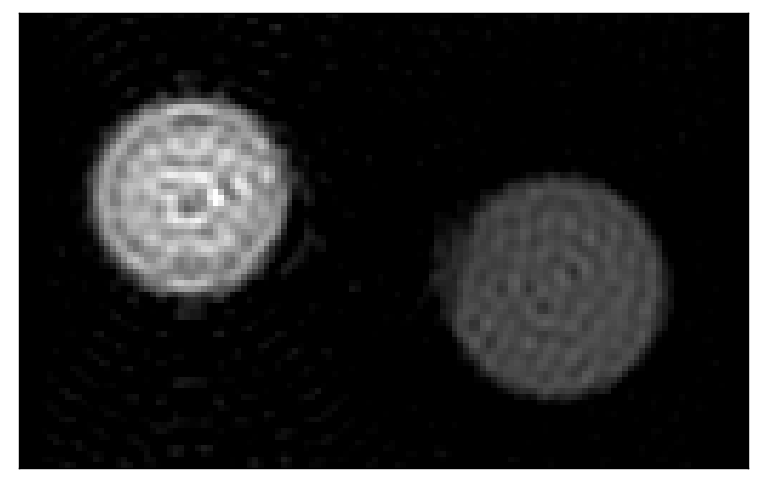} & 
\simprhspt
\includegraphics[simsz]{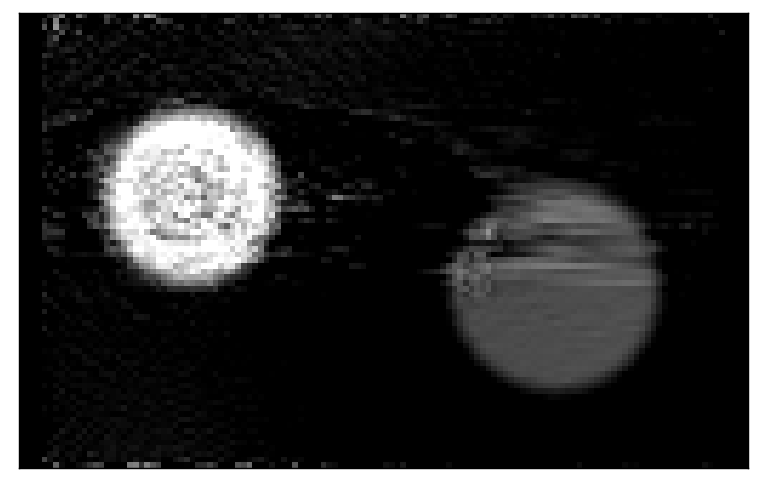} &
\hspace{-0.2in}
\includegraphics[simsz]{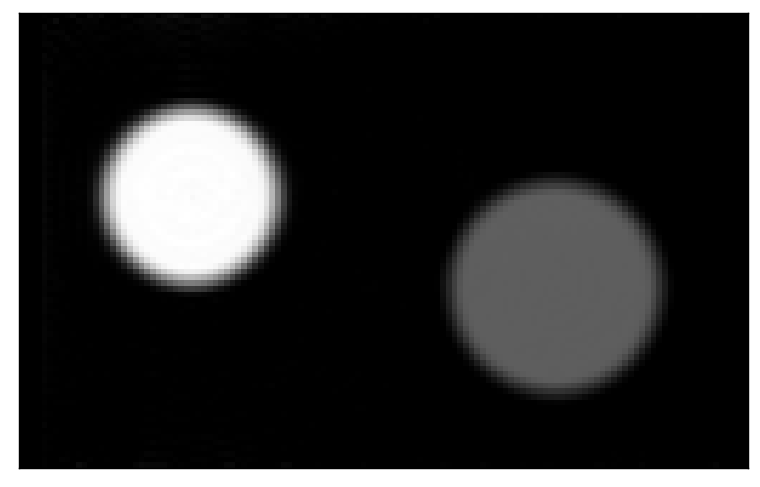} \\ 
\simprhspo (f) TIE & 
\simprhspt (g) CTF & 
\simprhspt (h) Mixed & 
\simprhspt (i) U-NLPR/Zero-Initial & 
\simprhspt (j) U-NLPR/CTF-Initial \\
\end{tabular}
\end{center}
\caption{\label{fig:supsimmultipr-nl} 
\textcolor{cgcol}{
Comparison of multi-distance phase-retrieval algorithms
using simulated phase-contrast CT
at propagation distances of 
$10\,mm$ ($FN=2.68$), $400\,mm$ ($FN=0.07$), 
and $800\,mm$ ($FN=0.03$) (see Fig. \ref{fig:supsimgtrad-nl}).
The object is multi-material with the same refractive
index decrements as SiC, Teflon, Alumina, and Polyimide.
The absorption indices were $10\times$ larger 
than those for SiC, Teflon, Alumina, and Polyimide.
(a-e) and (f-j) show the refractive index decrement reconstructions along the $u-v$ and $u-w$ slices 
respectively.
(a-c, f-h) are the reconstructions using the conventional
methods of TIE, CTF, and Mixed PR at the best 
regularization parameter that maximizes SSIM.
(d,e,i,j) are the reconstructions using U-NLPR. 
U-NLPR/CTF-Initial in (e,j) produces the best
reconstruction using the pre-determined fixed 
regularization (no tuning) from equation \eqref{eq:ctfregchoice}.}
}
\end{figure*}

\begin{figure*}[bht!]
\begin{center}
\begin{tabular}{cc}
\includegraphics[height=2in]{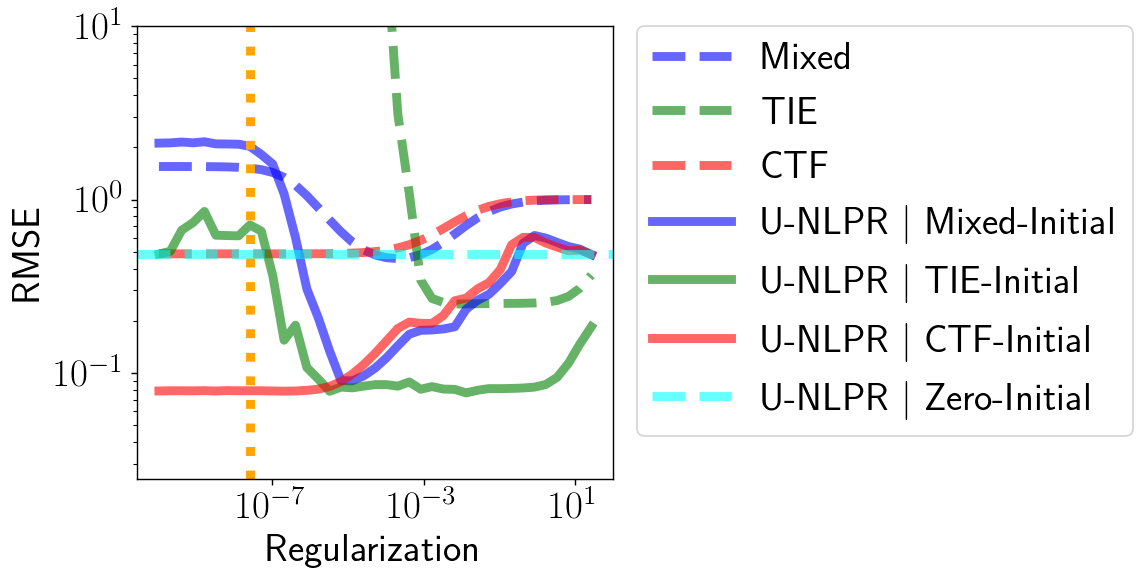} &
\includegraphics[height=2in]{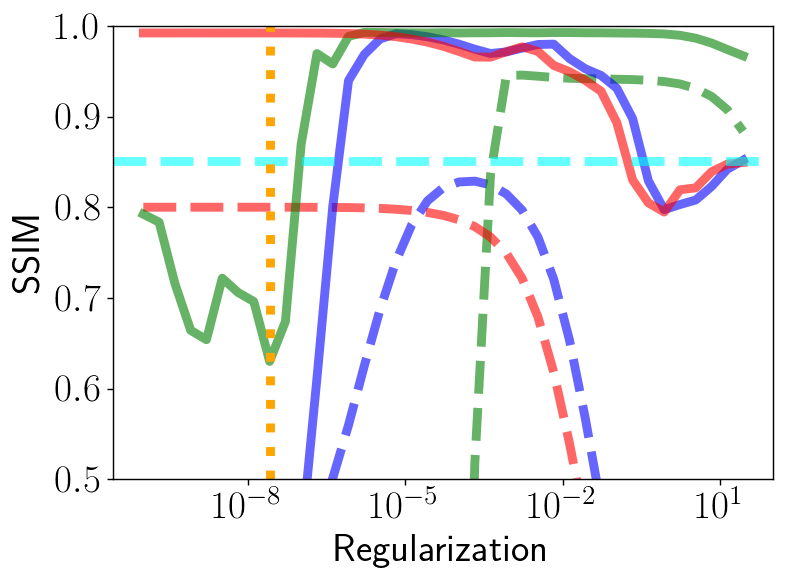} \\
(a) NRMSE vs. Regularization of Conventional PR  &
(b) SSIM vs. Regularization of Conventional PR \\
\end{tabular}
\end{center}
\caption{\label{fig:simswpmulti-nl} 
\textcolor{cgcol}{
Quantitative comparison of the refractive index 
reconstructions for the simulation experiment 
described in Fig. \ref{fig:supsimgtrad-nl} 
and \ref{fig:supsimmultipr-nl}.
(a) and (b) shows the NRMSE and SSIM as a function
of the regularization parameter for the conventional PR.
Since U-NLPR does not use regularization, the
parameter along the horizontal axis is for the conventional
PR methods used as initialization.
U-NLPR consistently out-performs the conventional PR
method used as initialization.
U-NLPR with CTF initialization (U-NLPR/CTF-Initial)
at the pre-determined regularization in equation \eqref{eq:ctfregchoice}
is the best result that also avoids parameter tuning 
(intersection of the vertical orange dotted line and the solid red line).
}}
\end{figure*}

\textcolor{cgcol}{For an analysis of the reconstruction performance
using simulated data with a higher degree of non-linearity,
we simulate phase-contrast images at propagation distances
of $10\,mm$, $400\,mm$, and $800\,mm$ 
as shown in Fig. \ref{fig:supsimgtrad-nl}.
This simulation is for a multi-material object 
with the same refractive indices as SiC, Teflon, Alumina, 
and Polymide, but with $10\times$ higher absorption indices
for the corresponding materials.
From Fig. \ref{fig:supsimmultipr-nl}
and Fig. \ref{fig:simswpmulti-nl}, 
we conclude that U-NLPR with CTF initialization is the 
best performing reconstruction without any artifacts
(for reference, ground-truth images are shown in Fig. \ref{fig:simgtrad}).
It also avoids tuning of the regularization 
parameter by using the pre-determined 
regularization from equation \eqref{eq:ctfregchoice}.}

\subsubsection{Single-distance phase-retrieval}
\begin{figure*}[htb!]
\begin{center}
\begin{tabular}{ccccc}
\multicolumn{3}{c}{Single Distance PR + FBP of Single-Material Object} & \multicolumn{2}{c}{Single Distance PR + FBP of Multi-Materials} \\
\simprhspo \includegraphics[simsz]{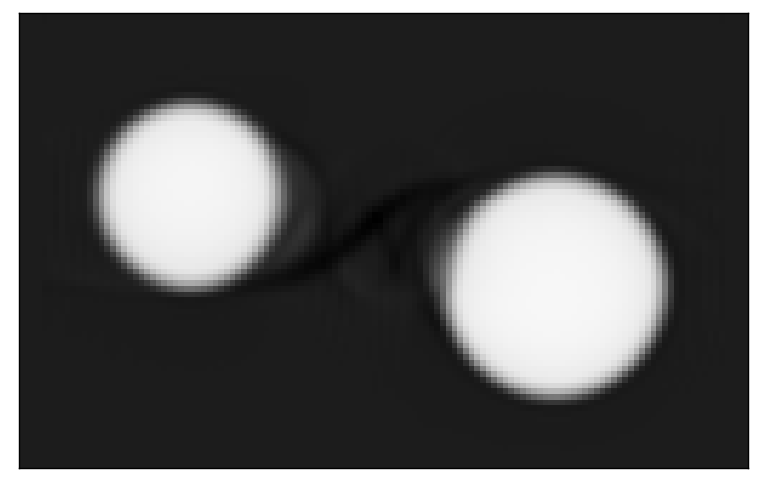} & 
\simprhspt \includegraphics[simsz]{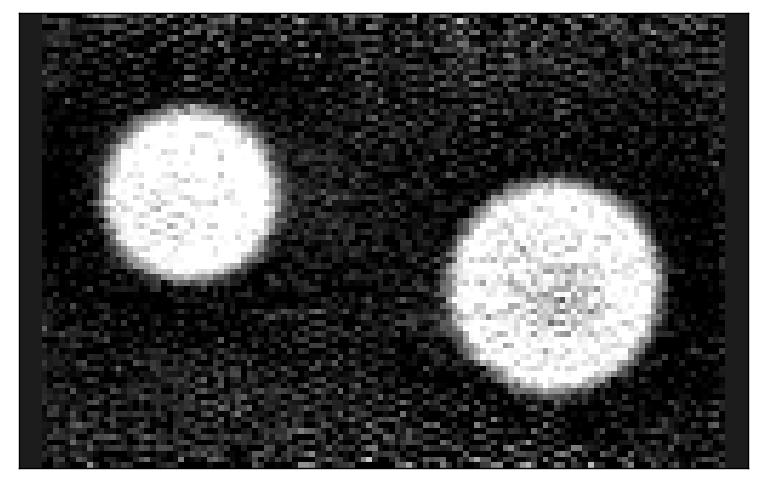} &
\simprhspt \includegraphics[simsz]{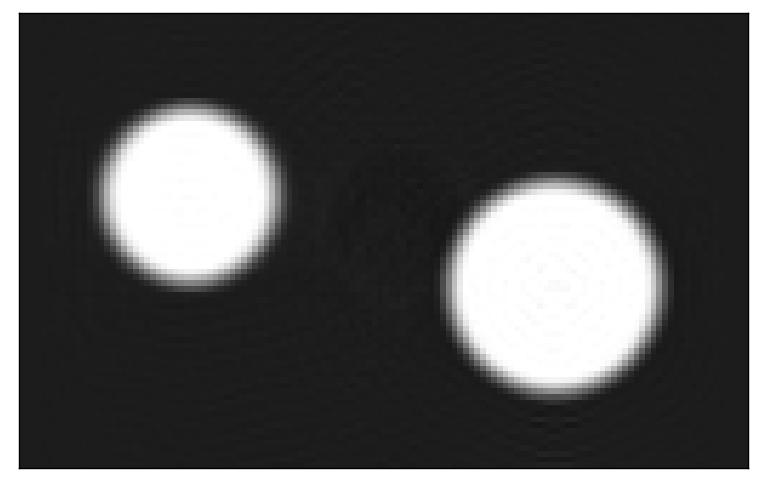} &
\simprhspt \includegraphics[simsz]{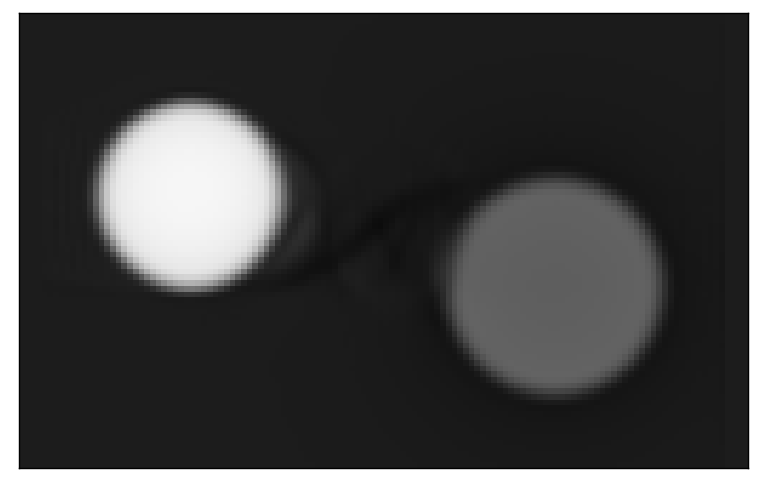} & 
\simprhspt \includegraphics[simsz]{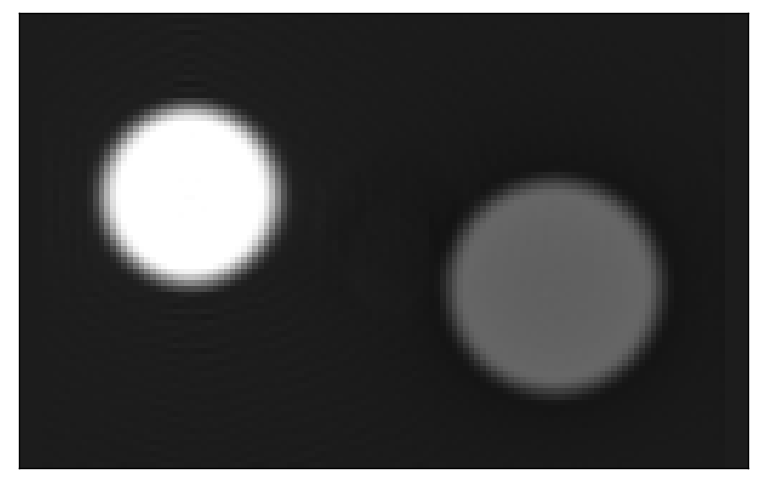} \\
\simprhspo (a) Paganin & 
\simprhspt (b) C-NLPR/0-Initial & 
\simprhspt (c) C-NLPR/Pag-Initial & 
\simprhspt (d) Paganin & 
\simprhspt (e) C-NLPR/Pag-Initial \\
\end{tabular}
\end{center}
\caption{\label{fig:supsimsinglepr} 
Tomographic reconstructions of the refractive index decrement from 
the phase images produced by the single-distance 
phase-retrieval (PR) algorithms of Paganin and C-NLPR/One-$\alpha$ (proposed).
(a-c) show reconstructions of the single material object.
(d, e) show reconstructions of the multi-material object.
(a-e) show planar slices along the 
$u-w$ axes that 
pass through the center of the reconstruction volume.
(a) and (d) show reconstructions using Paganin PR.
(b) shows the reconstruction using C-NLPR that is 
initialized with zeros for the phase image 
(label C-NLPR/0-Initial). 
(c) and (e) show reconstructions using C-NLPR 
that is initialized using Paganin PR (label C-NLPR/Pag-Initial). 
The gray values in (a-e) are scaled between
$-2.09\times 10^{-7}$ and $1.67\times 10^{-6}$.
C-NLPR with Paganin initialization produces the 
best reconstruction that minimize noise and artifacts.
}
\end{figure*}

\textcolor{cgcol}{
Fig. \ref{fig:supsimsinglepr} is a continuation of the
results in Fig. \ref{fig:simsinglepr}.
It demonstrates the superior reconstruction 
performance of C-NLPR with Paganin initialization
along the $u-w$ slice of the refractive index
reconstructions.}

\begin{figure*}[htb!]
\begin{center}
\begin{tabular}{ccc}
\hspace{-0.15in}
\includegraphics[width=1.7in]{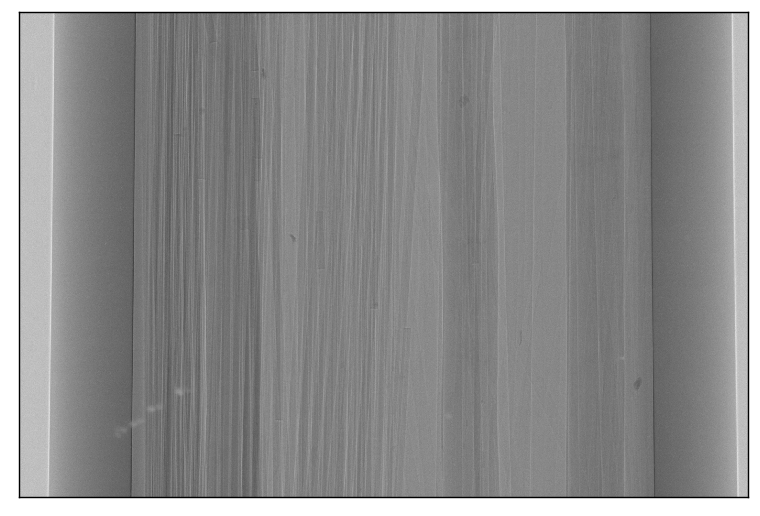} &
\hspace{-0.15in} 
\includegraphics[width=1.7in]{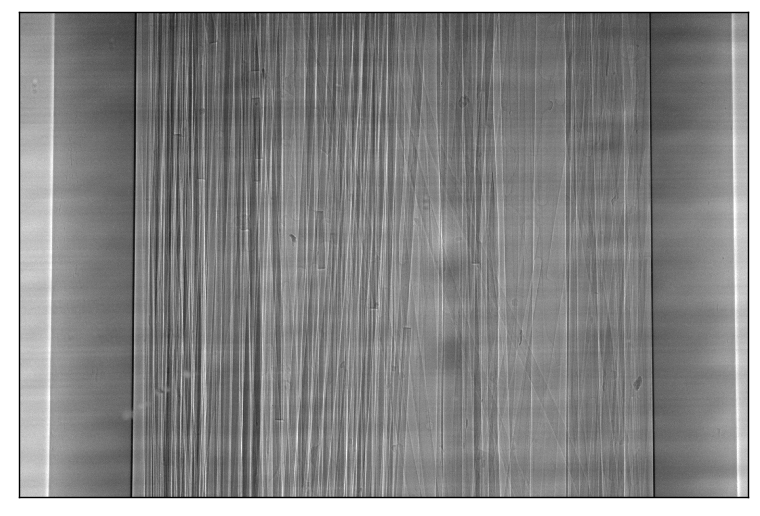} &
\hspace{-0.15in}
\includegraphics[width=1.7in]{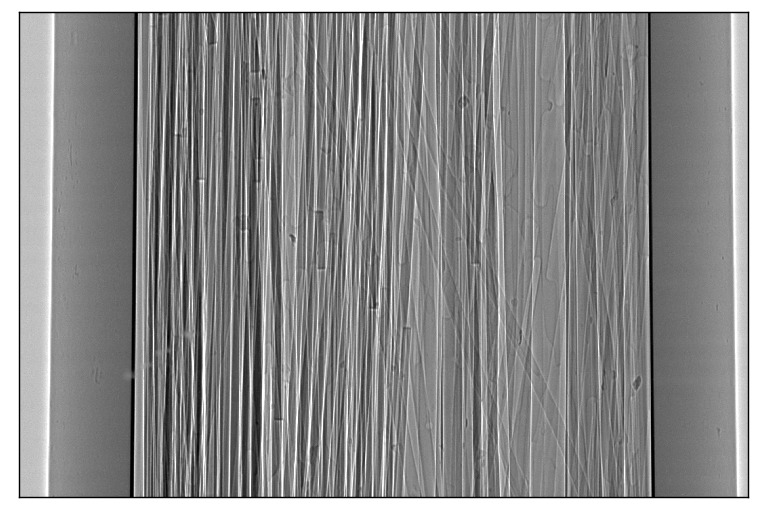} \\
(a) $R=30\,mm$ & (b) $R=100\,mm$ & (c) $R=250\,mm$ \\
\end{tabular}
\end{center}
\caption{\label{fig:supexpmuldata} 
Normalized X-ray images at the first view angle
from synchrotron phase-contrast CT of $Al$, $Al_2O_3$, $PP$, 
and $PET$ fibers.
(a), (b), and (c) show the X-ray images at propagation distances 
of $30\,mm$, $100\,mm$, and $250\,mm$ respectively. 
The intensity range for the gray-values is between $0.4$ and $1.2$.
Phase-contrast fringes are stronger in (c) than in (a) 
due to the larger propagation distance of (c).
}
\end{figure*}

\begin{figure*}[htb!]
\begin{center}
\begin{tabular}{cc}
\includegraphics[width=2.8in]{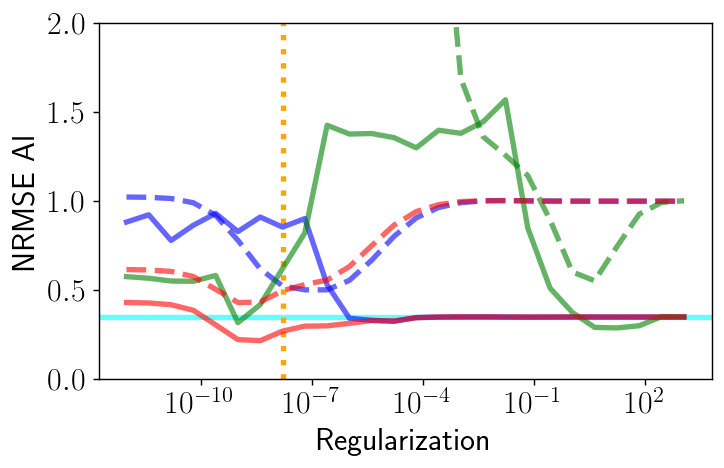} & 
\includegraphics[width=4in]{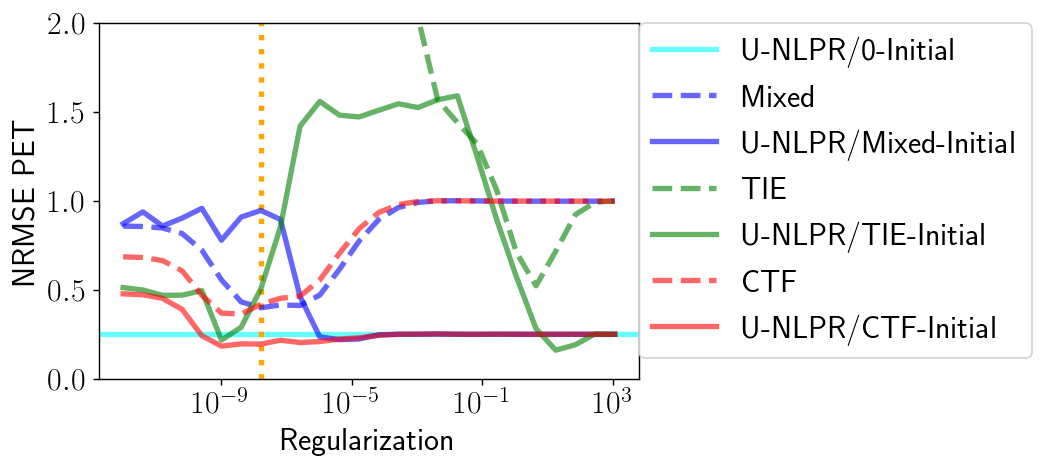} \\
\end{tabular}
\end{center}
\caption{\label{fig:supexpregsweep} 
NRMSE for Al and PET fibers vs. regularization 
of conventional PR methods
for the multi-distance experimental data. 
The dashed lines show the NRMSE for the conventional 
PR methods of TIE, CTF, and Mixed. 
The solid lines are the NRMSE plots for U-NPLR 
that is initialized using one of the conventional 
PR methods.
While U-NLPR does not use regularization, its performance
nevertheless varies with the regularization parameter of the 
conventional PR used for initialization.
U-NLPR out-performs all the conventional PR
methods irrespective of the initialization.
U-NLPR with CTF initialization 
(regularization from equation \eqref{eq:ctfregchoice} 
indicated by the dotted vertical orange line) 
has the best performance
without the need for parameter tuning.
}
\end{figure*}

\makeatletter
\define@key{Gin}{expsz}[true]{%
    \edef\@tempa{{Gin}{width=1.4in, keepaspectratio=true}}%
    \expandafter\setkeys\@tempa
}
\makeatother

\begin{figure*}[htb!]
\begin{center}
\begin{tabular}{cc}
\hspace{-0.15in}
\includegraphics[width=2in]{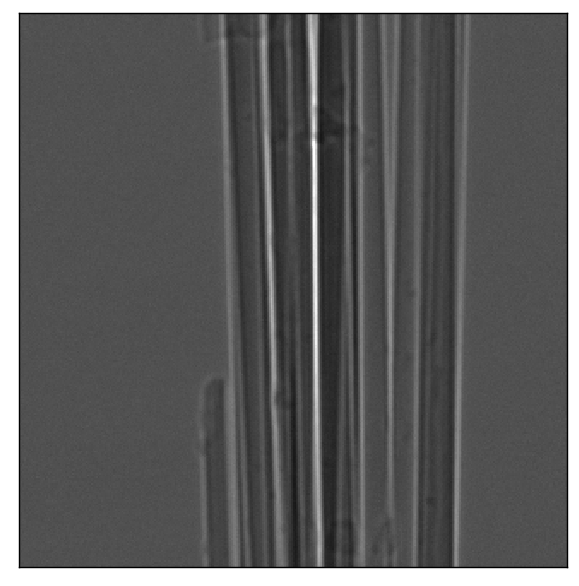} &
\hspace{-0.25in}
\includegraphics[width=2in, height=2in]{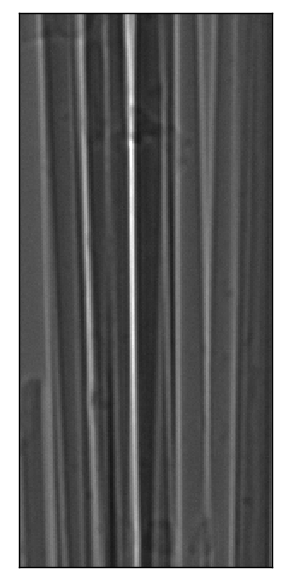} \\
(a) $R=98\,mm$ & (b) $R=98\,mm$/Zoomed \\
\end{tabular}
\end{center}
\caption{\label{fig:supexpsindata} 
(a) is the X-ray image of SiC fibers at the first tomographic 
view and propagation distance of $R=98\,mm$.
(b) zooms into the image in (a) to better visualize the 
phase-contrast fringes. 
The intensity range of gray-values 
is between $0.53$ and $1.97$.
}
\end{figure*}

\makeatletter
\define@key{Gin}{expsinsz}[true]{%
    \edef\@tempa{{Gin}{width=1.6in, keepaspectratio=true}}%
    \expandafter\setkeys\@tempa
}
\makeatother

\begin{figure*}[htb!]
\begin{center}
\begin{tabular}{cc}
\hspace{-0.2in}
\includegraphics[width=4in]{figs/exp/single/mtf_disc0_zsl160.png} &
\hspace{-0.2in}
\includegraphics[width=1.8in]{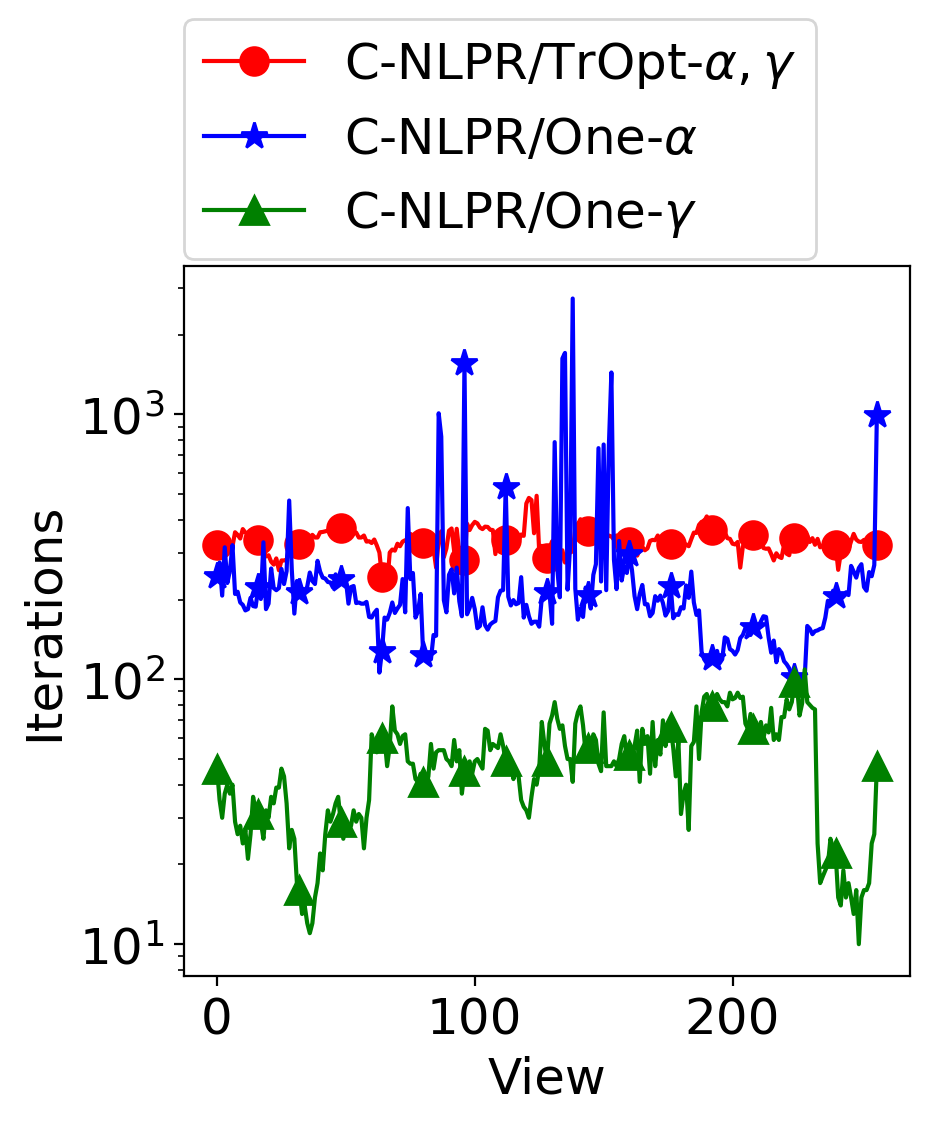} \\
(a) & (b) \\
\end{tabular}
\end{center}
\caption
{\label{fig:supexpsinmtfsweep}
(a) is a measure of sharpness using the modulation 
transfer function (MTF) for various single-distance PR methods.
The MTF is computed for the two circles inside the red box
in Fig. \ref{fig:expsinrec} (a, b).
C-NLPR/One-$\alpha$ and C-NLPR/TrOpt-$\alpha,\gamma$ 
has a higher MTF curve than Paganin and C-NLPR/One-$\gamma$,
which indicates sharper reconstructions. 
(b) is the number of LBFGS iterations 
as a function of the view index.
C-NLPR/One-$\gamma$ uses fewer iterations, 
but also suffers from lower sharpness (from (a)).
C-NLPR/One-$\alpha$ and C-NLPR/TrOpt-$\alpha,\gamma$ 
use similar number of iterations for convergence.
}
\end{figure*}

\subsection{Experimental Data}
\subsubsection{Multi-Distance Phase-Retrieval}
The discussion in this section is continued from 
section \ref{sec:expdatares}.
Fig. \ref{fig:supexpmuldata} shows the X-ray images
for the multi-material experimental sample. 
For quantitative analysis of the reconstructions, 
we compare the normalized root mean squared error (NRMSE) 
for Al and PET fibers in Fig. \ref{fig:supexpregsweep}.
The NRMSE is computed only using the reconstructed values 
in the regions within the Al and PET fibers 
shown in Fig. \ref{fig:supexpregsweep} 
that also exclude the edges of each fiber.
Importantly, to facilitate comparison with the theoretical values, 
we use background subtraction for the reconstructed 
values as explained in section \ref{ssec:quanteval}.
From the dashed plots in Fig. \ref{fig:supexpregsweep},
we can see that choosing a suitable regularization 
for the conventional phase-retrieval methods is challenging
due to the narrow range of parameters with best performance.
As the regularization for CTF phase-retrieval is reduced,
the NRMSE also reduces. 
For very low regularization values, all the conventional 
phase-retrieval methods see a moderate to large increase in NRMSE.
Compared to other conventional methods, 
CTF results in the lowest NRMSE at the lowest regularization parameters.
Hence, we still use equation \eqref{eq:ctfregchoice} 
to secure the best performance
from CTF without any manual regularization parameter tuning.
U-NLPR with zero initialization for the phase
results in an NRMSE that is lower than all conventional methods.
However, U-NLPR with CTF initialization provides even lower
NRMSE than U-NLPR with zero-initialization.

\subsubsection{Single-Distance Phase-Retrieval}
The discussion in this section is continued from
section \ref{sec:expdatares}.
The sharpness improvement using C-NLPR/One-$\alpha$ is also 
reflected in the modulation transfer function (MTF) plot
shown in Fig. \ref{fig:supexpsinmtfsweep} (a).
From the MTF plot, we do not see a sharpness benefit to using
C-NLPR/TrOpt-$\alpha,\gamma$ when compared to C-NLPR/One-$\alpha$.
The convergence speed of C-NLPR/One-$\alpha$ is similar to 
C-NLPR/TrOpt-$\alpha,\gamma$ from Fig. \ref{fig:supexpsinmtfsweep} (b).
While C-NLPR/One-$\gamma$ converges faster, it does not 
improve the sharpness as evidenced in Fig. \ref{fig:supexpsinmtfsweep} (a).
Hence, C-NLPR/One-$\alpha$ is our best choice for the C-NLPR 
reconstructions in Fig. \ref{fig:expsinrec}
since it only uses knowledge of the ratio $\delta/\beta$.

\end{document}